\documentclass[12pt]{iopart}
\usepackage{amssymb}
\usepackage[mathscr]{eucal}
\usepackage{epic}
\usepackage{eepic}
\usepackage[active]{srcltx}
\newtheorem{prop}{Proposition}

\newcommand{\nc}{\newcommand}
\nc{\JStP}{\emph{J. Stat. Phys. }} \nc{\IJMP}{\emph{Int. J. Mod.
Phys. }} \nc{\ds}{\displaystyle} \nc{\WB}{\overline{W}}
\nc{\sqo}{\omega^{1/2}} \nc{\sqoi}{\omega^{-1/2}}
\nc{\TRop}{\widetilde{\Rop}_{\widetilde{1}\widetilde{2}3}}
\nc{\qop}{\mathbf{q}}   \nc{\pop}{\mathbf{p}}
\nc{\mf}{\mathbf{f}_j} \nc{\vop}{\mathbf{v}} \nc{\uop}{\mathbf{u}}
\nc{\wop}{\mathbf{w}} \nc{\Weyl}{\mathfrak{W}}
\nc{\tuop}{\tilde{\mathbf{u}}} \nc{\twop}{\tilde{\mathbf{w}}}
\nc{\tu}{\tilde{u}} \nc{\tw}{\tilde{w}} \nc{\sop}{\mathbf{s}}
\nc{\xop}{\mathbf{x}} \nc{\zop}{\mathbf{z}} \nc{\Xop}{\mathbf{X}}
\nc{\Zop}{\mathbf{Z}} \nc{\A}{\mathcal{A}} \nc{\R}{\mathbf{R}}
\nc{\Rer}{\,\mathbb{R}} \nc{\Rop}{\mathcal{R}}
\nc{\CM}{\mathcal{M}} \nc{\CT}{\mathcal{T}} \nc{\CX}{\mathcal{X}}
\nc{\CY}{\mathcal{Y}} \nc{\CZ}{\mathcal{Z}} \nc{\CA}{\mathcal{A}}
\nc{\CB}{\mathcal{B}} \nc{\CL}{\mathcal{L}}
\nc{\AL}{{\mathcal{L}^{aux}}}
\nc{\Ropf}{\mathcal{R}^{(f)}}\nc{\Oc}{\mathcal{O}}
\nc{\beq}{\begin{equation}}          \nc{\eeq}{\end{equation}}
\nc{\bea}{\begin{eqnarray}}          \nc{\hx}{\hspace{3mm}}
\nc{\eea}{\end{eqnarray}} \nc{\bdm}{\begin{displaymath}}
\nc{\edm}{\end{displaymath}} \nc{\BAR}{\begin{array}}
\nc{\EAR}{\end{array}} \nc{\ny}{\nonumber} \nc{\om}{\omega}
\nc{\sig}{\sigma} \nc{\la}{\lambda} \nc{\hs}{\hspace{1cm}}
\nc{\hq}{\hspace{4mm}} \nc{\al}{\alpha} \nc{\ka}{\kappa}
\nc{\si}{\sigma} \nc{\ra}{\rightarrow}  \nc{\Lam}{\Lambda}
\nc{\lk}{\left(} \nc{\rk}{\right)}   \nc{\Rb}{\right]}
\nc{\T}{\Theta} \nc{\lb}{\left\{}  \nc{\rb}{\right\}}
\nc{\Lb}{\left[} \nc{\V}{\mathbf{v}\,+\,\mathbf{I}}
\nc{\vj}{\mathbf{v}} \nc{\I}{\mathbf{I}} \nc{\wf}{\,\mathfrak{w}}
\nc{\hal}{{\textstyle\frac{1}{2}}} \nc{\rN}{\right)^{\!N}\hn}
\nc{\rg}{\rangle}  \nc{\Lg}{\langle}
\nc{\Rmapf}{\mathcal{R}^{(f)}}

\def\>{\rangle} \def\<{\langle}

\nc{\norm}[1]{\left\Vert#1\right\Vert}
\nc{\abs}[1]{\left\vert#1\right\vert}
\nc{\set}[1]{\left\{#1\right\}}
\nc{\eeee}[4]{\left\{\!\begin{array}{cc}#1&#2\\#3&#4\end{array}\!\right\}}
\nc{\Real}{\mathbb{R}} \nc{\ZZ}{\mathbb{Z}} \nc{\pp}{\mathbf p}
\nc{\qq}{\mathbf q} \nc{\eps}{\varepsilon}
\nc{\To}{\longrightarrow} \nc{\BX}{\mathbf{B}(X)}
\def\>{\,\rangle}
\def\<{\langle\,}
\def\r#1{(\ref{#1})}
\def\a{\alpha}
\def\b{\beta}

\def\Uc{U}\def\Wc{W}\def\Utc{\tilde{U}}\def\Wtc{\tilde{W}}
\def\Kc{K}\def\Ktc{\tilde{K}}
\def\uc{u}\def\utc{\tilde{u}}

\def\fj{f_j}\def\sj{\sigma_j}
\def\P{{P'}}
\def\si{s}
\def\wopt{\tilde{\wop}}\def\uopt{\tilde{\uop}}
\def\pot{\bi{n}}  \nc{\zer}{\bi{\vec{0}}}
\nc{\one}{\bi{e}_1} \nc{\two}{\bi{e}_2} \nc{\thr}{\bi{e}_3}
\let\bi=\mathbf
\nc{\CC}{\mathbb{C}}

\begin{document}
\title{Bazhanov-Stroganov model from 3D approach}
\author{G von Gehlen$^\dag\neg$,~
S Pakuliak$^{\ddag\sharp\natural}$~
 and S Sergeev$^\flat$}
\address{$^\dag$\ Physikalisches Institut der Universit\"at Bonn,
Nussallee 12, D-53115 Bonn, Germany}
\address{$^\neg$ Mathematics Department, University of
Queensland, Brisbane, Qld 4072, Australia}
\address{$^\ddag$\ Bogoliubov Laboratory of Theoretical Physics,
Joint Institute for Nuclear Research, Dubna 141980, Moscow region,
Russia}
\address{$^\sharp$\  Max-Planck-Institut f\"ur Mathematik, Vivatsgasse 7,
D-53111 Bonn, Germany}
\address{$^\natural$\ Institute of Theoretical and Experimental Physics,
Moscow 117259, Russia}
\address{$^\flat$\ Department of Theoretical Physics, Building 59, Research
School of Physical Sciences and Engineering, The Australian
National University, Canberra ACT 0200, Australia}
\ead{gehlen@th.physik.uni-bonn.de, pakuliak@thsun1.jinr.ru,
sergey.sergeev@anu.edu.au}
\begin{abstract}
We apply a 3-dimensional approach to describe a new
parametrization of the $L$-operators for the 2-dimensional
Bazhanov-Stroganov (BS) integrable spin model related to the
chiral Potts model. This parametrization is based on the solution
of the associated classical discrete integrable system. Using a
3-dimensional vertex satisfying a modified tetrahedron equation,
we construct an operator which generalizes the BS quantum
intertwining matrix $\mathsf{S}$. This operator describes the
isospectral deformations of the integrable BS model.
\end{abstract}
%
\section*{Introduction}
The aim of this paper is to describe in detail the interrelation
between specific 3-dimensional (3D) and 2-dimensional (2D)
integrable lattice spin models. In particular, we shall make use
of 3D techniques in order to derive the isospectral
transformations of a 2D model, which would be difficult to find
directly. The 3D model we shall exploit is the generalized
Zamolodchikov-Bazhanov-Baxter (ZBB) model \cite{ZBB1,ZBB2,ZBB3} in
the vertex formulation \cite{SMS95,S,S1,S-Rev}. This model
describes chirally interacting $\mathbb{Z}_N$-spins placed on the
links of a 3D cubic lattice. The corresponding 2D model will be
the integrable Bazhanov-Stroganov (BS) model \cite{BS}, which is
related to the integrable Chiral Potts (CP) model
\cite{BPAuY,CPM-rev1,CPM-rev2}.

We shall use the approach to the generalized ZBB model developed
in \cite{S,S1}. The dynamical variables are affine Weyl pairs
$\uop_j,\wop_j$ which live on the links $j$ of an oriented 3D
lattice. The cornerstone of the approach is the explicit
construction of a canonical map of the triplet of Weyl pairs on
the three incoming links of a vertex to the Weyl pair triplet on
the three outgoing links. This map defines the Boltzmann weights
and by construction satisfies the tetrahedron equation. The
canonical map is uniquely determined postulating a Baxter
$Z$-invariance and a specific linear problem for the Weyl
variables.

The generalized ZBB model with $\mathbb{Z}_N$-spins is obtained
taking the Weyl variables to commute to the $N$-th root of unity.
Then the Weyl centers $\uop_j^N$ and $\wop_j^N$ are classical
dynamical variables of a classical discrete integrable system of
Hirota form which is determined by the canonical mapping and
boundary conditions. This system can be solved using standard
tools of algebraic geometry, i.e. $\Theta$ functions and
Fay-identities. Using the rational limit of the $\Theta$-functions
this is handled in a practical explicit way.

The Weyl operators at $N$-th root of unity are represented by
$N\times N$ matrices. Then the canonical mapping decomposes into a
functional mapping of the centers and a finite-dimensional
transformation given by $N^3\times N^3$ matrix. It was shown in
\cite{S} that the matrix element of this matrix coincide with the
Boltzmann weights of ZBB model. In this approach the ZBB model
appears in the special case that the functional mapping is
trivial, i.e. that trivial solutions to the Hirota-type equations
are chosen. Non-trivial solutions for the functional mapping mean
non-trivial classical dynamics of the Weyl centers, in particular,
solitonic solutions. The final aim is to find separated variables.

It is well-known that the few-layer ZBB model, with quasiperiodic
boundary conditions in the direction orthogonal to the layers, is
related to the integrable chiral Potts model. The main aim of this
paper is the explicit construction of various properties of the
Bazhanov-Stroganov quantum chain directly from the linear problem
and the canonical mapping operator of the generalized ZBB model.
The linear problem leads to the BS $L$-operator. The quantum
intertwining operator of the BS model is obtained as the product
of two 3D canonical mapping operators. In case of the trivial
functional mapping this gives the well-known BS
$\mathsf{S}$-matrix. However, this is generalized if non-trivial
classical dynamics is taken into account. Intertwining through the
whole BS chain leads to isospectrality transforms of the transfer
matrix. A special case of the BS model is the relativistic Toda
chain, for which isospectral transforms have been constructed
already in \cite{SP-RTCh}. An important advantage of the 3D
approach to 2D problems is the flexibility regarding the choice of
parametrization. The CP parametrization turns out to be less
convenient for the dynamical case than a parametrization using
simple cross-ratios and rational $\Theta$-functions.

The paper is organized as follows: In section \ref{Basic} we
summarize the main features and formulae of the models considered.
Then in section \ref{BS-from-3D} the $L$-operator and the quantum
intertwining relation for the BS-model will be derived using the
canonical mapping approach to the ZBB-model. A new parametrization
of the BS-intertwining matrix in terms of cross-ratios is
introduced. In the following section \ref{class-BS} we introduce a
classical counterpart of the BS-model and find the transformation
realizing the intertwining of two Lax-operators, using the
functional mapping of the 3D vertex ZBB-model. Section \ref{main}
starts stating the main result of the paper, the explicit formula
for the isospectral transformation of the BS-model. The proof of
this proposition is given in the following subsections. Section
\ref{discus} summarizes the results.

\section{The 3D and 2D models considered}\label{Basic}
We start with a summary of some basic features of the models
considered in the later sections. This will also serve to
establish the notation.
\subsection{Vertex formulation of the generalized ZBB-model}
In the vertex formulation of the ZBB-model \cite{SMS95} the
quantum variables are attached to the links $j$ of a 3D oriented
lattice. They are taken to be elements $\:(\uop_j,\wop_j)\:$ of an
ultra-local affine Weyl algebra at root of unity: \beq
\uop_j\cdot\wop_j=\om\:\wop_j\cdot\uop_j;\hs\hq\om^N=1;\hs\hq N\in
\mathbb{Z};\hs N\ge 2  \label{weylr}\eeq and
$\;\uop_i\cdot\wop_j=\wop_j\cdot\uop_i
                \hq\mbox{for}\hq i\neq j$.
                 We also attach a scalar variable $\kappa_j$ to each
link $j$ and denote these variables together as \beq
\wf_j\;=\;(\uop_j,\wop_j,\kappa_j\,)\,.\label{wkappa}\eeq
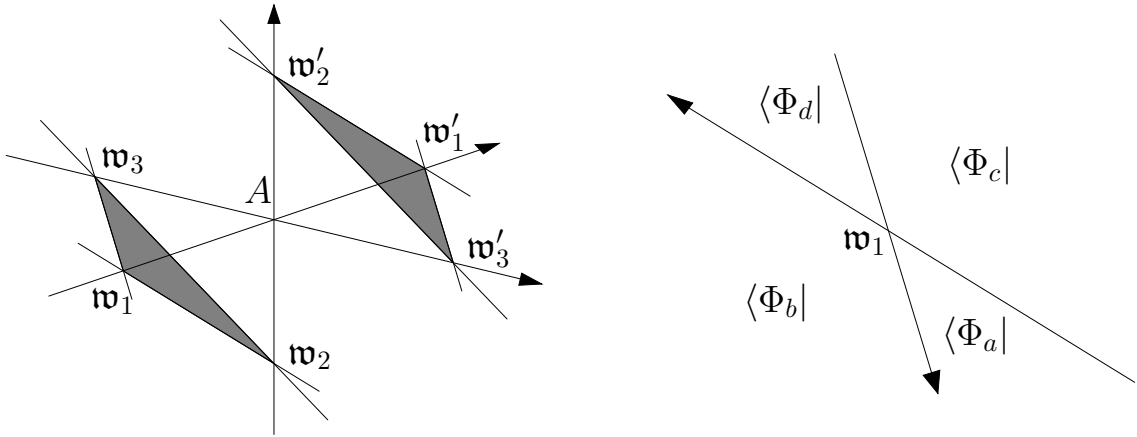
\begin{figure}[ht]
\begin{center}
\renewcommand{\dashlinestretch}{100}
\unitlength=0.07pt
\begin{picture}(4200,2200)(3000,1500)
\shade\path(2648, 2222)(2493, 2732)(3462, 1721)(2648, 2222)
\shade\path(4277, 2776)(3462, 3277)(4431, 2266)(4277, 2776)
\path(4678, 2912)(3462, 2499) \path(3462,1338)(3462,
 2499)(2016,2847)
\path(3462, 2499)(3462, 3660)
\path(2246,2086)(3462,2499)(4908,2151)
\put(2440,2020){\large$\wf_1$} \put(2493,2782){\large$\wf_3$}
\put(3505,1721){\large$\wf_2$}
\put(4230,2920){\large$\wf_1'$}\put(3500,3277){\large$\wf_2'$}
\put(4471,2290){\large$\wf_3'$} \put(3300,2585){\large$A$}
\blacken\path(3430,3560)(3462,3660)(3494,3560)(3430,3560) 
\blacken\path(4785,2145)(4908,2151)(4800,2210)(4785,2145) 
\blacken\path(4595,2855)(4678,2912)(4550,2905)(4595,2855) 
\path(2446.5, 2885.0)(2694.5, 2069.0) \path(2403.8,
2372.3)(3706.2, 1570.7) \path(3752.7, 1417.7)(2202.3, 3035.3)
\path(4477.2, 2113.0)(4230.8, 2929.0) \path(4521.5,
2625.7)(3217.5, 3427.3) \path(3171.3, 3580.3)(4721.7, 1962.7)
\path(6491.5, 3395.0)(7049.5, 1559.0) \path(5589.8,3173.3)(8113,
1620)
\blacken\path(5589.8,3173.3)(5725,3130)(5680,3070)(5589.8,3173.3)  
\blacken\path(7049.5, 1559.0)(6970, 1657)(7060, 1685)(7049.5, 1559.0) 
\put(6500,2340){\large$\wf_1$}
\put(7070,1820){\large$\langle\Phi_a|$}
\put(6000,2000){\large$\langle\Phi_b|$}
\put(7100,2740){\large$\langle\Phi_c|$}
\put(6070,3070){\large$\langle\Phi_d|$}
\end{picture}
\end{center}
\caption{\footnotesize{Left: The six links of the basic lattice
intersecting in the vertex A, intersected by auxiliary planes
(shaded) in two different positions: first passing through
$\wf_1,\,\wf_2,\,\wf_3$ and second through
$\wf_1',\,\wf_2',\,\wf_3'$. The second position is obtained from
the first by moving the auxiliary plane parallel through the
vertex $A$. The Weyl variables, elements of $\wf_i,\,\wf_i',$ live
on the links of the basic lattice. $\Rop_{123}$ can be considered
to be attached to the vertex $A$, it maps the left auxiliary
triangle onto the upper right one. Right: the auxiliary plane in
the neighborhood of $\wf_1$, showing the "co-currents" $\langle
\Phi_a|,\;\ldots\;,\langle\Phi_d|$ in the four sectors cut out by
the directed lines $\stackrel{\longrightarrow}{\wop_2\wop_1}$ and
$\stackrel{\longrightarrow}{\wop_3\wop_1}$. The Linear Problem
relates these four adjacent co-currents according to the values of
$\wf_1=(\uop_1,\wop_1,\kappa_1)$.}} \label{tri}\end{figure}
Fig. \ref{tri} shows on the left the three Weyl variables $\wf_j$
on the ingoing links of a vertex A and the three variables
$\wf_j'$ on the corresponding outgoing links.

In the approach of \cite{S} the basic object of the generalized
vertex ZBB-model is the operator $\Rop_{123}$ mapping canonically
the triple affine Weyl algebra on the ingoing links to the
corresponding triple Weyl algebra on the outgoing links. This
mapping is an invertible rational mapping: For any rational
function $\;\Psi\;$ of the $\;\uop_1,\ldots,\wop_3,$ we define
\beq  \lk\Rop_{123}\circ\Psi\rk
(\uop_1,\wop_1,\uop_2,\ldots,\wop_3)\;\stackrel{\rm def}{=}\;
\Psi(\uop_1',\wop_1',\uop_2',\ldots,\wop_3').  \label{Rop}\eeq
In \cite{S,S1}, $\Rop_{123}$ has been determined uniquely
postulating a Baxter $Z$-invariance (lines may be shifted with
respect to each other) and a linear relation ("Linear Problem")
between the four "co-currents" attached to the four adjacent
sectors around each $\wf_j$ in the auxiliary plane. The right hand
part of Fig. \ref{tri} shows the auxiliary plane in the
neighborhood of $\wf_1$ and the corresponding four co-currents.
The "Linear Problem" at $\wf_1$ is taken to be \beq 0\;=\;\langle
\Phi_a|\;+\om^{1/2}\langle\Phi_b|\:\uop_1\;
 +\langle \Phi_c|\:\wop_1\;+\kappa_1\langle \Phi_d|\:\uop_1\wop_1\,,
 \label{lipro} \eeq
analogously at all links $j$. The lines in the auxiliary plane
have a direction (we shall not go into the rule here) so that e.g.
the co-current appearing in (\ref{lipro}) multiplied by
$\om^{1/2}\uop_1$ is the one {\it between} the arrows.

For the 2D auxiliary plane the relation (\ref{lipro}) contains
analog information as does in the standard 1D quantum chain case
the QIS $L$-operator relation
$$\langle\Phi^{(k)}|\,L^{(k)}(x)\;=\;\langle\Phi^{(k+1)}|\;\ell^{(k)}.$$

Before giving the explicit formula for the mapping operator, we
use the fact that for $\om$ a $N$-th root of unity, the affine
Weyl operators (\ref{weylr}) can be represented by $N\times N$
matrices. Omitting for a moment the index $j$ (because of the
ultralocality the full representation space is just a direct
product), we write \beq \label{Weyl} \uop\:\equiv
u\:\Xop\,;\quad\quad \; \wop \equiv\: w\:\Zop\,; \hs u,\:w\,\in
\mathbb{C};
     \hs\Xop\:\Zop\;=\;\om\,\Zop\:\Xop,\label{Nrep}\eeq
and shall use the natural basis \beq\label{Weyl-finite} \ds
\Xop\;|\beta\>\;=\;\om^\beta\;|\beta\>\;;\hs \Zop\;|\beta\> \;=\;
|\beta+1\>\;;\hs \<\al|\,\beta\> \;=\; \delta_{\al,\beta}\;.\eeq
Clearly $\Xop^N\:=\:\Zop^N\:=\:1\,.$ The $N$-th powers of the Weyl
elements are centers and we shall denote them by $\:U_j,\;W_j\,$:
\beq \uop_j^N\;=\;u_j^N\;\equiv\;
U_j;\hs\wop_j^N\;=\;w_j^N\;\equiv\; W_j. \label{Capcent}\eeq
Now  $\:(\uop_j+\wop_j\,)^N\,=\: U_j\,+\, W_j\,,$ and the mapping
$\Rop_{123}$ implies a purely {\it functional} mapping
$\;\Rop_{123}^{(f)}\;$ of the centers $\;U_j,\;W_j\:,$ or taking
$N$-th roots, of the $\;u_j,\;w_j\,$:
\beq \lk\Rop_{123}^{(f)}\circ\psi\rk
(u_1,\,w_1,\,u_2,\,\ldots,\,w_3)\;\stackrel{def}{=}\;
\psi(u_1',\:w_1',\:u_2',\:\ldots\:,w_3')\,. \label{Rop-f}\eeq
The remarkable feature (observed in \cite{BaBoRe})~ arises that
$\:\Rop_{123}\:$ decomposes into a matrix conjugation
$\:\R_{123}\:$ (this is a $\:N^3\times N^3$-matix)~ and a purely
functional mapping $\Rop_{123}^{(f)}\:$: \beq
\Rop_{123}\:\circ\:\Psi\;=\;\R_{123}^{}\;\lk
\Rop^{(f)}_{123}\circ\Psi\rk\;\R_{123}^{-1}\:.\label{frfr}\eeq

The matrix $\R_{123}$ can be given compactly in terms of the
Bazhanov-Baxter cyclic functions $\:w_p(n)\:$ (not to be confused
with $w$ in (\ref{Nrep})) which depend on the
$\mathbb{Z}_N$-variable $n$ and on a point $\,p=(x,\,y)\,$
restricted to a Fermat curve:
\beq  \fl
\frac{w_p(n)}{w_p(n-1)}\;=\;\frac{y}{1\,-\,\om^n\,x}\,;\hq\;\;\;\;
x^N\,+\,y^N\,=\,1\,;\;\;
       \hq\;\; n\in\ZZ_N;\hq n\ge 1;\hq w_p(0)=1\,. \label{Fermat} \eeq
The cyclic property $\;w_{p}(n+N)=w_{p}(n)\;$ is guaranteed by the
Fermat curve restriction in \r{Fermat}. The functions $\:w_p(n)\:$
are root of unity analogs of $q$-gamma functions and can be used
to develop the theory of the corresponding $q$-hypergeometric
functions (see e.g. \cite{SMS95,Hyper-Yang-Perk}). One can show
\cite{S,S1,S-Rev,gps} that $\:\R_{123}\:$ can be written as a
weighted cross-ratio of four of these cyclic functions. In
components: \beq\label{R-mat}
\mathbf{R}_{i_1i_2i_3}^{j_1j_2j_3}\;=\;\delta_{i_2+i_3,j_2+j_3}
\om^{(j_1-i_1)j_3}
\frac{w_{p_1}(i_2-i_1)w_{p_2}(j_2-j_1)}{w_{p_3}(j_2-i_1)w_{p_4}(i_2-j_1)}\,.
\eeq Here  $p_i=(x_i,y_i)$, $i=1,2,3,4\;$ are four points on the
Fermat curve (\ref{Fermat}) which are related by the constraint
\cite{SMS95} \beq x_1\,x_2\,=\,\om\,x_3\,x_4\,.\label{xxoxx} \eeq
These Fermat points can be expressed in terms of the parameters
$\;u_j,\,w_j,\,\kappa_j,\:u_j',\,w_j'\;$ of the initial and final
Weyl pairs $\,\wf_j,\,\wf_j'\;(j=1,2,3):$ \bea
x_1=\sqoi\frac{u_2}{\kappa_1u_1}\;;\;\;
x_2=\sqoi\frac{\kappa_2u_2'}{u_1'}\;;\;\;
x_3=\omega^{-1}\frac{u_2'}{u_1}\;;\ny\\
\frac{y_3}{y_1}=\frac{\kappa_1w_1}{u_3'}\;;\;\;\;
\frac{y_4}{y_1}=\omega^{-1/2}\frac{\kappa_3
w_3}{w_2}\,.\label{fer-u} \eea
${u_1'}^N,\,{u_2'}^N,\,{u_3'}^N$ are related to the initial
variables by the functional transformation (\ref{Rop-f}). Defining
$\;K_j\equiv \kappa_j^N\,,$ the mapping $\;\Rop_{123}^{(f)}\;$ is
explicitly given by
%
\beq \label{clas7} \BAR{rcl} \ds \frac{\Uc'_1}{\Uc_1}=
\frac{\Wc'_3}{\Wc_3}&=&\ds
\frac{\Kc_2\Uc_2\Wc_2}{\Kc_1\Uc_1\Wc_2+\Kc_3\Uc_2\Wc_3+\Kc_1\Kc_3\Uc_1\Wc_3}
\,;\\[4mm]
\ds \frac{\Wc_1}{\Wc'_1}= \frac{\Wc'_2}{\Wc_2}&=&\ds
\frac{\Wc_1\Wc_3}{\Wc_1\Wc_2+\Uc_3\Wc_2+\Kc_3\Uc_3\Wc_3}
\,;\\[4mm]
\ds \frac{\Uc'_2}{\Uc_2}\:=\: \frac{\Uc_3}{\Uc'_3}&=&\ds
\frac{\Uc_1\Uc_3}{\Uc_2\Uc_3+\Uc_2\Wc_1+\Kc_1\Uc_1\Wc_1}\,. \EAR
\eeq
%
If we need the $u_j'$ rather than the $U_j'$, which is the case
when we calculate the Fermat points via \r{fer-u}, we have to take
$N$-th roots. The choice of phases is restricted by the fact that
the complete mapping $\Rop_{123}$ leaves the following three
products invariant \cite{gps}: \beq
\wop_1\,\wop_2=\wop_1'\,\wop_2';\hs\uop_2\,\uop_3=\uop_2'\,\uop_3';\hs
                  \uop_1\,\wop_3^{-1}=\uop_1'{\wop_3'}^{-1}.
                  \label{R-center}\eeq
Considering a 3D model of $\,\mathbb{Z}_N\,$ spins on the links of
the lattice, the $\:\mathbf{R}_{i_1i_2i_3}^{j_1j_2j_3}\:$ can be
taken to be the (generally not positive) Boltzmann weights of the
vertices. Via the Fermat parameters, these depend on the scalar
parameters $u_1,\,u_2,\ldots,\kappa_3$. Each solution of the
functional equations gives rise to an integrable 3D model.

It can be seen, that by construction, the mapping $\:\Rop_{123}\:$
satisfies the tetrahedron equation and that $\R_{123}\:$ solves
the modified tetrahedron equation, see e.g. \cite{gps1}.

\subsection{Integrable Chiral Potts model}

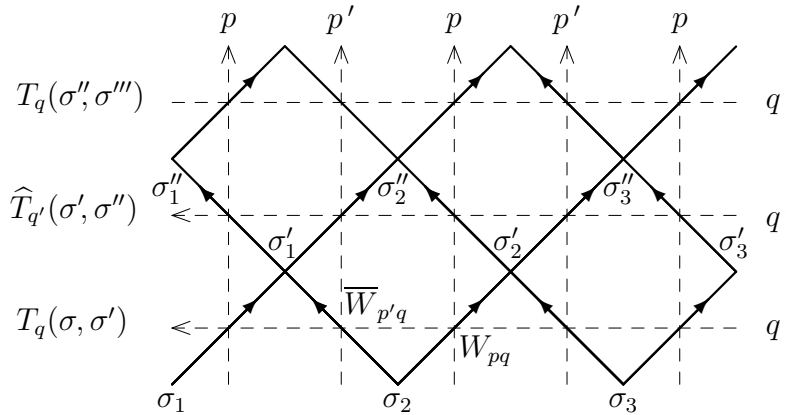
\begin{figure}[ht]
\setlength{\unitlength}{0.3mm}
\begin{picture}(420,156)(-160,-5)
\renewcommand{\dashlinestretch}{30}
\thicklines \multiput(0,0)(100,0){2}{\vector(1,1){138}}
\multiput(0,0)(100,0){2}{\vector(1,1){88}}
\multiput(0,0)(100,0){2}{\vector(1,1){38}}
\put(200,0){\vector(1,1){38}}
\put(100,0){\line(-1,1){100}} \put(0,100){\line(1,1){50}}
\put(0,100){\vector(1,1){38}} \put(200,0){\line(1,1){50}}
\put(250,50){\vector(-1,1){88}} \put(200,0){\vector(-1,1){88}}
\put(200,0){\vector(-1,1){88}} \put(250,50){\vector(-1,1){38}}
\put(100,0){\vector(-1,1){88}} \put(100,0){\vector(-1,1){88}}
\put(200,0){\vector(-1,1){38}}  \put(100,0){\vector(-1,1){38}}
\put(100,0){\vector(-1,1){38}} \multiputlist(0,
-8)(100,0){$\sigma_1$,$\sigma_2$,$\sigma_3$}
\multiputlist(49,63)(100,0){$\sigma_1'$,$\sigma_2'$,$\sigma_3'$}
\multiputlist(-2,87)(100,0){$\sigma_1''$,$\sigma_2''$,$\sigma_3''$}
      \put(76.2,31){$\overline{W}_{p'q}$} \put(127,12){$W_{pq}$}
\multiputlist(25,160)(100,0)[c]{$p$,$p$,$p$}
\multiputlist(75,161.8)(100,0)[c]{${p\,}'$,${p\,}'$}
\multiputlist(263,24)(0,50)[l]{$q$,${q\,}'$,$q$}
\put(-69,24){$T_q(\sigma,\sigma')$}
\put(-69,124){$T_{q}(\sigma''\!\!,\sigma''')$}
\put(-72,74){$\widehat{T}_{q'}(\sigma'\!,\sigma'')$}
\multiput(0,0)(100,0){2}{\line(1,1){150}}
\put(200,0){\line(-1,1){150}}  \put(200,100){\line(-1,1){50}}
\thinlines \multiput(0,0)(50,0){5}{\dashline{6}(25,0)(25,150)}
\multiput(0,0)(50,0){5}{\path(22,142.5)(25,150)(28,142.5)}
\multiput(0,25)(0,50){3}{\dashline{6}(0,0)(250,0)}
\multiput(0,25)(0,50){2}{\path(8,-3)(0,0)(8,3)}
\end{picture}
\caption{\footnotesize{The square diagonal directed lattice with
the
 $\mathbb{Z}_N$-variables at the vertices and the Boltzmann weights $W_{pq}$ on
 the right pointing links and $\overline{W}_{pq}$ on the left pointing links.
 The weight corresponding to the line from $\sig_2$ to $\sig_2'$ is
 $W_{pq}(\sig_2-\sig_2')$, analogous for the
 link from $\sig_2$ to $\sig_1'$ it is $\overline{W}_{p'q}(\sig_2-\sig_1')$.
 Dashed lines are the rapidity lines which indicate the parameter dependence of
 the Boltzmann weights. For simplicity, we show only the special case of
 alternating rapidities $\:p,p'$ and $q,q'$. There are two different transfer
 matrices $T$ and $\widehat{T}$.}}
\label{Wlattice}\end{figure}

The integrable chiral Potts model (CP) is defined on a 2D lattice
with $\:\mathbb{Z}_N$ spins $\:\sig_j\:$ attached to the vertices.
There are two sets of directed rapidity lines $p,\:p',$ and
$q,\:q'$ which cross on the links of the lattice, see Fig.
\ref{Wlattice}. If the edge linking the spins is between the
rapidity directions, the pair interaction Boltzmann weight (which
depends on both rapidities crossing on the link) is
$\WB_{pq}(\sig_j-\sig_{j'})$. If the edge is to the right of the
rapidity directions, it is $W_{pq}(\sig_j-\sig_{j'})$. Such
Boltzmann weights which satisfy the star-triangle relation (for
the explicit form of $\:R_{pqr}\,$ see e.g. \cite{M-S}) \beq
\fl\sum_d\WB_{qr}(d-b)W_{pr}(d-a)\WB_{pq}(c-d)\;=\;
           R_{pqr}\:W_{pq}(b-a)\WB_{pr}(c-b)W_{qr}(c-a)\label{StTri}\eeq
have been constructed in \cite{BPAuY} and are defined by the
difference relations ($n\in\ZZ_N$) \beq \fl {W_{pq}(n)\over
W_{pq}(n-1)} \;=\; \lk{\mu_p\over\mu_q}\rk\;{y_q \,-\, \om^n\, x_p
\over y_p \,-\, \om^n\, x_q}\,;\hs
 {\overline{W}_{pq}(n)\over \overline{W}_{pq}(n-1)} \;=\;
  \lk\mu_p\,\mu_q\rk\;{\om\, x_p \,-\, \om^n\, x_q \over y_q
\,-\, \om^n\, y_p}\,.\label{Wpq}\eeq We shall use the
normalization $\;W_{pq}(0)\;=\;\WB_{pq}(0)\;=\:1.$ The parameters
appearing in (\ref{Wpq}) are constrained to the high-genus curve
\beq \fl x_q^N+y_q^N\,=\,k\,(x_q^N\,y_q^N\,+\,1);\hs\hx k\,
x_q^N=1-k'\mu_q^{-N};\hs\hx
        k\,y_q^N=1-k'\mu_q^N, \label{Bxteq}\eeq
\noindent (same for $\:x_p,\:y_p,\:\mu_p$) where $k$ and $k'$ are
fixed temperature-like parameters related by $k^2+k'^2=1\,.\;\:$
The constraints (\ref{Bxteq}) guarantee the cyclic property \beq
\label{cycW} W_{pq}(n\,+N)=W_{pq}(n);\hs\hs
\WB_{pq}(n\,+N)=\WB_{pq}(n)\,.\eeq The star-triangle relations
(\ref{StTri}) are quite special since the three rapidities
involved appear not as differences as usual, but each separately.
Due to this feature, many standard techniques cannot be applied
straightforwardly to the CP model. Functional relations involving
the BS-model discussed in the following have been crucial for
obtaining analytic solutions for the CP-model, see e.g.
 \cite{BBP, CPM-sol, CPM-sol1, Baxt-sol, Roan} and references therein.

\subsection{Bazhanov-Stroganov model}
The CP model has been found to be intimately related to the
six-vertex model in a seminal paper by Bazhanov and Stroganov
\cite{BS}. They first noticed that the twisted six-vertex R-matrix
\begin{equation}
\ds R(\la,\nu)\;=\;\left(\begin{array}{cccc}
\la-\om\nu & 0 & 0 & 0 \\[1mm]
0 & \om(\la-\nu) & \la (1-\om) & 0 \\[1mm]
0 & \nu (1-\om) & \la-\nu & 0 \\[1mm]
0 & 0 & 0 & \la-\om\nu \end{array}\right)\;.
\end{equation}
at root of unity $\;\om=e^{2\pi i/N}\:$ intertwines not only the
six-vertex $L$-operator, but also the following $L$-operators
containing Weyl elements $\Xop,\:\Zop\,$ (see also \cite{Kor}):
\beq L(\la,\mathbf{a})\;=\;\lk\BAR{cc} 1+\la\,b_1\Zop;&
 \la\,\Xop^{-1}(a_1-b_2\Zop)\\&\\[-2mm]
 \Xop(a_2-b_3\Zop);& \la\, a_1a_2+b_2b_3b_1^{-1}\Zop\EAR \rk;
 \quad \Xop\,\Zop=\om\,\Zop\,\Xop\,,  \label{SEL}\eeq
 where $\la,\:a_1,\:\ldots\,,\,b_3\:\in \mathbb{C}\,.\;$
  We
 collectively denote the parameters $\:a_1,\ldots, \:b_3\:$ by $\:\mathbf{a}$.
In the representation (\ref{Weyl-finite}) the intertwining relation is\\[-1mm]
\beq \fl \sum_{j_1,j_2,\beta}R_{i_1j_1,i_2j_2}(\la,\nu)\:
                 L_{j_1k_1}^{\al_1\beta}(\la,\mathbf{a})
               \; L_{j_2k_2}^{\beta\al_2}(\nu,\mathbf{a})
               \,=\sum_{j_1,j_2,\beta}L_{i_2j_2}^{\al_1\beta}(\nu,\mathbf{a})
 \;L_{i_1j_1}^{\beta\al_2}(\la,\mathbf{a})\:R_{j_1k_1,j_2k_2}(\la,\nu).\;\:\eeq
where greek indices run over the values $\:0,1,\ldots,N-1\:$ and
the latin indices take the values $\:0,1\,$,
$\;R_{11,22}=\om(\la-\nu),\;\:R_{12,21}=\la(1-\om),\;etc.$

Moreover, Bazhanov and Stroganov found that there is also an
intertwining relation with respect to the $\:\mathbb{Z}_N$ (greek)
indices, i.e. in the Weyl quantum space, if the parameters
$\:\mathbf{a}$ are chosen as \beq a_1=x_q;\hq\; a_2=\om
x_{q'};\hq\; b_1=\frac{y_qy_{q'}}{\mu_q\mu_{q'}};\hq\;
b_2=\frac{y_{q'}}{\mu_q\mu_{q'}};\hq\;
b_3=\frac{y_q}{\mu_q\mu_{q'}}\,,
  \label{abax}\eeq where the $x_q,\:y_q,\:\mu_q\,$ etc. satisfy the CP
conditions (\ref{Bxteq}) with fixed $k$. Writing (\ref{SEL}) with
the parameters (\ref{abax}) as $\:L(\la;\,q,q')\:,$ we get \beq
\label{BS-L-oper} \ds L(\la;\,q,q') = \left(\begin{array}{ccc} \ds
1\,+\,\la {y_{q} y_{q'}\over \mu_{q} \mu_{q'}}\,\Zop && \ds
\la\,\Xop^{-1}\,\left(x_{q} - {y_{q'}\over \mu_{q}
\mu_{q'}}\,\Zop\right)   \\[4mm]
\ds \Xop\,\left(\om x_{q'}-{y_{q} \over \mu_{q}
\mu_{q'}}\,\Zop\right) && \ds \la\,\om\, x_{q} x_{q'} + {1\over
\mu_{q} \mu_{q'} } \,\Zop\end{array}\right)\!\!. \eeq Apart from
the spectral parameter $\la$, this $\:L(\la;\,q,q')\:$ depends on
three independent continous variables, e.g. $x_q,\:x_{q'}$ and the
modulus $\,k\,$. We shall not write the latter explicitly as an
argument.
The quantum space intertwining relation is \bea
\lefteqn{\sum_{\beta_1\beta_2,k}\!
   \mathsf{S}_{\al_1\al_2;\beta_1\beta_2}(p,p',q,q')
   \;L_{i_1k}^{\beta_1\gamma_1}(\la;p,p')\;L_{k\,i_2}^{\beta_2,\gamma_2}
   (\la;q,q')}\ny\\[-2mm] \hs\hx=\;\sum_{\beta_1\beta_2,k}\!\!\!
    \; L_{i_1k}^{\al_2\beta_2}(\la;q,q')\;L_{k\,i_2}^{\al_1\beta_1}
    (\la;p,p')\;\:\mathsf{S}_{\beta_1,\beta_2;\gamma_1,\gamma_2}(p,p',q,q').
    \label{SLL}\eea
The matrix $\:\mathsf{S}\:$ turns out to be the product of four
CP-Boltzmann weights (\ref{Wpq}): \beq
 \fl\mathsf{S}_{\al_1\al_2,\beta_1\beta_2}(p,p',q,q')
\;=\;W_{pq'}(\al_1-\al_2)
   W_{p'q}(\beta_2-\beta_1)\WB_{pq}(\beta_2-\al_1)\WB_{p'q'}
   (\beta_1-\al_2).\label{BSS}\eeq
One can verify the relations \r{SLL},\,\r{BSS} by explicit
calculations, using \r{Weyl-finite} and \r{Wpq} several times,
e.g.
%
\bdm \fl
\sum_{\beta_1}\Zop_{\al_1\beta_1}\mathsf{S}_{\beta_1,\beta_2;\gamma_1,\gamma_2}\!=
   \mathsf{S}_{\al_1-1,\beta_2;\gamma_1,\gamma_2}\!
 =\mu_q\mu_{q'}\:
   \frac{y_p\!-\om^{\al_1-\al_2}x_{q'}}{y_{q'}-\om^{\al_1-\al_2}x_p}\;
   \;\frac{\om x_p\!-\om^{\beta_2-\al_1+1}x_q}{y_q\!-\om^{\beta_2-\al_1+1}y_p}\;
   \mathsf{S}_{\al_1,\beta_2;\gamma_1,\gamma_2}.\edm

The Bazhanov-Stroganov periodic quantum chain of length $Q$ is
defined via its $L$-operator $L_{ik}^{\al\beta}(\la;\,q,q')$ given
in (\ref{BS-L-oper}) and the corresponding monodromy matrix
\beq   \fl \mathsf{M}\lk
\la,\{q_i,q_i'\}_{i=0}^{Q-1}\rk\;=\;L(\la;q_0,q_0')\,
L(\la;q_1,q_1')\,L(\la;q_2,q_2')\;\ldots\;
L(\la;q_{Q-1},q_{Q-1}'), \label{transfer} \eeq where each $L$ has
its pair of rapidities $q_i,q'_i$ and all these rapidities may be
different while keeping the Baxter modulus $k$ to be the same for
all $L$. The transfer matrix is \beq
\mathbf{t}\:=\:\mbox{Tr}_{\mathbb{C}^2}\;\mathsf{M}\,.\label{qtrans}\eeq
Baxter \cite{CPM-sol,CPM-sol1} calls this model the
$\:\tau_2(t_q)$-model after the notation $\tau_2$ introduced for
the transfer matrix in equation (5.33) of \cite{BS}. Mostly in
Baxter's work not the fully inhomogenous model is used, but the
rapidities take two alternating values. In \r{BS-tau} we give the
relation to Baxter's notation. Fusion of the BS-transfer matrices
leads to the functional relations mentioned in the last
subsection.

\section{3D approach to the BS model}\label{BS-from-3D}
\subsection{$L$-operator}

We now show how the BS-$L$-operator can be obtained from the
Linear Problem (\ref{lipro}) of the 3D approach, imposing a
periodicity condition. We follow the general argument of
\cite{IS}, which also gives a quantum group background of the
construction.

\begin{figure}[h]
\setlength{\unitlength}{0.00055in}
{\renewcommand{\dashlinestretch}{30}
\begin{picture}(3649,3864)(-2000,-10)
\path(4062,462)(4062,3612)
\path(4092.000,3492.000)(4062.000,3612.000)(4032.000,3492.000)
\path(5412,2712)(237,2712)
\path(357.000,2742.000)(237.000,2712.000)(357.000,2682.000)
\path(5412,1362)(237,1362)
\path(357.000,1392.000)(237.000,1362.000)(357.000,1332.000)
\dottedline{95}(12,3837)(12,12)(5637,12)
        (5637,3837)(12,3837)
\dottedline{95}(2800,3837)(2800,12)
\put(4650,1900){\makebox(0,0)[lb]{$\<\chi_0|$}}
\put(4650,867){\makebox(0,0)[lb]{$\<\chi_1|$}}
\put(4550,3162){\makebox(0,0)[lb]{$\xi\<\chi_1|$}}
\put(2410,3162){\makebox(0,0)[lb]{$\xi\<\phi_1|$}}
\path(1587,462)(1587,3612)
\path(1617.000,3492.000)(1587.000,3612.000)(1557.000,3492.000)
\put(2480,1900){\makebox(0,0)[lb]{$\<\phi_0|$}}
\put(3950,100){\makebox(0,0)[lb]{$L_2$}}
\put(2480,822){\makebox(0,0)[lb]{$\<\phi_1|$}}
\put(587,3162){\makebox(0,0)[lb]{$\xi\<\psi_1|$}}
\put(687,1900){\makebox(0,0)[lb]{$\<\psi_0|$}}
\put(687,822){\makebox(0,0)[lb]{$\<\psi_1|$}}
\put(1710,2802){\makebox(0,0)[lb]{$\tilde{\wf}_1$}}
\put(1710,1510){\makebox(0,0)[lb]{$\wf_1$}}
\put(4160,2802){\makebox(0,0)[lb]{$\tilde{\wf}_2$}}
\put(4152,1510){\makebox(0,0)[lb]{$\wf_2$}}
\put(1480,100){\makebox(0,0)[lb]{$L_1$}}
\end{picture}
} \caption{\footnotesize Piece of the auxiliary plane which
corresponds to the product of two $L$-operators. The Weyl pairs
together with parameters $\kappa_1$, $\kappa_2$, $\tilde\kappa_1$,
$\tilde\kappa_2$ are associated with the corresponding vertices in
this plane.} \label{LL}
\end{figure}
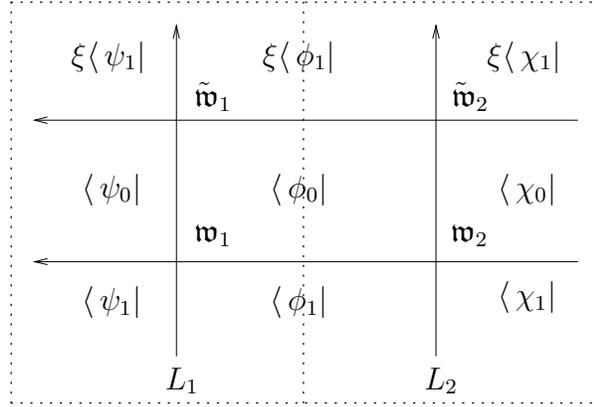

Consider the domain of the auxiliary plane containing the four
variables $\wf_1,\:\tilde{\wf_1},\:\wf_2,\:\tilde{\wf_2}$, see
Fig. \ref{LL}. The co-currents around each Weyl variable are taken
to be related by the Linear Problem (\ref{lipro}). We impose the
periodicity condition \beq\langle \psi_{-1}|=\xi\,\langle
\psi_1|\,;\hs\hq \langle \phi_{-1}|
 =\xi\langle \phi_1|\,;\hs\hq \langle \chi_{-1}|=\xi\langle \chi_1| \label{qperi}\eeq
in the vertical direction, with $\xi$ a quasi-momentum.
Then the conditions (\ref{lipro}) at $\wf_1$ and $\tilde{\wf_1}$
(those in the left hand dotted box of Fig. \ref{LL} denoted $L_1$)
are \beq\label{lin1}
\begin{array}{rcl}
0&=&\<\psi_0|+ \xi\omega^{1/2}\<\psi_1|\tilde{\uop}_1+
\<\phi_0|\tilde{\wop}_1+ \xi\tilde{\kappa}_1
\<\phi_1|\tilde{\uop}_1\tilde{\wop}_1\;,\\[2mm]
0&=&\<\psi_1|+ \;\;\omega^{1/2}\<\psi_0|{\uop}_1+
\<\phi_1|{\wop}_1\;+\;\kappa_1\<\phi_0|{\uop}_1{\wop}_1\;.\end{array}\eeq
These linear relations can be rewritten in matrix form as follows
\beq\label{lin2} \<\psi|\,(\om\,\xi\,\uop_1\,\tilde{\uop}_1\,-\,1
)\tilde{\wop}_1^{-1} = \<\phi|\cdot L_1(\xi)\,, \eeq where $
\<\phi|$ and $ \<\psi|$ denote the rows $(\<\phi_0|, \<\phi_1|)$
and  $(\<\psi_0|, \<\psi_1|)$, respectively, and $L_1(\xi)$ is the
following $2\times2$ matrix with operator valued elements \beq
\label{lin3} \ds L_{1}(\xi)\;=\ \left(\begin{array}{lcl} \ds
   1 -\om^{1/2}\xi \uop_1\tilde{\uop}_1\kappa_1
{\wop}_1  \tilde{\wop}^{-1}_1&& \ds -\uop_1\lk \om^{1/2}-\kappa_1
{\wop}_1  \tilde{\wop}^{-1}_1 \rk
\\&&\\ \xi \tilde{\uop}_1 \lk\tilde\kappa_1-\om^{1/2}
 {\wop}_1  \tilde{\wop}^{-1}_1 \rk  &&
-\om^{1/2}\xi\uop_1\tilde{\uop}_1\tilde\kappa_1  +
 {\wop}_1\tilde{\wop}^{-1}_1   \end{array}\right)\,.\eeq
We want to use a matrix representation for the Weyl elements.
Observing that only the three elements
$\:\wop_1\twop_1^{-1},\;\uop_1,\;\tuop_1\:$ appear, we may use a
$N$-dimensional representation, writing
\beq\label{newWeyl} \wop_1\tilde{\wop}^{-1}_1=\frac{w_1}{\tilde
w_1}\:\Zop\,;\hs \uop_1=u_1\,\Xop\,;\hs
\tilde{\uop}_1=\tilde{u}_1\,\Xop^{-1}\, \eeq
in the basis (\ref{Weyl-finite}). In order to bring this into a
form comparable with \r{BS-L-oper} we put \beq
\kappa_1=\sqo\frac{x_q}{y_{q'}};
    \hs \tilde{\kappa}_1=\sqoi\frac{y_q}{x_{q'}}; \hs
\frac{w_1}{\tw_1}=\om^{-1}\frac{y_qy_{q'}}{x_qx_{q'}\mu_q\mu_{q'}}\,,
 \label{condi-ident}\eeq
so that \beq \label{lin3a}\fl  L_{1}(\xi)\;=\;\lk\BAR{cc}\ds 1-\xi
u_1\tilde{u}_1\frac{y_q}{x_{q'}\mu_q\mu_{q'}}\Zop\:;&\ds\hspace*{-6mm}
-u_1\Xop\lk\om^{1/2}-\om^{-1/2}\frac{y_q}{x_{q'}\mu_q\mu_{q'}}\Zop\rk\!\!\\[5mm]
\ds \xi\,\tilde{u}_1\om^{-1/2}\frac{y_q}{x_{q'}}\Xop^{-1}\lk
1-\frac{y_{q'}}{x_q\mu_q\mu_{q'}}\Zop\rk;& \ds -\xi
u_1\tilde{u}_1\frac{y_q}{x_{q'}}+\frac{y_qy_{q'}}{\om
x_qx_{q'}\mu_q\mu_{q'}}\Zop\EAR\rk\!. \eeq \vspace*{1mm}
Conjugating with the Pauli matrix $\:\sig_2\:$ and introducing a
new spectral parameter $\la$ by \beq \la\;=\;-\:\frac{1}{\om\,
u_1\tilde{u}_1x_q\,y_q\,\xi} \label{xi-la}\eeq we obtain
 \bea \label{lin3b}\fl \sig_2\:
L_{1}(\xi)\:\sig_2&=&\frac{1}{\la\,\om\, x_qx_{q'}}\lk\BAR{cc}\ds
1\,+\,\la\frac{y_qy_{q'}}{\mu_q\mu_{q'}}\Zop\,;& \ds\hspace*{-2mm}
\frac{\Xop^{-1}}{\om^{1/2}u_1x_q}\lk
x_q-\frac{y_{q'}}{\mu_q\mu_{q'}}\Zop\rk\!\!\! \\
\ds\!\!\!
\la\,\om^{1/2}u_1x_q\Xop\lk\om\,x_{q'}-\frac{y_q}{\mu_q\mu_{q'}}\Zop\!\rk\,;
& \ds \la\,\om\,x_qx_{q'}+\frac{1}{\mu_q\mu_{q'}}\Zop \EAR\rk\!.\ny\\
\fl && \eea
A gauge transformation with \beq \label{gaug} P_1\;=\;\lk\BAR{cc}
\sqrt{\om^{1/2}u_1x_q\la}&0\\0&1/\sqrt{\om^{1/2}u_1x_q\la}\EAR\rk\eeq
 leads
to \beq P_1\:\sig_2\,L_1(\xi)\,\sig_2\:P_1^{-1}\;=\;
\frac{1}{\la\,\om \,x_q\,y_q}\:L(\la;\,q,\,q') \label{Lxila} \eeq
with $\;L(\la;\,q,\,q')\;$ defined in \r{BS-L-oper}. We shall see
in \r{ff} that the identification \r{condi-ident} will also appear
when we express the BS intertwining matrix $\mathsf{S}$ within the
3D framework. Using the parametrization introduced later in
section \ref{parametrization} we will show in \r{indep} that
considering the monodromy matrix, the factor
$\;u_1\tilde{u}_1x_qy_q\;$ relating the spectral parameters
$\:\xi\:$ and $\:\la\:$ does not depend on the site considered.
The same holds for the combination $\;u_1x_q\:$ appearing in
\r{gaug}. So the BS-monodromy \r{transfer} can be written as \beq
\fl\mathsf{M}\lk \la,\{q_i,q_i'\}_{i=0}^{Q-1}\rk \;=\;
P_1\sig_2\:L_0(\xi)\, L_1(\xi)\,\ldots\, L_{Q-1}(\xi)\:\sig_2
P_1^{-1}\;\prod_{i=0}^{Q-1}\lk\la\,\om\, x_{q_i}y_{q_i}\rk.
\label{transferU} \eeq

We finally remark that demanding periodicity \r{qperi} not after
{\it two} vertical steps as done here, but after $N$ steps, one
obtains $\,N\times N\,$ $L$-matrices,
see \cite{IS}.

\subsection{3D interpretation of the relation
                    $\;\mathsf{S}LL=LL\,\mathsf{S}\,.$}\label{BSSD}

Consider the product of the successive action two $L$-operators
(the simplest monodromy) \beq\label{lin4} \<\psi|\;
(\om\,\xi\,\uop_2\,\tilde{\uop}_2\,-1)\,(\om\,\xi\,\uop_1\,
\tilde{\uop}_1\,-1)\,\tilde{\wop}_2^{-1}\tilde{\wop}_1^{-1}\, =
\<\chi|\cdot L_2(\xi)\, L_1(\xi)\,.  \eeq
%
%
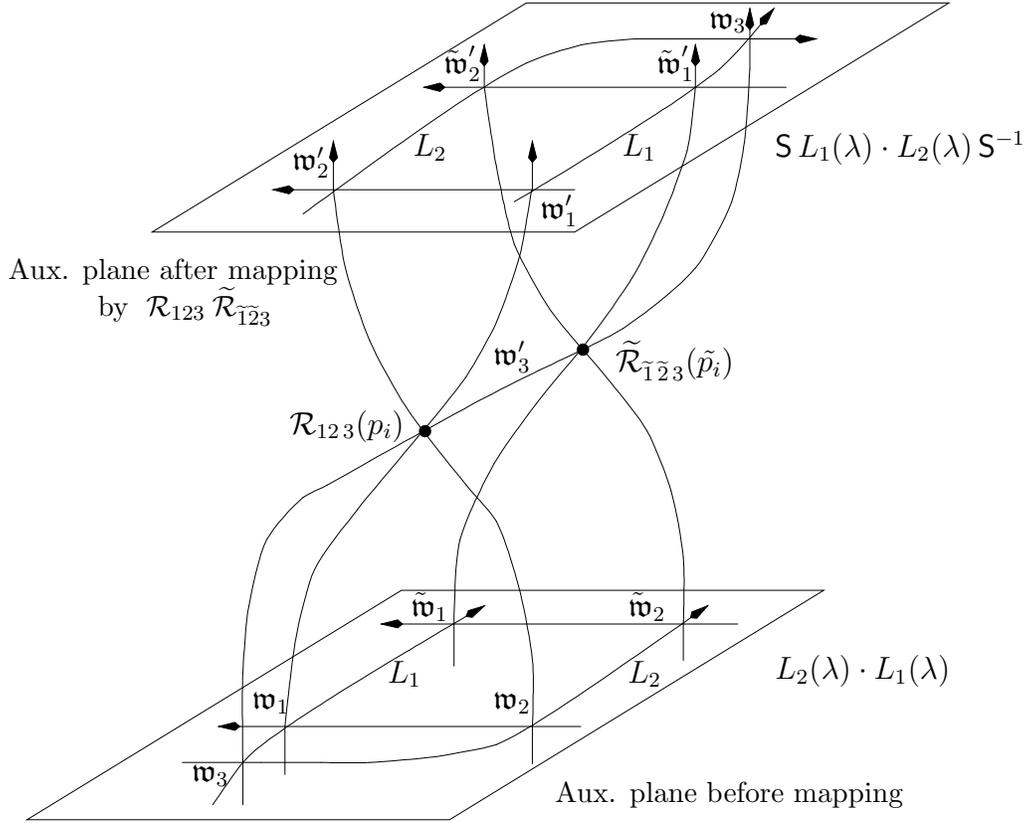
\begin{figure}[ht]
\setlength{\unitlength}{0.00070in}
{\renewcommand{\dashlinestretch}{30}
\begin{center}\begin{picture}(6897,6200)(0,0)
\path(3780,707) (3781.348,758.789)
    (3782.517,806.952)
    (3783.505,851.670)
    (3784.314,893.124)
    (3784.944,931.496)
    (3785.393,966.967)
    (3785.753,1029.931)
    (3785.393,1083.466)
    (3784.314,1129.023)
    (3782.517,1168.050)
    (3780.000,1202.000)
\path(3780,1202)    (3774.324,1251.783)
    (3765.579,1312.265)
    (3754.504,1379.825)
    (3741.836,1450.839)
    (3728.314,1521.685)
    (3714.675,1588.743)
    (3701.658,1648.388)
    (3690.000,1697.000)
\path(3690,1697)    (3675.666,1751.647)
    (3657.461,1818.095)
    (3647.147,1854.474)
    (3636.154,1892.279)
    (3624.577,1931.002)
    (3612.514,1970.135)
    (3600.059,2009.170)
    (3587.309,2047.599)
    (3574.361,2084.914)
    (3561.310,2120.608)
    (3535.284,2185.096)
    (3510.000,2237.000)
\path(3510,2237)    (3464.761,2301.136)
    (3434.173,2337.965)
    (3401.055,2375.878)
    (3367.492,2413.305)
    (3335.572,2448.680)
    (3285.000,2507.000)
\path(3285,2507)    (3239.270,2564.014)
    (3183.335,2633.842)
    (3152.666,2672.279)
    (3120.795,2712.380)
    (3088.173,2753.632)
    (3055.249,2795.521)
    (3022.473,2837.535)
    (2990.297,2879.160)
    (2959.168,2919.882)
    (2929.538,2959.190)
    (2901.856,2996.568)
    (2876.573,3031.505)
    (2835.000,3092.000)
\path(2835,3092)    (2805.240,3139.518)
    (2770.289,3198.359)
    (2732.172,3264.912)
    (2692.916,3335.566)
    (2654.546,3406.713)
    (2619.086,3474.740)
    (2588.562,3536.040)
    (2565.000,3587.000)
\path(2565,3587)    (2543.525,3639.451)
    (2519.153,3704.135)
    (2506.289,3739.825)
    (2493.188,3777.086)
    (2480.014,3815.422)
    (2466.930,3854.338)
    (2454.098,3893.336)
    (2441.681,3931.921)
    (2429.841,3969.599)
    (2418.742,4005.871)
    (2399.414,4072.220)
    (2385.000,4127.000)
\path(2385,4127)    (2376.036,4166.671)
    (2366.947,4212.473)
    (2357.467,4266.119)
    (2347.335,4329.324)
    (2341.941,4365.046)
    (2336.286,4403.800)
    (2330.335,4445.801)
    (2324.056,4491.262)
    (2317.416,4540.399)
    (2310.382,4593.424)
    (2302.921,4650.553)
    (2295.000,4712.000)
\path(3195,1472)    (3196.277,1537.867)
    (3197.682,1599.125)
    (3199.231,1656.004)
    (3200.941,1708.737)
    (3202.829,1757.552)
    (3204.910,1802.681)
    (3207.201,1844.356)
    (3209.719,1882.805)
    (3215.500,1950.953)
    (3222.387,2008.971)
    (3230.509,2058.705)
    (3240.000,2102.000)
\path(3240,2102)    (3258.144,2161.692)
    (3284.495,2232.203)
    (3300.062,2270.302)
    (3316.858,2309.650)
    (3334.607,2349.760)
    (3353.036,2390.149)
    (3371.870,2430.329)
    (3390.835,2469.817)
    (3409.656,2508.126)
    (3428.059,2544.770)
    (3462.513,2611.126)
    (3492.000,2665.000)
\path(3492,2665)    (3520.344,2711.831)
    (3556.897,2767.684)
    (3599.131,2829.394)
    (3644.516,2893.795)
    (3690.525,2957.721)
    (3734.627,3018.008)
    (3774.295,3071.489)
    (3807.000,3115.000)
\path(3807,3115)    (3846.088,3164.900)
    (3895.065,3225.064)
    (3950.634,3292.071)
    (4009.500,3362.500)
    (4068.366,3432.929)
    (4123.935,3499.936)
    (4172.912,3560.100)
    (4212.000,3610.000)
\path(4212,3610)    (4244.620,3653.586)
    (4284.083,3707.249)
    (4327.932,3767.761)
    (4373.711,3831.891)
    (4418.964,3896.410)
    (4461.234,3958.088)
    (4498.065,4013.694)
    (4527.000,4060.000)
\path(4527,4060)    (4551.751,4103.456)
    (4580.665,4157.259)
    (4612.143,4218.064)
    (4644.585,4282.525)
    (4676.391,4347.295)
    (4705.962,4409.028)
    (4731.698,4464.379)
    (4752.000,4510.000)
\path(4752,4510)    (4772.211,4558.197)
    (4796.127,4617.415)
    (4822.222,4684.097)
    (4848.971,4754.688)
    (4874.849,4825.629)
    (4898.330,4893.366)
    (4917.889,4954.342)
    (4932.000,5005.000)
\path(4932,5005)    (4946.264,5074.042)
    (4953.306,5117.345)
    (4960.534,5168.347)
    (4968.133,5228.431)
    (4976.287,5298.979)
    (4980.630,5338.610)
    (4985.181,5381.375)
    (4989.964,5427.447)
    (4995.000,5477.000)
\path(4905,1472)    (4906.348,1523.789)
    (4907.517,1571.952)
    (4908.505,1616.670)
    (4909.314,1658.124)
    (4909.944,1696.496)
    (4910.393,1731.967)
    (4910.753,1794.931)
    (4910.393,1848.466)
    (4909.314,1894.023)
    (4907.517,1933.050)
    (4905.000,1967.000)
\path(4905,1967)    (4899.466,2016.792)
    (4891.065,2077.296)
    (4880.416,2144.883)
    (4868.134,2215.921)
    (4854.835,2286.782)
    (4841.135,2353.835)
    (4827.652,2413.451)
    (4815.000,2462.000)
\path(4815,2462)    (4800.659,2508.138)
    (4781.604,2563.800)
    (4759.028,2625.667)
    (4734.120,2690.424)
    (4708.072,2754.752)
    (4682.076,2815.336)
    (4657.321,2868.857)
    (4635.000,2912.000)
\path(4635,2912)    (4611.289,2950.361)
    (4580.132,2995.422)
    (4543.820,3044.780)
    (4504.639,3096.035)
    (4464.878,3146.786)
    (4426.826,3194.631)
    (4392.770,3237.169)
    (4365.000,3272.000)
\path(4365,3272)    (4320.857,3325.799)
    (4265.554,3390.568)
    (4234.921,3425.964)
    (4202.941,3462.780)
    (4170.097,3500.574)
    (4136.869,3538.906)
    (4103.739,3577.335)
    (4071.188,3615.419)
    (4039.697,3652.719)
    (4009.749,3688.791)
    (3956.403,3755.494)
    (3915.000,3812.000)
\path(3915,3812)    (3884.338,3859.084)
    (3848.553,3917.411)
    (3809.766,3983.472)
    (3770.096,4053.759)
    (3731.665,4124.763)
    (3696.591,4192.977)
    (3666.996,4254.892)
    (3645.000,4307.000)
\path(3645,4307)    (3626.001,4364.149)
    (3606.590,4434.634)
    (3596.888,4473.507)
    (3587.275,4514.069)
    (3577.814,4555.772)
    (3568.568,4598.068)
    (3559.601,4640.408)
    (3550.976,4682.246)
    (3542.759,4723.031)
    (3535.012,4762.217)
    (3521.183,4833.597)
    (3510.000,4892.000)
\path(3510,4892)    (3502.634,4931.783)
    (3494.686,4977.665)
    (3485.892,5031.358)
    (3475.988,5094.579)
    (3470.537,5130.297)
    (3464.710,5169.039)
    (3458.474,5211.020)
    (3451.795,5256.454)
    (3444.642,5305.554)
    (3436.980,5358.536)
    (3428.777,5415.613)
    (3420.000,5477.000)
\path(2070,4532)    (2113.400,4563.979)
    (2156.051,4595.367)
    (2197.963,4626.171)
    (2239.145,4656.399)
    (2279.607,4686.055)
    (2319.358,4715.147)
    (2358.409,4743.681)
    (2396.767,4771.663)
    (2434.444,4799.101)
    (2471.448,4825.999)
    (2507.788,4852.365)
    (2543.476,4878.206)
    (2612.928,4928.335)
    (2679.881,4976.438)
    (2744.410,5022.567)
    (2806.592,5066.773)
    (2866.503,5109.107)
    (2924.219,5149.621)
    (2979.816,5188.366)
    (3033.371,5225.395)
    (3084.960,5260.757)
    (3134.659,5294.506)
    (3182.544,5326.693)
    (3228.691,5357.368)
    (3273.178,5386.583)
    (3316.079,5414.391)
    (3357.471,5440.842)
    (3397.431,5465.988)
    (3436.034,5489.881)
    (3473.357,5512.571)
    (3544.468,5554.552)
    (3611.372,5592.342)
    (3674.679,5626.354)
    (3735.000,5657.000)
\path(3735,5657)    (3774.058,5674.555)
    (3822.366,5693.667)
    (3876.951,5713.459)
    (3934.841,5733.057)
    (3993.064,5751.586)
    (4048.646,5768.170)
    (4098.615,5781.933)
    (4140.000,5792.000)
\path(4140,5792)    (4179.969,5799.516)
    (4228.965,5806.820)
    (4283.996,5813.749)
    (4342.073,5820.136)
    (4400.202,5825.819)
    (4455.394,5830.634)
    (4504.657,5834.415)
    (4545.000,5837.000)
\path(4545,5837)    (4585.164,5838.470)
    (4634.314,5839.160)
    (4689.450,5839.250)
    (4747.575,5838.920)
    (4805.691,5838.350)
    (4860.799,5837.720)
    (4909.901,5837.210)
    (4950.000,5837.000)
\path(4950,5837)    (4999.022,5837.000)
    (5059.060,5837.000)
    (5126.444,5837.000)
    (5197.500,5837.000)
    (5268.556,5837.000)
    (5335.940,5837.000)
    (5395.978,5837.000)
    (5445.000,5837.000)
\path(5445,5837)    (5511.102,5837.000)
    (5552.457,5837.000)
    (5601.105,5837.000)
    (5658.365,5837.000)
    (5725.555,5837.000)
    (5763.286,5837.000)
    (5803.994,5837.000)
    (5847.844,5837.000)
    (5895.000,5837.000)
\blacken\path(5775.000,5807.000)(5895.000,5837.000)(5775.000,5867.000)
(5739.000,5837.000)(5775.000,5807.000) \path(3645,4622)
(3686.907,4647.289)
    (3728.062,4672.150)
    (3768.473,4696.587)
    (3808.150,4720.607)
    (3847.100,4744.215)
    (3885.333,4767.417)
    (3922.857,4790.218)
    (3959.682,4812.624)
    (3995.815,4834.642)
    (4031.266,4856.275)
    (4100.156,4898.415)
    (4166.421,4939.088)
    (4230.132,4978.341)
    (4291.359,5016.218)
    (4350.172,5052.765)
    (4406.640,5088.027)
    (4460.834,5122.050)
    (4512.824,5154.879)
    (4562.680,5186.558)
    (4610.472,5217.135)
    (4656.270,5246.653)
    (4700.144,5275.158)
    (4742.164,5302.695)
    (4782.400,5329.310)
    (4820.922,5355.048)
    (4857.800,5379.955)
    (4893.105,5404.075)
    (4959.273,5450.138)
    (5019.986,5493.599)
    (5075.805,5534.822)
    (5127.289,5574.168)
    (5175.000,5612.000)
\path(5175,5612)    (5244.255,5677.903)
    (5310.000,5747.000)
\path(5310,5747)    (5350.852,5792.302)
    (5405.250,5854.981)
    (5439.507,5895.144)
    (5479.523,5942.420)
    (5526.091,5997.731)
    (5580.000,6062.000)
\blacken\path(5525.943,5950.744)(5580.000,6062.000)(5479.947,5989.272)
(5479.828,5942.411)(5525.943,5950.744) \path(1620,437)
(1619.079,507.705)
    (1618.281,573.455)
    (1617.605,634.495)
    (1617.053,691.074)
    (1616.623,743.439)
    (1616.316,791.836)
    (1616.131,836.514)
    (1616.070,877.719)
    (1616.131,915.698)
    (1616.316,950.699)
    (1617.053,1012.754)
    (1618.281,1065.862)
    (1620.000,1112.000)
\path(1620,1112)    (1622.577,1161.215)
    (1626.346,1221.368)
    (1631.147,1288.800)
    (1636.819,1359.849)
    (1643.200,1430.854)
    (1650.132,1498.155)
    (1657.452,1558.090)
    (1665.000,1607.000)
\path(1665,1607)    (1674.436,1657.118)
    (1686.777,1717.946)
    (1701.635,1785.809)
    (1718.618,1857.031)
    (1737.336,1927.937)
    (1757.399,1994.851)
    (1778.417,2054.097)
    (1800.000,2102.000)
\path(1800,2102)    (1821.857,2138.422)
    (1850.985,2179.951)
    (1885.447,2224.456)
    (1923.307,2269.805)
    (1962.629,2313.866)
    (2001.476,2354.506)
    (2070.000,2417.000)
\path(2070,2417)    (2108.408,2440.489)
    (2159.025,2464.966)
    (2210.130,2487.961)
    (2250.000,2507.000)
\path(2250,2507)    (2319.851,2545.911)
    (2362.711,2570.452)
    (2407.815,2596.483)
    (2452.880,2622.578)
    (2495.621,2647.313)
    (2565.000,2687.000)
\path(2565,2687)    (2606.468,2710.223)
    (2653.774,2736.665)
    (2706.090,2765.855)
    (2762.587,2797.326)
    (2822.435,2830.607)
    (2884.808,2865.230)
    (2948.875,2900.725)
    (3013.807,2936.622)
    (3078.778,2972.454)   (3142.956,3007.750)
    (3205.515,3042.041)
    (3265.625,3074.858)
    (3322.457,3105.731)
    (3375.182,3134.193)
    (3422.973,3159.772)
    (3465.000,3182.000)
\path(3465,3182)    (3500.427,3200.289)
    (3543.986,3222.338)
    (3592.979,3246.869)
    (3644.707,3272.604)
    (3696.472,3298.263)
    (3745.575,3322.567)
    (3789.317,3344.239)
    (3825.000,3362.000)
\path(3825,3362)    (3887.680,3392.478)
    (3924.741,3410.070)
    (3964.797,3428.943)
    (4007.231,3448.883)
    (4051.428,3469.673)
    (4096.773,3491.099)
    (4142.651,3512.945)
    (4188.446,3534.996)
    (4233.543,3557.036)
    (4277.326,3578.851)
    (4319.180,3600.224)
    (4358.490,3620.941)
    (4394.640,3640.786)
    (4455.000,3677.000)
\path(4455,3677)    (4512.952,3716.614)
    (4581.917,3766.741)
    (4619.235,3794.847)
    (4657.753,3824.502)
    (4696.953,3855.344)
    (4736.318,3887.015)
    (4775.328,3919.154)
    (4813.467,3951.402)
    (4850.217,3983.397)
    (4885.059,4014.781)
    (4946.951,4074.274)
    (4995.000,4127.000)
\path(4995,4127)    (5029.117,4172.933)
    (5066.558,4230.690)
    (5105.610,4296.637)
    (5144.557,4367.139)
    (5181.685,4438.561)
    (5215.278,4507.268)
    (5243.621,4569.626)
    (5265.000,4622.000)
\path(5265,4622)    (5278.055,4661.554)
    (5291.438,4710.273)
    (5304.727,4765.156)
    (5317.500,4823.203)
    (5329.336,4881.411)
    (5339.812,4936.781)
    (5348.508,4986.311)
    (5355.000,5027.000)
\path(5355,5027)    (5362.405,5084.837)
    (5369.657,5155.704)
    (5373.171,5194.670)
    (5376.580,5235.270)
    (5379.861,5276.962)
    (5382.994,5319.204)
    (5385.954,5361.455)
    (5388.721,5403.175)
    (5391.271,5443.820)
    (5393.583,5482.851)
    (5397.402,5553.904)
    (5400.000,5612.000)
\path(5400,5612)    (5400.832,5645.071)
    (5401.110,5690.079)
    (5400.832,5752.297)
    (5400.486,5791.509)
    (5400.000,5837.000)
\put(1650,800){\makebox(0,0)[lb] {$\wf_1$}}
\put(3450,800){\makebox(0,0)[lb] {$\wf_2$}}
\put(3950,100){\makebox(0,0)[lb] {\small Aux. plane before
mapping}} \put(1200, 260){\makebox(0,0)[lb] {$\wf_3$}}
\put(2835,1490){\makebox(0,0)[lb]{$\tilde{\wf}_1$}}
\put(4455,1490){\makebox(0,0)[lb]{$\tilde{\wf}_2$}}
\put(5590,1000){\makebox(0,0)[lb]{$L_2(\lambda)\cdot
L_1(\lambda)$}} \put(3800,4440){\makebox(0,0)[lb]{$\wf_1'$}}
\put(1950,4800){\makebox(0,0)[lb]{$\wf_2'$}}
       \put(2890,4930){\makebox(0,0)[lb]{$L_2$}}
       \put(4450,4930){\makebox(0,0)[lb]{$L_1$}}
       \put(2700,1000){\makebox(0,0)[lb]{$L_1$}}
       \put(4490,1000){\makebox(0,0)[lb]{$L_2$}}
\put( -130,4000){\makebox(0,0)[lb]{\small Aux. plane after
mapping}} \put(540,3700){\makebox(0,0)[lb]
     {\small by  $\;\Rop_{123}\:\widetilde{\Rop}_{\widetilde{1}\widetilde{2}3}$ }}
\put(3080,5520){\makebox(0,0)[lb]{$\tilde{\wf}_2'$}}
\put(4670,5520){\makebox(0,0)[lb]{$\tilde{\wf}_1'$}}
\put(5590,4900){\makebox(0,0)[lb]
              {$\mathsf{S}\,L_1(\la)\cdot L_2(\la)\,\mathsf{S}^{-1}$}}
\put(5060,5882){\makebox(0,0)[lb]{$\wf_3$}}
\put(3450,3350){\makebox(0,0)[lb]{$\wf_3'$}} {
\put(1970,2830){\makebox(0,0)[lb]{$\Rop_{1\,\!2\,3}(p_i)$}}
\put(4400,3285){\makebox(0,0)[lb]{$\widetilde{\Rop}_{\widetilde{1}\,\widetilde{2}\,3}(\tilde{p_i})$}}
} \put(2920,2850){\makebox(0,0)[lb]{$\bullet$}}
\put(4100,3460){\makebox(0,0)[lb]{$\bullet$}}
 \path(1170,437)
(1231.304,437.000)
    (1288.310,437.000)
    (1341.231,437.000)
    (1390.282,437.000)
    (1435.677,437.000)
    (1477.630,437.000)
    (1516.356,437.000)
    (1552.069,437.000)
    (1615.311,437.000)
    (1669.071,437.000)
    (1715.063,437.000)
    (1755.000,437.000)
\path(1755,437) (1799.563,437.000)
    (1854.143,437.000)
    (1915.402,437.000)
    (1980.000,437.000)
    (2044.598,437.000)
    (2105.857,437.000)
    (2160.437,437.000)
    (2205.000,437.000)
\path(2205,437) (2240.646,436.847)
    (2284.298,436.475)
    (2333.290,436.017)
    (2384.955,435.601)
    (2436.625,435.361)
    (2485.634,435.426)
    (2529.315,435.929)
    (2565.000,437.000)
\path(2565,437) (2614.216,439.740)
    (2674.373,443.906)
    (2741.808,449.197)
    (2812.860,455.311)
    (2883.867,461.950)
    (2951.167,468.811)
    (3011.099,475.594)
    (3060.000,482.000)
\path(3060,482) (3105.001,488.770)
    (3159.908,497.543)
    (3221.376,507.958)
    (3286.057,519.654)
    (3350.607,532.267)
    (3411.678,545.437)
    (3465.924,558.802)
    (3510.000,572.000)
\path(3510,572) (3551.217,589.201)
    (3601.695,613.970)
    (3651.325,640.254)
    (3690.000,662.000)
\path(3690,662) (3738.075,690.522)
    (3788.981,721.501)
    (3843.229,755.283)
    (3901.330,792.215)
    (3963.794,832.641)
    (4031.133,876.909)
    (4066.789,900.592)
    (4103.856,925.365)
    (4142.397,951.271)
    (4182.476,978.354)
    (4224.157,1006.656)
    (4267.503,1036.222)
    (4312.579,1067.095)
    (4359.448,1099.317)
    (4408.174,1132.931)
    (4458.821,1167.982)
    (4511.454,1204.513)
    (4566.135,1242.566)
    (4622.928,1282.185)
    (4681.898,1323.414)
    (4743.108,1366.294)
    (4806.623,1410.871)
    (4872.505,1457.187)
    (4940.820,1505.285)
    (5011.630,1555.208)
    (5047.991,1580.868)
    (5085.000,1607.000)
\blacken\path(5004.291,1513.266)(5085.000,1607.000)(4969.676,1562.274)
(4957.579,1517.001)(5004.291,1513.266)
\path(12,12)(2802,1722)(5952,1722)
    (3162,12)(12,12)
\path(4140,707)(1440,707)
\blacken\path(1560.000,737.000)(1440.000,707.000)(1560.000,677.000)
(1596.000,707.000)(1560.000,737.000) \path(5310,1472)(2655,1472)
\blacken\path(2775.000,1502.000)(2655.000,1472.000)(2775.000,1442.000)
(2811.000,1472.000)(2775.000,1502.000)
\path(945,4397)(3735,6107)(6885,6107)
    (4095,4397)(945,4397)
\path(4095,4712)(1845,4712)
\blacken\path(1965.000,4742.000)(1845.000,4712.000)(1965.000,4682.000)
(2001.000,4712.000)(1965.000,4742.000) \path(5670,5477)(2970,5477)
\blacken\path(3090.000,5507.000)(2970.000,5477.000)(3090.000,5447.000)
(3126.000,5477.000)(3090.000,5507.000) \path(3195,1472)(3195,1157)
\path(3780,707)(3780,427) \path(4905,1472)(4905,1200)
\path(1935,707)(1935,347) \path(1620,437)(1620,122)
\path(2295,4712)(2295,5072)
\blacken\path(2325.000,4952.000)(2295.000,5072.000)(2265.000,4952.000)
(2295.000,4916.000)(2325.000,4952.000) \path(3420,5477)(3420,5792)
\blacken\path(3450.000,5672.000)(3420.000,5792.000)(3390.000,5672.000)
(3420.000,5636.000)(3450.000,5672.000) \path(3780,4712)(3780,5072)
\blacken\path(3810.000,4952.000)(3780.000,5072.000)(3750.000,4952.000)
(3780.000,4916.000)(3810.000,4952.000) \path(4995,5477)(4995,5792)
\blacken\path(5025.000,5672.000)(4995.000,5792.000)(4965.000,5672.000)
(4995.000,5636.000)(5025.000,5672.000) \path(5400,5837)(5400,6062)
\blacken\path(5430.000,5942.000)(5400.000,6062.000)(5370.000,5942.000)
(5400.000,5906.000)(5430.000,5942.000) \path(1935,707)
(1942.339,777.949)
    (1949.255,843.916)
    (1955.781,905.149)
    (1961.946,961.896)
    (1967.781,1014.403)
    (1973.317,1062.918)
    (1978.585,1107.690)
    (1983.615,1148.965)
    (1988.439,1186.991)
    (1993.086,1222.016)
    (2001.977,1284.052)
    (2010.533,1337.053)
    (2019.000,1383.000)
\path(2019,1383)    (2029.206,1432.899)
    (2042.385,1493.607)
    (2058.001,1561.430)
    (2075.524,1632.675)
    (2094.418,1703.648)
    (2114.151,1770.655)
    (2134.190,1830.004)
    (2154.000,1878.000)
\path(2154,1878)    (2180.710,1927.047)
    (2216.950,1983.867)
    (2259.871,2045.585)
    (2306.625,2109.325)
    (2354.363,2172.212)
    (2400.237,2231.371)
    (2441.399,2283.925)
    (2475.000,2327.000)
\path(2475,2327)    (2518.333,2381.561)
    (2572.433,2447.483)
    (2602.416,2483.508)
    (2633.764,2520.935)
    (2666.034,2559.287)
    (2698.785,2598.084)
    (2731.574,2636.847)
    (2763.958,2675.097)
    (2795.496,2712.355)
    (2825.744,2748.142)
    (2880.604,2813.387)
    (2925.000,2867.000)
\path(2925,2867)    (2961.285,2911.026)
    (3005.910,2964.757)
    (3055.950,3025.087)
    (3108.480,3088.910)
    (3160.575,3153.119)
    (3209.310,3214.608)
    (3251.760,3270.271)
    (3285.000,3317.000)
\path(3285,3317)    (3309.904,3355.797)
    (3339.038,3403.915)
    (3370.747,3458.383)
    (3403.376,3516.230)
    (3435.271,3574.486)
    (3464.776,3630.180)
    (3490.238,3680.342)
    (3510.000,3722.000)
\path(3510,3722)    (3530.423,3770.163)
    (3553.732,3829.581)
    (3578.673,3896.604)
    (3603.994,3967.581)
    (3628.440,4038.860)
    (3650.759,4106.791)
    (3669.696,4167.721)
    (3684.000,4218.000)
\path(3684,4218)    (3692.367,4251.513)
    (3701.030,4290.229)
    (3710.253,4335.599)
    (3720.300,4389.071)
    (3731.435,4452.098)
    (3737.492,4487.646)
    (3743.920,4526.128)
    (3750.752,4567.722)
    (3758.021,4612.612)
    (3765.759,4660.977)
    (3774.000,4713.000)
\path(1395,122) (1438.975,186.670)
    (1477.104,242.267)
    (1510.044,289.714)
    (1538.456,329.934)
    (1584.330,392.384)
    (1620.000,437.000)
\path(1620,437) (1663.466,483.477)
    (1710.000,527.000)
\path(1710,527) (1763.266,568.721)
    (1820.757,612.033)
    (1883.100,657.332)
    (1950.922,705.014)
    (1987.082,729.872)
    (2024.847,755.474)
    (2064.294,781.870)
    (2105.503,809.108)
    (2148.550,837.238)
    (2193.515,866.310)
    (2240.475,896.374)
    (2289.510,927.477)
    (2340.697,959.671)
    (2394.114,993.005)
    (2449.840,1027.527)
    (2507.954,1063.287)
    (2568.532,1100.336)
    (2631.655,1138.721)
    (2697.399,1178.494)
    (2765.843,1219.702)
    (2837.066,1262.396)
    (2873.744,1284.315)
    (2911.146,1306.624)
    (2949.282,1329.330)
    (2988.161,1352.438)
    (3027.793,1375.954)
    (3068.189,1399.885)
    (3109.357,1424.236)
    (3151.309,1449.015)
    (3194.052,1474.227)
    (3237.598,1499.878)
    (3281.956,1525.975)
    (3327.136,1552.523)
    (3373.147,1579.530)
    (3420.000,1607.000)
\blacken
\path(3331.635,1520.446)(3420.000,1607.000)(3301.300,1572.212)
(3285.407,1528.128)(3331.635,1520.446)
\end{picture} \end{center}
} \caption{\footnotesize  Graphical image of the intertwining
 relation for the BS model showing the origin of the Bazhanov-Stroganov intertwining
matrix $\mathsf{S}$.
 The two elements $\wf_1$ and $\tilde{\wf}_1$ form the operator
 $L_1(\la)$, the elements $\wf_2$ and $\tilde{\wf}_2$ the operator $L_2(\la)$,
 see the earlier Fig. \ref{LL}. Periodicity in the third direction gives that
  the two $\wf_3$ at the top right and bottom left are the same.
  The two points where $\Rop_{123}$ and
 $\widetilde{\Rop}_{\widetilde{1}\widetilde{2}3}$ act are vertices of the
 physical 3-dim lattice.} \label{figureS} \end{figure}

We are interested in the relation of the action of $\;L_2(\xi)\:
L_1(\xi)\,$ to the action of $L_1(\xi)\: L_2(\xi)$. In the 3D
approach, $\Rop_{123}$ maps a triangle into a reflected one,
recall the left hand picture of \mbox{Fig. \ref{tri}}. Let us
consider the four Weyl operators in the auxiliary plane as in Fig.
\ref{LL} (the scalar $\kappa_j$ is included in $\wf_j$, see
(\ref{wkappa})) and a further variable $\wf_3$, see the bottom
auxiliary plane in Fig. \ref{figureS}.

We consider the special case of $\Rop_{123}$ mapping the triple
Weyl algebra $(\wf_1,\wf_2,\wf_3)$ into $(\wf_1',\wf_2',\wf_3')$
with the functional part of the transformation
$\:\Rop_{123}^{(f)}\;$ taken to be trivial (Bazhanov-Baxter case):
$u_i'=u_i,\;\:w_i'=w_i, \;\:i=1,2,3$. We act with a similar
mapping $\TRop$ on the initial triple Weyl algebra
$(\widetilde{\wf}_1,\widetilde{\wf}_2,\wf_3')$ and obtain the
triple algebra
$(\widetilde{\wf}_1',\widetilde{\wf}_2',\wf_3'')\,.$ As in
Fig.~\ref{LL} we demand periodicity after the second step in the
third direction: $\:\wf_3''=\wf_3.\:$ Fig. \ref{figureS}
illustrates these mappings.

In the upper auxiliary plane the action of the operators $L_1$ and
$L_2$ appears in the reversed order if we keep track of the
direction of the lines.

So, taking into account the property (\ref{frfr}) we expect an
intertwining relation of the form (\ref{SLL}), with $\:\mathsf{S}$
being a bilinear expression in $\R_{123}$ and
$\widetilde{\R}_{\widetilde{1}\widetilde{2}3}$.

\subsection{BS matrix $\;\mathsf{S}\;$ from ZBB model}

In order to derive the precise relation, let us calculate the
trace of the product of two matrices $\mathbf{R}$ and
$\widetilde{\mathbf{R}}$, both of the form (\ref{R-mat}), but the
first with the Fermat parameters $p_i=(x_i,y_i)$, the second with
$\widetilde{p}_i=(\widetilde{x}_i,\widetilde{y}_i)\,$,
$\:i=1,2,3,4\,$, each satisfying the restriction (\ref{xxoxx}):
\begin{equation}\label{S-mat}
\mathsf{S}_{i_1i_2,\ell_1\ell_2}^{j_1j_2,k_1k_2}=
\sum_{m,n\in\ZZ_N}\mathbf{R}_{i_1i_2m}^{j_1j_2n}\;\:
\widetilde{\mathbf{R}}_{\ell_1\ell_2n}^{k_1k_2m}.
\end{equation}
Actually, we shall not need the full $N^4\times N^4$ matrix
(\ref{S-mat}) but only those matrix elements in \r{S-mat} with
\begin{equation}\label{reduc}
i_1+\ell_1=i_2+\ell_2=j_1+k_1=j_2+k_2=0\,,
\end{equation}
forming a $N^2\times N^2$ matrix.
Inserting the explicit expressions (\ref{R-mat}) and renaming the
discrete variables we obtain \beq\label{S-mat1}\fl
\mathsf{S}_{\a_1\a_2}^{\b_1\b_2}\;=\;\omega^{(\a_1-\b_1)(\b_2-\a_2)}
\;\;\frac{w_{p_1}(\a_2-\a_1)w_{\tilde p_1}(\a_1-\a_2)}
{w_{p_3}(\b_2-\a_1)w_{\tilde p_3}(\a_1-\b_2)}\
\frac{w_{p_2}(\b_2-\b_1)w_{\tilde p_2}(\b_1-\b_2)}
{w_{p_4}(\a_2-\b_1)w_{\tilde p_4}(\b_1-\a_2)}\,. \eeq In order to
rewrite \r{S-mat1} in the form (\ref{BSS})
we use the following property of the cyclic functions
$w_p(n)$\footnote{Relation \r{property} is an analog of well known
relation $\;\Gamma(x)\Gamma(1-x)=\frac{\pi}{\sin\,\pi x}$, ~~see
e.g. \cite{SMS95,Hyper-Yang-Perk}.}:
\begin{equation}\label{property}
w_p(n)=\frac{1}{w_{Op}(-n)\:\Phi(n)}\,,
\end{equation}
where $O$ is an automorphism of the Fermat curve such that
\begin{equation}\label{auto}
p=(x,y) \mapsto Op=(\omega^{-1}x^{-1},\omega^{-1/2}x^{-1}y)
\end{equation}
and \\[-8mm]
\begin{equation}\label{Phi}
\Phi(n)=(-)^n\omega^{n^2/2}\,.
\end{equation}
So we get
\begin{equation}\label{S-mat2}
\begin{array}{l}
\mathsf{S}_{\a_1\a_2}^{\b_1\b_2}=\ds
\omega^{(\a_1-\b_1)(\b_2-\a_2)}\;
\frac{\Phi(\a_1-\b_2)\Phi(\a_2-\b_1)}{\Phi(\a_1-\a_2)\Phi(\b_2-\b_1)}
\\[5mm]
\quad \times\ds \frac{w_{\tilde p_1}(\a_1-\a_2)}{w_{O
p_1}(\a_1-\a_2)}\ \frac{w_{p_2}(\b_2-\b_1)}{w_{O\tilde
p_2}(\b_2-\b_1)}\ \frac{w_{O\tilde
p_3}(\b_2-\a_1)}{w_{p_3}(\b_2-\a_1)}\
\frac{w_{Op_4}(\b_1-\a_2)}{w_{\tilde p_4}(\b_1-\a_2)}\
\end{array}
\end{equation}
Now we  can identify
\begin{equation}\label{identify}
\begin{array}{l}
W_{pq'}(\al_1-\al_2)\equiv\ds \frac{w_{\tilde
p_1}(\a_1-\a_2)}{w_{O p_1}(\a_1-\a_2)}\,;\ \
 W_{p'q}(\beta_2-\beta_1)\equiv\ds
 \frac{w_{p_2}(\b_2-\b_1)}{w_{O\tilde p_2}(\b_2-\b_1)}\,;\\[5mm]
\overline{W}_{pq}(\beta_2-\al_1)\equiv\ds \frac{w_{O\tilde
p_3}(\b_2-\a_1)}{w_{p_3}(\b_2-\a_1)}\,;\ \
\overline{W}_{p'q'}(\beta_1-\al_2)\equiv\ds
\frac{w_{Op_4}(\b_1-\a_2)}{w_{\tilde p_4}(\b_1-\a_2)}\,,
\end{array}
\end{equation}
because of the trivial identity \beq\label{idi}
\;\;\om^{(\a_1-\b_1)(\b_2-\a_2)}\;\frac{\Phi(\a_1-\b_2)\Phi(\a_2-\b_1)}
          {\Phi(\a_1-\a_2)\Phi(\b_2-\b_1)}\;=\;1 .
\eeq The notation for the rapidities $p,q,p',q'$ which
parameterize the point on the Baxter curve \r{Bxteq} should not be
mixed up with that of the eight Fermat curve points $p_i,\;\tilde
p_i\;(i=1,2,3,4)\,.$
Using  \r{identify} and \r{Wpq}, \r{Fermat} we can write
\begin{equation}\label{ident1}\fl
\begin{array}{l}
\ds\frac{W_{p,q'}(n)}{W_{p,q'}(n-1)}\;=\;\ds \frac{\mu_p
y_{q'}}{\mu_{q'} y_p}\;
\frac{1-\omega^n(x_p/y_{q'})}{1-\omega^n(x_{q'}/y_p)}\; \equiv\;
\ds \frac{\omega^{1/2}\tilde y_1
x_1}{y_1}\;\:\frac{1-\omega^n(\omega x_1)^{-1}}
{1-\omega^n \tilde x_1 }\\[4mm]
\ds\frac{W_{p',q}(n)}{W_{p',q}(n-1)}\;=\;\ds \frac{\mu_{p'}
y_q}{\mu_q y_{p'}}\;
\frac{1-\omega^n(x_{p'}/y_q)}{1-\omega^n(x_q/y_{p'})}\; \equiv\;
\ds \frac{\omega^{1/2} y_2\tilde x_2}{\tilde y_2}\;\:
\frac{1-\omega^n(\omega \tilde x_2)^{-1}} {1-\omega^n  x_2 }
\end{array}
\end{equation}
and
\begin{equation}\label{ident2}\fl
\begin{array}{l}
\ds\frac{\overline W_{p,q}(n)}{\overline W_{p,q}(n-1)}\;\;\,=\;
\ds \frac{\om \mu_p\mu_q }{ x^{-1}_p y_q}\
\frac{1-\om^{n-1}(x_q/x_p)}{1-\om^n(y_p/y_q)} \;\;\;\:\equiv\;
\frac{\om^{-1/2}\tilde y_3}{y_3\tilde
x_3}\;\:\frac{1-\om^{n-1}(\om x_3)}
{1-\om^n (\omega\tilde x_3)^{-1} }\\[4mm]
\ds\frac{\overline W_{p',q'}(n)}{\overline W_{p',q'}(n-1)}\;=
\;\ds \frac{\om \mu_{p'}\mu_{q'}}{  x^{-1}_{p'}y_{q'}}\
\frac{1-\om^{n-1}(x_{q'}/x_{p'})}{1-\om^n(y_{p'}/y_{q'})}\;
\equiv\; \frac{\om^{-1/2}y_4}{\tilde y_4
x_4}\;\:\frac{1-\om^{n-1}(\om
  \tilde x_4)}
{1-\om^n (\om x_4)^{-1} }\,.
\end{array}
\end{equation}
From \r{ident1} and \r{ident2} we can read off the relation of the
Fermat parameters to the CP-variables on the Baxter curve:
\begin{equation}\label{ident4}
\begin{array}{llll}
\ds  x_1=\frac{y_{q'}}{\om x_{p}}\;;& \ds
x_2=\frac{x_{q}}{y_{p'}}\;;& \ds  x_3=\frac{x_q}{\om x_{p}}\;;&
\ds  x_4=\frac{y_{q'}}{\om y_{p'}}\;;\\
&\\
\ds  \tilde{x}_1=\frac{x_{q'}}{y_{p}}\;;& \ds
\tilde{x}_2=\frac{y_{q}}{\om x_{p'}}\;;& \ds
\tilde{x}_3=\frac{y_q}{\om y_{p}}\;;& \ds
\tilde{x}_4=\frac{x_{q'}}{\om x_{p'}}\,.
\end{array}
\end{equation}
Note that both  restrictions \r{xxoxx} are valid automatically for
the Fermat points $p_i$ and $\tilde p_i$. Comparing the
coefficients in front of the ratios  in \r{ident1} and \r{ident2}
we can identify the ratios $\tilde y_i/y_i$ with ratios of the
Baxter curve parameters \bea\label{ident5} \frac{\tilde
y_1}{y_1}=\omega^{1/2} \frac{\mu_{p} x_{p}}{\mu_{q'}
y_{p}}\;;\quad \frac{\tilde y_2}{y_2}=\omega^{-1/2}
\frac{\mu_{q} y_{p'}}{\mu_{p'} x_{p'}}\;;\ny\\
\frac{\tilde y_3}{y_3}=\omega^{1/2} \frac{\mu_{p}\mu_{q} x_{p}}{
y_{p}}\;;\quad \frac{\tilde y_4}{y_4}=\omega^{-1/2} \frac{
y_{p'}}{\mu_{p'}\mu_{q'} x_{p'}}\;. \eea From \r{ident5} we obtain
an important relation which connects the Fermat points $p_i$ and
$\tilde p_i$:
\begin{equation}\label{restriction1}
\frac{y_1y_2}{y_3y_4}=\frac{\tilde{y}_1\tilde{y}_2}{\tilde{y_3}\tilde{y}_4}\;.
\end{equation}

By taking the $N$-th powers of the formulas \r{ident5} one can see
that the cyclic property of the functions $w_p(n)$ \r{Fermat}
implies
the cyclic property \r{cycW} of the Boltzmann weights \r{Wpq}.\\[-6mm]

We conclude this discussion of the emergence of the
BS-$\mathsf{S}$-matrix within the ZBB-model with showing that the
parametrization of the $L$-operator postulated in
(\ref{condi-ident}) agrees with the identifications
(\ref{ident4}),(\ref{ident5}) made here. First we pass from the
parameters
$u_i,\,\tu_i,\,w_i,\,\tw_i,\,\kappa_i,\,\tilde{\kappa}_i\;$ used
in (\ref{lin3}) to the Fermat parameters by (\ref{fer-u}). Since
the functional transformation is taken to be trivial, we can omit
all primes on the $\;u_i\:$ and $\:w_i.$
 \bea\label{fer-utt}
x_1=\sqoi\frac{u_2}{\kappa_1u_1}\;;\quad
x_2=\sqoi\frac{\kappa_2u_2}{u_1}\;;\quad
x_3=\om^{-1}\frac{u_2}{u_1}\;;\ny\\
\frac{y_3}{y_1}=\frac{\kappa_1w_1}{u_3}\;;\quad
\frac{y_4}{y_1}=\sqoi\frac{\kappa_3 w_3}{w_2}\,. \eea The
counterpart with tildes is: \bea
\tilde{x}_1=\sqoi\frac{\tu_2}{\tilde{\kappa}_1\tu_1}\;;\quad
     \tilde{x}_2=\sqoi\frac{\tilde{\kappa}_2\tu_2}{\tu_1}\;;\quad
     \tilde{x}_3=\om^{-1}\frac{\tu_2}{\tu_1}\;;\ny\\
\frac{\tilde{y}_3}{\tilde{y}_1}=\frac{\tilde{\kappa}_1\tw_1}{u_3}\;;\quad
\frac{\tilde{y}_4}{\tilde{y}_1}=\sqoi\frac{\kappa_3 w_3}{\tw_2}\,.
      \label{tfer-u}\eea
Observe that there are no tildes on $\kappa_3,\;u_3$ and $w_3$.
(\ref{fer-utt}) and (\ref{tfer-u}) immediately give \bea
\label{ff}    \kappa_1=\sqo\frac{x_3}{x_1}\;;\quad
\tilde{\kappa}_1=\sqo\frac{\tilde{x}_3}{\tilde{x}_1}\;;\quad
\frac{w_1}{\tw_1}=
        \frac{\tilde{\kappa}_1\,\tilde{y}_1\,y_3}{\kappa_1\,y_1\,\tilde{y}_3}\;;\ny\\
\frac{u_2}{u_1}=\om x_3\;;\quad\frac{\tu_2}{\tu_1}=\om
\tilde{x}_3.          \label{LCP}\eea The first three equations of
(\ref{condi-ident}) follow by simply inserting from
   (\ref{ident4}) into the first three equations of (\ref{ff}).
The last two equations of (\ref{ff}) with (\ref{ident4}) lead to
\beq x_pu_2\,=\,x_qu_1\hq \mbox{and}\hq
y_p\tu_2\,=\,y_q\tu_1\,\;\hs \mbox{or}\hq
u_1\tu_1x_qy_q\:=\:u_2\tu_2x_py_p\,.\label{rescinv}\eeq
\r{rescinv} shows that the rescaling of the spectral parameter
$\;\xi^{-1}=-\om u_1\tu_1x_qy_q\,\la\:$ is the same for
$\:L_2(\la;p,p')\;$ as it is for $\:L_1(\la;q,q')$.

Summarizing, the relation of the parameters
$\:\kappa_1,\,\tilde{\kappa}_1,\,
\kappa_2,\,\tilde{\kappa}_2,\,\ds\frac{w_1}{\tilde{w}_1},\,\frac{w_2}{\tilde{w}_2}\;$
to the \ CP-parameters is: \bea \label{weylCP}
\kappa_1=\sqo\frac{x_q}{y_{q'}};\quad\tilde{\kappa}_1=\sqoi\frac{y_q}{x_{q'}};\hx
  &&\kappa_2=\sqo\frac{x_p}{y_{p'}};\quad\tilde{\kappa}_2=\sqoi\frac{y_p}{x_{p'}};\ny\\
\frac{w_1}{\tilde{w}_1}=\om^{-1}\frac{y_qy_{q'}}{x_qx_{q'}\mu_q\mu_{q'}};\hx&&
\frac{w_2}{\tilde{w}_2}=\om^{-1}\frac{y_py_{p'}}{x_px_{p'}\mu_p\mu_{p'}}\,.\eea

\subsection{Parametrization in terms of cross-ratios}\label{parametrization}

The intertwining matrix $\mathsf{S}$ defined by (\ref{BSS})
depends on 5 independent continuous parameters. One may use
several equivalent parameterizations:
\begin{enumerate} \item CP-parametrization: $\hs q,\ q',\ p,\ p',\ k\,.$
\item Fermat parametrization: $\;\;\;
x_1,\:x_2,\:x_3,\:\tilde{x}_1,\:\tilde{x_2},\:\tilde{x_3}\hq$ with
the constraint \r{restriction1}. \item Weyl-parametrization:$
\hq\!\kappa_1,\,\tilde{\kappa_1},\,\kappa_2,
\,\tilde{\kappa}_2,\,w_1/\tilde{w}_1,\,w_2/\tilde{w}_2,\hx\!$
again with one constraint.
\end{enumerate}
One obtains (ii) from (i) by \r{ident4} and \r{ident5}, while
(iii) follows from (i) by \r{weylCP}. In (iii) one may choose
$\;u_2/u_1, \:\tu_2/\tu_1\;$ instead of
$\;w_1/\tilde{w}_1,\:w_2/\tilde{w}_2\;$ as is seen from
\r{fer-utt},~\r{tfer-u}.
\\[1mm]
\indent In the following let us concentrate on the {\it
functional} transformations, which deal with only the $N$-th
powers of the variables. We start considering the Weyl
parametrization (iii). The functional transformations have been
solved in \cite{gps2}, first rewriting them in Hirota form and
then using standard methods of algebraic geometry. The results can
be read off from (65),~(66) of \cite{gps} and Table~1 of
\cite{gps2}. In the moment we are dealing with only the trivial
version of the functional transformations. Their solutions can be
obtained either by specializing Table~1 of \cite{gps2}, or
directly from \r{clas7} solving the system \bea
\;\;\;(K_2U_2-K_1U_1)\,W_2&=&(K_1U_1+U_2)\,K_3W_3\:;\ny\\[1mm]
\fl (W_1-K_3U_3)\,W_3\;=\;(W_1+U_3)\,W_2\:;&&\hs
(U_1-U_2)\,U_3\;=\;(K_1U_1+U_2)\,W_1\,\eea in terms of three pairs
of complex points which we shall call
$\;X',\;X,\;\;Y',\;Y;\;\;Z_0',\;Z_0.$ The solution is \bea
U_1\;=\,-\eps\,\frac{Y-Z_0'}{Y-Z_0};\hs\!
U_2\;=\,-\eps\,\frac{X-Z_0'}{X-Z_0};\hs\!\!
U_3\;=\,-\eps\,\frac{X-Y'}{X-Y};\ny\\[1mm]
W_1\;=\;\eps\,\frac{Y'-Z_0}{Y-Z_0};\hs
W_2\;=\;\eps\,\frac{X'-Z_0}{X-Z_0};\hs
W_3\;=\;\eps\,\frac{X'-Y}{X-Y};\ny\\
K_1\,=-\eeee{Y'}{Y}{Z_0'}{Z_0};\hx
K_2\,=-\eeee{X'}{X}{Z_0'}{Z_0};\hx
K_3\,=-\eeee{X'}{X}{Y'}{Y}\!,\label{Weylxyz}\eea where \beq
\eps\;=\;(-1)^N\,;\hs\hs\hs K_i\,=\;\kappa_i^N,\;\hs i=1,2,3,\eeq
and for cross-ratios we use the notation \beq
\eeee{A}{B}{C}{D}\;\equiv\;\frac{(A-C)\:(B-D)}{(A-D)\:(B-C)}\:.\eeq
There are analogous equations for the variables with tilde. We
solve these introducing another pair of complex points
$\;Z_1',\;Z_1\,.\;$ The expressions are the same as \r{Weylxyz},
only everywhere $\:\:Z_0',\:Z_0\,\:$ replaced by
$\:\:Z_1',\:Z_1\,,\:$ e.g.
$\;\widetilde{U}_1\;=\,-\eps\,(Y-Z_1')/(Y-Z_1)\,,\;etc.$ Observe
that the ratios appearing in (iii) can be written as cross-ratios:
\beq \frac{W_1}{\widetilde{W}_1}\;=\;\eeee{Y'}{Y}{Z_0}{Z_1};\hq\hs
     \frac{W_2}{\widetilde{W}_2}\;=\:\eeee{X'}{X}{Z_0}{Z_1},\eeq
so that all six variables in (iii) are expressible as cross-ratios
of eight complex points. Frequently, we shall use calligrafic
letters to denote {\it pairs} of points: \beq
\CX\:\equiv\:(X',X);\hq \CY\:\equiv\:(Y',Y);\hq
\CZ_0\:\equiv\:(Z_0',Z_0),\hq
\CZ_1\:\equiv\:(Z_1',Z_1).\label{pair} \eeq
Due to projective invariance, from these eight points five
independent cross ratios can be formed. Observe the index 1 is
associated to the pairs $\CY,\:\CZ$, the index 2 to the pairs
$\CZ,\:\CX$ and index 3 to $\CX,\:\CY$.

The Fermat parametrization (ii) is obtained from \r{Weylxyz} by
\r{fer-utt},~\r{tfer-u}: \bea \fl
x_1^N=\eeee{X}{Y'}{Z_0'}{Z_0}\,;\hx
x_2^N=\eeee{Y}{X'}{Z_0}{Z_0'}\,;\hx
x_3^N=\eeee{Y}{X}{Z_0}{\!Z_0'}\,;\hx
x_4^N=\eeee{X'}{\!Y'}{Z_0'}{\!Z_0}\,;\ny\\ \fl
y_1^N=\eeee{X}{Z_0'}{Y'}{Z_0}\,;\hx
y_2^N=\eeee{Y}{Z_0}{X'}{Z_0'}\,;\hx
y_3^N=\eeee{Y}{Z_0}{X}{Z_0'}\,;\hx
y_4^N=\eeee{X'}{Z_0'}{Y'}{Z_0}\,,\label{Fermat-div}\eea and
analogous formulas for $\tilde{x}_i$ and $\tilde{y}_i$ where the
pair $\CZ_0$ is replaced by $\CZ_1$. Observe that the Fermat curve
conditions \r{Fermat} and \r{xxoxx} are trivially satisfied.

We shall see now that also Baxter's modules $k^2$, $k'^2$ and the
$N$-th powers of the CP parameters $x_p$, $y_p$, etc. have similar
good expressions in terms of points $X,\,X',\ldots\,$ Since
\r{ident4} and \r{ident5} involve only ratios of the CP variables,
to solve e.g. for $\:x_p^N\:$ and $\:y_p^N\:$ we have to use
the Baxter curve relation \r{Bxteq}.\\
One may proceed as follows: From \r{weylCP} and \r{Weylxyz} get
$$ \lk\frac{\tilde{\kappa}_2\:\tw_2}{\kappa_2\:w_2}\rk^N=\lk\mu_p\mu_{p'}\rk^N
    =\eeee{X'}{X}{Z_1'}{Z_0'}.$$
Then use \r{Bxteq} to write
$$   k\,y_p^N\:=\:1\,-\frac{k'}{\mu_{p'}^N}\:\mu_p^N\mu_{p'}^N\:=\:
     1\,-\,\lk\mu_p\mu_{p'}\rk^N\lk 1\,-k\,x_{p'}^N\rk\,.    $$
Substitute here from \r{weylCP} $\;x_{p'}=\sqoi
y_p/\tilde{\kappa}_2\;$ and solve for $\:y_p^N\,:$
$$ k\,y_p^N\;=\;\frac{1\,-\,\lk\mu_p\,\mu_{p'}\rk^N}
         {1\,+\,\lk\mu_p\,\mu_{p'}\rk^N\!\tilde{K}_2^{-1}}. $$
and insert, using also \r{Weylxyz}. To obtain $\:x_p^N\:$ the same
procedure works. So we find all CP-variables:

\begin{equation}\fl\label{potts-divisors}
\begin{array}{c}
\ds k\,x_{p}^N=\eeee{X}{Z_1'}{Z_0}{Z_0'}\;;\hq
k\,y_{p}^N=\eeee{X}{Z_0'}{Z_1}{Z_1'}\;;\hq
k'\mu_p^N=\eeee{X}{Z_1}{Z_0'}{Z_1'}\;;\\[5mm]
\ds k\,x_{p'}^N=\eeee{X'}{Z_0'}{Z_1}{Z_1'}\;;\hq
k\,y_{p'}^N=\eeee{X'}{Z_1'}{Z_0}{Z_0'}\,;\hq
k'\mu_{p'}^N=\eeee{X'}{Z_0}{Z_1'}{Z_0'}\;;\\[5mm]
\ds k\,x_{q}^N=\eeee{Y}{Z_1'}{Z_0}{Z_0'}\;;\hq
\:k\,y_{q}^N=\eeee{Y}{Z_0'}{Z_1}{Z_1'}\;;\hq
k'\mu_q^N=\eeee{Y}{Z_1}{Z_0'}{Z_1'}\;;\\[5mm]
\ds k\,x_{q'}^N=\eeee{Y'}{Z_0'}{Z_1}{Z_1'}\;;\hq
k\,y_{q'}^N=\eeee{Y'}{Z_1'}{Z_0}{Z_0'}\;;\hq
\,k'\mu_{q'}^N=\eeee{Y'}{Z_0}{Z_1'}{Z_0'}\;.
\end{array}
\end{equation}
Each line of \r{potts-divisors} yields the same expression for
$\;\;k^2\,=k\,x_p^N\,+k\,y_p^N-k^2x_p^Ny_p^N\,:$
\begin{equation}\label{modules}
k^2=\eeee{Z_0}{Z_0'}{Z_1}{Z_1'}\hs \mbox{or}\hs
{k'}^2=\eeee{Z_0'}{Z_1'}{Z_0}{Z_1}\,.
\end{equation}
We also note that Baxter's rapidities $p$, $p'$, $q$, $q'$
correspond directly to the points $X$, $X'$, $Y$, $Y'\,,$
respectively,
while Baxter's module $k$ does not depend on these points.

Returning to the parametrization of $L$-operator \r{lin3}, in
\r{newWeyl} we found that it depends on three parameters
$\kappa_1$, $\tilde\kappa_1$ and $w_1/\tilde{w}_1$. These are
cross-ratios of six points $\:Y$, $Y'$, $Z_0$, $Z_0'$, $Z_1$ and
$Z_1'$. Due to projective invariance six points give rise to only
three independent cross-ratios. These correspond to the rapidities
$q$, $q'$ and the module $k$ in the Bazhanov-Stroganov
parametrization \r{BS-L-oper}.

Finally note that in this parametrization the combinations
encountered in \r{xi-la} and \r{gaug} depend only on $\,\CZ_0\,$
and $\,\CZ_1\;$: \beq U_1\,\widetilde{U}_1\:x_q^N\,y_q^N\;=\;
\frac{Z_1'-Z_0'}{Z_0-Z_1}\,;\hs
U_1\:x_q^N\;=\;-\epsilon\,\frac{Z_1'-Z_0'}{k\,(Z_1'-Z_0)}\,.\label{indep}
\eeq

\section{Classical BS-model and intertwining of its
                                 $L$-operators}\label{class-BS}

\subsection{Functional mapping on $N$-powers of Weyl
operators}\label{intertwining}

As has been mentioned before, for $\om^N=1$ the $N$th powers of
the Weyl operators are central and can be considered classical
variables. As in \r{Capcent} we use $\;(i=1,2,3)\,$: \beq
\label{clas0}\fl \Uc_i=\uop_i^N\,;\hq\;\; \Wc_i=\wop_i^N\,;\hq\;\;
\Utc_i=\tilde{\uop}_i^N\,;\hq\;\;
\Wtc_i=\tilde{\wop}_i^N\,;\hq\;\; \Kc_i=\kappa_i^N\,; \hq\;\;
\Ktc_i=\tilde{\kappa}_i^N\,,\eeq and $\;\Lambda\:=\:\xi^N\,. $

According to the approach developed in \cite{S,ps} we introduce a
classical analog of the linear problem \r{lin1} \beq\label{clas1}
\begin{array}{rcl}
0&=&\Psi_0- \Psi_1 \Lambda \Utc_1+
\Phi_0 \Wtc_1+ \Phi_1\Lambda \Ktc_1\Utc_1\Wtc_1\;\\[2mm]
0&=&\Psi_1- \Psi_0\Uc_1\;+\,\Phi_1 \Wc_1\,+\,\Phi_0 \Kc_1\Uc_1
\Wc_1\;
\end{array}\eeq
which can be rewritten in the matrix form \beq\label{clas2}
\Psi\:(\Lambda\,\Uc_1\,\Utc_1\,-\,1 )= \Phi\cdot
\mathcal{L}_1(\Lambda) \eeq and defines the classical $L$-operator
\beq\label{clas3} \ds \mathcal{L}_{1}(\Lambda) \;=\;
\left(\begin{array}{ccc} \ds \Wtc_1 + \Lambda
\Uc_1\Utc_1\Kc_1\Wc_1
&& \ds \Uc_1\lk \Wtc_1+\Kc_1 \Wc_1  \rk  \\[4mm]
\Lambda \Utc_1 \lk\Wc_1+ \Ktc_1 \Wtc_1  \rk &&  \Wc_1 +\Lambda
\Ktc_1\Uc_1\Utc_1\Wtc_1
\end{array}\right)\eeq
acting in the space of the linear variables $\Psi=(\Psi_0,\Psi_1)$
and $\Phi=(\Phi_0,\Phi_1)$. We take \r{clas3} to define a discrete
classical analog of the Bazhanov-Stroganov model. We can also
define \r{clas3} by an averaging prescription from \r{lin3}:
Define \cite{Tarasov}  \beq  \langle A(\,\xi^N\,)\,\rangle\;=\;
\prod_{i\,\in\:\mathbb{Z}_N}\:A(\,\xi\,\om^i\,)\,;\hq \left<
\lk\BAR{cc} A&B\\C&D\EAR\rk \right>\,=\,\lk\BAR{cc}
\langle A\rangle&\langle B\rangle\\
\langle C\rangle&\langle D\rangle\EAR\rk\,.\eeq
Then \beq \mathcal{L}_1(\Lambda)\;=\;\tilde{W}_1\,\cdot\,\left<\,
L_1(\xi)\,\right>.\label{znav}\eeq

Analogously, we introduce an operator $\mathcal{L}_2(\Lambda)$
such that the classical variables and parameters
$\;\Uc_1,\:\Utc_1,\:\Wc_1,\:\Wtc_1,\:K_1,\:\Ktc_1\;$ are replaced
by $\;\Uc_2,\:\Utc_2,\:\Wc_2,\:\Wtc_2,\:K_2,\:\Ktc_2\,.\;$ Let
$\mathcal{L}^\star_1(\Lambda)\:$ and
$\:\mathcal{L}^\star_2(\Lambda)\:$ again be $L$-operators of the
form \r{clas3}, but with the variables
\begin{equation}\label{clas4}
\Uc_i^\star\,,\quad \Wc_i^\star\,,\quad \Utc_i^\star\,,\quad
\Wtc_i^\star\,,\quad \Kc_i\,,\quad \Ktc_i\,,\quad\quad i=1,2.
\end{equation}
Our aim is now to find the transformation
$(\Uc_1,\Utc_1,\Wc_1,\ldots,\Wtc_2) \mapsto
(\Uc_1^\star,\,\Utc_1^\star,\Wc_1^\star,\ldots,\Wtc_2^\star)$
which solves the intertwining relation \beq\label{clas5}
\CL_2(\Lam)\:\CL_1(\Lam)\;=\;\CL^\star_1(\Lam)\:\CL^\star_2(\Lam)\,.\eeq
%
However, trying to find the nonlinear 8-variable mapping by direct
calculation without further guidance looks quite hopeless.
Fortunately, the 3D approach will provide a solution to this
problem \cite{S}.

\subsection{Solving the classical BS-intertwining relation via the
       3D functional transformation}\label{CLIT}

The mapping \r{clas5} we are looking for can be found using the
functional mapping of the vertex ZBB-model given in \r{clas7}
\cite{S}. We introduce two additional variables $\Uc_3$, $\Wc_3$
and the additional parameter $\Kc_3$ and consider the rational
mapping $\Rop_{123}^{(f)}$
\begin{equation}\label{clas6}
\Rop_{123}^{(f)}\;:\;\;\;\;\Uc_1,\Wc_1^{},\Uc_2^{},\Wc_2^{},\Uc_3^{},\Wc_3^{}
\;\;\mapsto\;\; \Uc_1',\Wc_1',\Uc_2',\Wc_2',\Uc_3',\Wc_3'\;
\end{equation}
given explicitly in \r{clas7}.
We define the composition of two of these rational transformations
\r{clas6}
\begin{equation}\label{clas8}
\Rop_{123}^{(f)}\;:\;\;\;\Uc_1,\Wc_1^{},\Uc_2^{},\Wc_2^{},\Uc_3^{},\Wc_3^{}
\;\;\mapsto\;\;
\Uc_1^\star,\Wc_1^\star,\Uc_2^\star,\Wc_2^\star,\Uc'_3,\Wc'_3\;,
\end{equation}
\begin{equation}\label{clas9}
\Rop_{\widetilde{1}\,\widetilde{2}\,3}^{(f)}
\;:\;\;\;\Utc_1,\Wtc_1^{},\Utc_2^{},\Wtc_2^{},\Uc'_3,\Wc'_3
\;\;\mapsto\;\;
\Utc_1^\star,\Wtc_1^\star,\Utc_2^\star,\Wtc_2^\star,
\Uc_3^\star,\Wc_3^\star\;,
\end{equation}
together with a periodic condition
\begin{equation}\label{clas10}
\Uc_3^\star=\Uc_3\,,\quad \Wc_3^\star=\Wc_3
\end{equation}
and denote this composition by \bea\label{clas11} S_{12}^{(f)}
\;:\;\;\;\Uc_1,\Wc_1^{},\Uc_2^{},\Wc_2^{},
\Utc_1,\Wtc_1^{},\Utc_2^{},\Wtc_2^{} \ny\\
\hspace*{5cm} \mapsto\;\;
\Uc_1^\star,\Wc_1^\star,\Uc_2^\star,\Wc_2^\star,
\Utc_1^\star,\Wtc_1^\star,\Utc_2^\star,\Wtc_2^\star\;. \eea In
\r{clas8} the constants $K_1,K_2,K_3$ have to be used, while in
\r{clas9} the constants are $\Ktc_1,\Ktc_2$ and $K_3$. With these
definitions we have
\begin{prop}\label{S-fun}
The rational transformation $S_{12}^{(f)}$ \r{clas11} solves the
mapping defined by the intertwining relations \r{clas5}.
\end{prop}

\noindent The {\em Proof}\ \ is provided by straightforward
calculation. We first determine $\Uc_3^\star,\;\Wc_3^\star$ in
terms of the variables $U_3,\;W_3$ and other variables from the
successive application of first \r{clas9} and then \r{clas8}.
Imposing the periodicity condition \r{clas10} gives two equations
which can be solved easily for the auxiliary variables $U_3$ and
$W_3$, leading to \beq
\begin{array}{rcl}
\Uc_3&=&\ds\frac {\Uc_1(\Ktc_1\Utc_1\Wtc_1 + \Kc_1\Utc_2\Wc_1)+
\Utc_2(\Uc_2\Wc_1 + \Uc_1\Wtc_1) } {\Uc_1 \Utc_1   - \Uc_2\Utc_2}
\\[4mm]
\Wc_3&=&\ds\frac{\Wc_2 \Wtc_2 (\Ktc_2 \Kc_2 \Utc_2\Uc_2 - \Ktc_1
\Kc_1\Utc_1  \Uc_1)} {\Kc_3 (\Uc_2(\Ktc_1 \Utc_1 \Wtc_2 +\Kc_2
\Utc_2 \Wc_2) + \Ktc_1\Utc_1 ( \Kc_1\Uc_1\Wtc_2  +  \Kc_2\Uc_2
\Wc_2))}
\end{array}
\label{double3}\eeq
 Then \r{double3} is used to eliminate $\:U_3\,$ and $\,W_3\,$ and we get
$$ \frac{\Uc_1^\star}{U_1}\;=\;\frac{\Utc_1}{\Utc_1^\star}\;=\;
\frac{\Ktc_1\Utc_1(K_1U_1\Wtc_2+K_2U_2W_2)
    +U_2(\Ktc_1\Utc_1\Wtc_2+K_2\Utc_2W_2)}
{\Kc_1\Uc_1(\Ktc_1\Utc_1\Wc_2+\Ktc_2\Utc_2\Wtc_2)
    +\Utc_2(\Kc_1\Uc_1\Wc_2+\Ktc_2\Uc_2\Wtc_2)}\,;$$
$$ \frac{\Uc_2^\star}{U_2}\;=\;\frac{\Utc_2}{\Utc_2^\star}\;=\;
\frac{\Uc_1(\Ktc_1\Utc_1\Wtc_1 +
\Kc_1\Utc_2\Wc_1)+\Utc_2(\Uc_2\Wc_1 + \Uc_1\Wtc_1) }
{U_2\Wtc_1(\Ktc_1\Utc_1+\Utc_2) + \Utc_1W_1(K_1U_1+U_2)}\,;
$$
\beq\frac{\Wc_1^\star}{W_1}\;=\;\frac{W_2}{W_2^\star}\;=\;-\frac{K_3}{W_1\Wtc_2}
  \:\frac{V_1}{V_0}\,;\hs
\frac{\Wtc_1^\star}{\Wtc_1}\;=\;\frac{\Wtc_2}{\Wtc_2^\star}\;=\;-\frac{K_3}{\Wtc_1
W_2}
  \:\frac{V_2}{V_0}\,; \label{double12}\eeq
where \bea \fl V_0&=&(K_1\Ktc_1
U_1\Utc_1-K_2\Ktc_2U_2\Utc_2)\:(U_1\Utc_1-U_2\Utc_2)\,;
\ny\\[3mm] \fl
 V_1&=&\Ktc_1 U_1\Utc_1^2\Lb U_2(W_1+\Ktc_1\Wtc_1)
          (\Wtc_2+K_2W_2)+K_1U_1W_1\Wtc_2\Rb \ny\\
\fl &+& K_2U_2\Utc_2^2\Lb U_1(\Wtc_1+K_1W_1)
    (W_2+\Ktc_2\Wtc_2)+\Ktc_2U_2W_1\Wtc_2\Rb \ny\\
\fl &+& U_1\Utc_1U_2\Utc_2\Lb
K_2W_1W_2(1+K_1\Ktc_1)+\Ktc_1\Wtc_1\Wtc_2(1+K_2\Ktc_2)\:+\,2\,\Ktc_1K_2\Wtc_1W_2\Rb,
\eea and $\;V_2\;$ is obtained from $\;V_1\;$ interchanging the
variables with tildes and without tildes.\\[2mm]
Finally, we simply insert into \r{clas5}. These are $2\times 2$
matrices with four entries and we compare the coefficients of
$\Lambda^0,\;\Lambda^1,\;\Lambda^2$ each, giving 12 equations
which turn out to be correct.
\hfill$\Box$\\[2mm]
Because of the periodic boundary conditions \r{clas10} the mapping
\r{clas11} has the invariants \beq\label{class12} \fl
\Uc_1\Utc_1=\Uc_1^\star\Utc_1^\star\,;\hq\;
\Uc_2\Utc_2=\Uc_2^\star\Utc_2^\star\,;\hq\;
\Wc_1\Wc_2\;=\;\Wc_1^\star\Wc_2^\star\,;\hq\;
\Wtc_1\Wtc_2\;=\;\Wtc_1^\star\Wtc_2^\star\,.\eeq Note that the
first two invariants in \r{class12} reflect the fact that the
combinations $\Uc_i\Utc_i$ are scale factors of the spectral
parameter, see (\ref{rescinv}). The last two invariants show that
a difference in the normalization of the classical and quantum
$L$-operators \r{clas3} and \r{lin3} is not important.

\section{Main result: Isospectral transform of the
                       BS transfer matrix.}\label{main}

We now consider the BS-quantum chain of length $Q$ defined by the
monodromy \bdm
\mathbf{M}(\xi)\;=\;L_0(\xi,u_0,\ldots,\tilde{\kappa}_0)\:
L_1(\xi,u_1,\ldots,\tilde{\kappa}_1)\:\ldots\:
L_{Q-1}(\xi,u_{Q-1},\ldots,\tilde{\kappa}_{Q-1})\edm and the
transfer matrix \beq \mathbf{T}(\xi)\;=\;{\rm
tr_{\CC^2}}\;\mathbf{M}(\xi) \label{transu}\eeq
where the Lax operators are defined by \r{lin3},~\r{newWeyl}: \bea
\fl \label{lina3} \ds
L_{n}(\xi;u_n,\tilde{u}_n,\frac{w_n}{\tilde{w}_n},\kappa_n,\tilde{\kappa}_n)&\!=\!&
\!\!\lk\begin{array}{lcl} \ds \!\!1 -\om^{1/2}\xi\,
u_n\,\tilde{u}_n\,\kappa_n\, \frac{w_n}{\tilde{w}_n}\:\Zop_n; &&
\ds \hspace*{-7mm}-u_n\Xop_n\lk
\!\!\om^{1/2}\!\!-\kappa_n\,\frac{w_n}{\tilde{w}_n}\,\Zop_n \rk\!
\\[4mm] \ds \!\!\xi\, \tilde{u}_n\,\Xop_n^{-1}\!\!
\lk\!\tilde{\kappa}_n\!-\om^{1/2}\frac{w_n}{\tilde{w}_n}\,\Zop_n\!
 \rk\!;  && \ds\hspace*{-5mm}
-\om^{1/2}\xi\, u_n\,\tilde{u}_n\,\tilde\kappa_n  +
 \frac{w_n}{\tilde{w}_n}\,\Zop_n\! \end{array}\!\!\rk\!\!.\label{LUW}\eea
The arguments
$\;u_n,\tilde{u}_n,w_n/\tilde{w}_n,\kappa_n,\tilde{\kappa}_n\,$ of
all $\:L_n\;\;(n=0,\ldots,Q-1)\;$ may be different and shall be
parameterized following \r{Weylxyz} in terms of $\:Q+2\:$ pairs of
points $\:\CY_n\:$ and $\:\CZ_0,\;\CZ_1\,:$
\bea u_n^N&=&-\eps\frac{Y_n-Z_0'}{Y_n-Z_0};\hs
\tilde{u}^N_n\;=\;-\eps\frac{Y_n-Z_1'}{Y_n-Z_1};\hs
\frac{w_n^N}{\tilde{w}_n^N}\;=\;\eeee{Y_n'}{Y_n}{Z_0}{Z_1};\ny\\
\kappa_n^N&=&-\eeee{Y_n'}{Y_n}{Z_0'}{Z_0};\hq\:
\tilde{\kappa}_n^N\;=\;-\eeee{Y_n'}{Y_n}{Z_1'}{Z_1}.\label{uwk-rat}\eea
In the equivalent formulation of the transfer matrix in \r{qtrans}
in terms of CP variables $\:x_{q_0},\:x_{q'_0},\ldots,k^2\:$ this
means that we take the Baxter modulus $\,k\,$ to be the same for
all $\,L_n\,$ but the two rapidities in each $\,L_n\,$ may be
different. The normalization adopted in \r{transu} differs from
~\r{transfer},~\r{qtrans}, see \r{transferU}.

The main result of this paper, which will be proven in the
remaining part of this section, is the following:
\begin{prop}\label{P2}
For a given Baxter modulus $k$ and a fixed set of\ $2Q$
rapidities, there exists a $Q-1$ parametric family of transfer
matrices with the same spectrum as the initial one. This family of
transfer matrices is defined by the same formulas
~\r{transu},~\r{LUW},~\r{uwk-rat}~ where the
matrices satisfy as usual
$\;\Xop_{n_1}\Zop_{n_2}\:=\:\om^{\delta_{n_1n_2}}\Zop_{n_2}\Xop_{n_1}\;,$
but without being normalized to $\:\Xop_n^N\:=\:\Zop_n^N\,=\,1\,$.
The form of the inhomogeneous centers is given by~ (\ref{inhom}).
\end{prop}
In order to state the results for $\:\Xop_n^N\,,\;\Zop_n^N\,,\;$
we have to consider the following system of algebraic equations
for the unknowns $\,P,\,P'\,$, with the pairs
$\,\CY_n,\,\CZ_0,\,\CZ_1$ being given: \bea
\prod_{n=0}^{Q-1}\;\eeee{P'}{P}{Y_n'}{Y_n}&=& 1\:;\label{psys1}\\
\prod_{i=0,1}\;\eeee{P'}{P}{Z_i'}{Z_i}&=& 1\:.\label{psys2} \eea
The system \r{psys1},~\r{psys2} has exactly $g=Q-1$ nonequivalent
solutions $\{(P_j',P_j^{})\}$, $j=0,...,g-1$, the pair
$(P_j',P_j^{})$ taken to be equivalent to the pair $(P_j^{},P_j')$.\\[-4mm]

Now, given a fixed set of $g$ pairs of complex numbers
$(P_0',P_0),\,\ldots,\,(P_{g-1}',P_{g-1})\,,$ we define the
function $H$ of a $\:g$-dimensional vector
$\;(f_0,\,f_1,...,f_{g-1})\,\equiv\, \lb f_j\rb\:,$ denoted
$H\lk\lb f_j\rb\rk\:$  (this arises as the rational limit of the
$\Theta$-function on a genus $g$ generic algebraic curve, see the
Appendix of \cite{SP-RTCh}) \footnote{These rational functions
appear in soliton theory (see, for example, \cite{Hirota}, $g$:
number of solitons), and in
3D integrable models \cite{S-Soliton}:\\[2mm]
$H(\;)=1;\hs H(f_0)=1-f_0;\hs
H(f_0,f_1)=1-\frac{P_1-P_0'}{P_1-P_0}f_0-\frac{P_0-P_1'}{P_1-P_0}f_1
+\frac{P_1'-P_0'}{P_1-P_0}\,f_0f_1;   \\[2mm]
H(f_0,f_1,f_2)=1-\frac{(P_0'-P_2)(P_0'-P_1)}{(P_0-P_1)(P_0-P_2)}\,f_0+\ldots
+\frac{(P_1'-P_2)(P_0'-P_2)(P_1'-P_0')}{(P_2-P_1)(P_2-P_0)(P_1-P_0)}\,f_0f_1+\ldots
\frac{(P_0'-P_1')(P_1'-P_2')(P_2'-P_0)}{(P_2-P_1)(P_2-P_0)(P_1-P_0)}\,f_0f_1f_2.$}
by \beq\label{fay2} H(\{f_j\})\;=\;\frac{\det|\,P_j^i\,-\,f_j
P_j^{\prime\, i}\,|_{i,j=0}^{g-1}}{\prod_{i>j}(P_i-P_j)}\;, \eeq
This is seen to be normalized to $H(\{0\})=1$. Let further
\beq\label{mfsig} \mf(Y)\:=\:\frac{P_j-Y}{P_j'-Y}\:\:f_j;\hs\hx
\sig_j(\CY)\:\equiv\:\sig_j(Y',Y)\:=\:\eeee{P_j'}{P_j}{Y'}{Y}\!,
\eeq recall \r{pair} $\:\;\CY\,=\,(Y',Y)\:$, and define
\beq\label{I-q}
I_j(n)\,=\,\prod_{i=0}^{n-1}\eeee{P_j'}{P_j}{Y_i}{Y_i'}\:
         =\:\prod_{i=0}^{n-1}\:\sig_j^{-1}(\CY_i)\,;\hs
I_j(0)\:=\:I_j(Q)\:=\:1\,. \eeq
Then, the spectrum of the transfer matrix defined by
\r{transu},~\r{lina3},~\r{uwk-rat}~ with
\begin{equation}\label{inhom}
\begin{array}{l}
\ds \Xop_n^N \;=\; \frac{H\lk\:\{\mf(Y_n)\; I_j(n)\}\:\rk}
{H\lk\:\{\mf(Y_n)\;I_j(n)\;\sig_j(\CZ_0)\}\:\rk}\;\:;\\
\\
\ds \Zop_n^N \;=\; \frac{ H\lk\:\{\mf(Z_0)\; I_j(n+1)\}\:\rk}{
H\lk\:\{\mf(Z_1)\; I_j(n+1)\}\:\rk} \:\cdot\:
\frac{H\lk\:\{\mf(Z_1)\;I_j(n)\}\:\rk}
{H\lk\:\{\mf(Z_0)\;I_j(n)\}\:\rk}
\end{array}
\end{equation}\vspace*{3mm}
does not depend on the set $\{f_j\}$. If $\{f_j\}=\{0\}$, then
(\ref{inhom}) reduces to $\:\Xop_n^N=\Zop_n^N=1\:$, and so the
spectrum is the same as for the initial transfer matrix.

In order to prove these statements we shall make extensive use of
the 3D-formalism which will allow an easy description of the
necessary intertwining operations. We shall derive a
generalization of the parametrization \r{uwk-rat} which will
involve the functions $H$ which depend on the parameters $\lb
f_j\rb\,.$ A key point will be that the product
$\;u_n\tilde{u}_n\;$ and the $\,\kappa_n,\;\tilde{\kappa}_n\;$
will not involve the $\lb f_j\rb,$ see eqs.\r{cent1}, and that in
\r{transu} $\,\Zop_n\,$ is always multiplied by (the $\lb
f_j\rb$-dependent) factor $\,w_n/\tilde{w}_n\,$ and $\,\Xop_n\,$
is multiplied by the $\lb f_j\rb$-dependent $u_n$. Details should
become clear as we proceed.

\subsection{Uniformization of the classical maps}

The map \r{clas11} describes the explicit relation between the
final ``star'' variables and the initial ``non-star'' variables.
According to \cite{S,gps2} this map, as well as map \r{clas7},
  can be parameterized in terms of algebraic geometry data.
  In this paper we will  consider the uniformization of these maps
using a specific set of the rational functions and identities
between them. This construction was exploited previously in
\cite{SP-RTCh}.

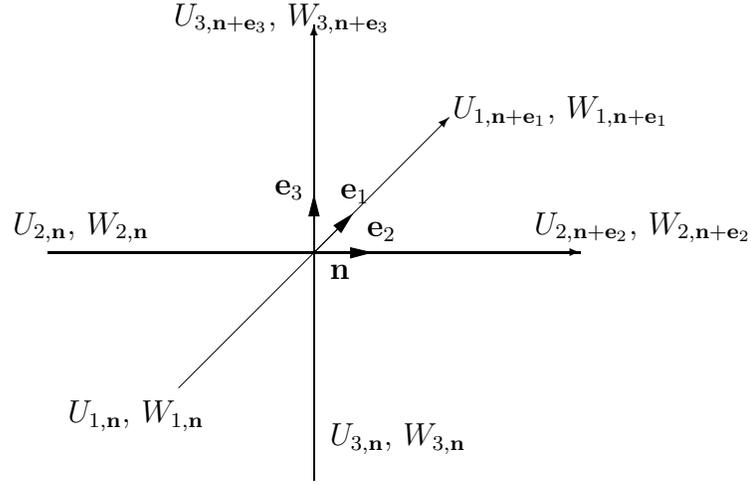
\begin{figure}
\setlength{\unitlength}{0.00045in}
\begin{center}
{\renewcommand{\dashlinestretch}{30}
\begin{picture}(6324,5960)(0,-10)
\path(3162,12)(3162,5412) \put(3162,12){\vector(0,1){5400}}
\put(12,2712){\vector(1,0){6300}}
\put(1562,1112){\vector(1,1){3200}} \path(3162,2712)(3162,3387)
\blacken\path(3222.000,3147.000)(3162.000,3387.000)
(3102.000,3147.000)(3222.000,3147.000) \path(3162,2712)(3837,2712)
\blacken\path(3597.000,2652.000)(3837.000,2712.000)
(3597.000,2772.000)(3597.000,2652.000) \path(3162,2712)(3612,3162)
\blacken\path(3484.721,2949.868)(3612.000,3162.000)
(3399.868,3034.721)(3484.721,2949.868)
{ \put(3350,2400){\makebox(0,0)[lb]{${\pot}$}}
\put(3792,2847){\makebox(0,0)[lb]{${\two}$}}
\put(3477,3297){\makebox(0,0)[lb]{${\one}$}}
\put(2700,3342){\makebox(0,0)[lb]{${\thr}$}}
\put(3350,327){\makebox(0,0)[lb] {${\Uc_{3,\pot},\,
\Wc_{3,\pot}}$}} \put(250,600){\makebox(0,0)[lb]
{${\Uc_{1,\pot},\,\Wc_{1,\pot}}$}}
\put(-400,2830){\makebox(0,0)[lb] {${\Uc_{2,\pot},\,
\Wc_{2,\pot}}$}} \put(5772,2830){\makebox(0,0)[lb]
{${\Uc_{2,\pot+\two},\, \Wc_{2,\pot+\two}}$}}
\put(4800,4200){\makebox(0,0)[lb] {${\Uc_{1,\pot+\one},\,
\Wc_{1,\pot+\one}}$}} \put(1500,5300){\makebox(0,0)[lb]
{${\Uc_{3,\pot+\thr},\, \Wc_{3,\pot+\thr}}$}} }
\end{picture}
} \caption{\label{fig4} {\footnotesize Association of the
dynamical variables to the edges of the cubic lattice. The vector
$\one$ is pointing into the paper plane.}}
\end{center}
\end{figure}

Consider a general three-dimensional lattice with classical
variables placed on its edges. Let $\pot=n_1\one+n_2\two+n_3\thr$
be marks for the vertices in this 3D lattice with discrete
coordinates $\,n_1,n_2,n_3.\;$ The assignment of the classical
variables to the links around a given vertex $\:\pot\,$ is shown
in Fig.~\ref{fig4}. In this notation the map \r{clas7} relating
the neighboring classical variables is
\begin{equation}\label{clas77}
\begin{array}{l}
\ds \frac{\Uc_{1,\pot+\one}}{\Uc_{1,\pot}}=
\frac{\Wc_{3,\pot+\thr}}
{\Wc_{3,\pot}}\\
\quad=\ds \frac{\Kc_{2:n_1,n_3}\Uc_{2,\pot}\Wc_{2,\pot}}
{\Kc_{1:n_2,n_3}\Uc_{1,\pot}\Wc_{2,\pot}+
\Kc_{3:n_1,n_2}\Uc_{2,\pot}\Wc_{3,\pot}+\Kc_{1:n_2,n_3}\Kc_{3:n_1,n_2}
\Uc_{1,\pot}\Wc_{3,\pot}}
\,;\\[4mm]
\ds \frac{\Wc_{1,\pot}}{\Wc_{1,\pot+\one}}=
\frac{\Wc_{2,\pot+\two}}{\Wc_{2,\pot}}=\ds
\frac{\Wc_{1,\pot}\Wc_{3,\pot}}
{\Wc_{1,\pot}\Wc_{2,\pot}+\Uc_{3,\pot}\Wc_{2,\pot}+
\Kc_{3:n_1,n_2}\Uc_{3,\pot}\Wc_{3,\pot}}
\,;\\[4mm]
\ds \frac{\Uc_{2,\pot+\two}}{\Uc_{2,\pot}}=
\frac{\Uc_{3,\pot}}{\Uc_{3,\pot+\thr}} =\ds
\frac{\Uc_{1,\pot}\Uc_{3,\pot}}
{\Uc_{2,\pot}\Uc_{3,\pot}+\Uc_{2,\pot}\Wc_{1,\pot}+\Kc_{1:n_2,n_3}
\Uc_{1,\pot}\Wc_{1,\pot}}\,.
\end{array}
\end{equation}
One may think about these relations as discrete equations which
describe the interrelation of the classical variables along the 3D
lattice.

The next step is to observe that after the change of variables
\bea\label{block1} \Uc_{1,\pot}&=&
-\eps\frac{Y_{n_2}-Z_{n_3}'}{Y_{n_2}-Z_{n_3}}\,
\frac{\tau_{2,\pot}}{\tau_{2,\pot+\thr}}\;;\quad
\Wc_{1,\pot}\;=\;\eps\; \frac{Z_{n_3}-Y'_{n_2}}{Z_{n_3}-Y_{n_2}}\,
\frac{\tau_{3,\pot+\two}}{\tau_{3,\pot}}\;;\ny \\[3mm]
\ds \Uc_{2,\pot}&=&-\eps
\frac{X_{n_1}-Z_{n_3}'}{X_{n_1}-Z_{n_3}}\,
\frac{\tau_{1,\pot}}{\tau_{1,\pot+\thr}}\;;\quad
\Wc_{2,\pot}\;=\;\eps\; \frac{Z_{n_3}-X_{n_1}'}{Z_{n_3}-X_{n_1}}\,
\frac{\tau_{3,\pot}}{\tau_{3,\pot+\one}}\;;\ny \\[3mm]
\ds \Uc_{3,\pot}&=&
-\eps\frac{X_{n_1}-Y_{n_2}'}{X_{n_1}-Y_{n_2}}\,
\frac{\tau_{1,\pot+\two}}{\tau_{1,\pot}}\;;\quad
\Wc_{3,\pot}\;=\;\eps\; \frac{Y_{n_2}-X_{n_1}'}{Y_{n_2}-X_{n_1}}\,
\frac{\tau_{2,\pot}}{\tau_{2,\pot+\one}}\;, \eea
together with the cross-ratio parametrization of the
$K_{i:n_jn_k}$ \beq\label{kappas}\fl
\Kc_{1:n_2,n_3}\!=\!-\eeee{Y'_{n_2}}{Y_{n_2}}{Z'_{n_3}}{Z_{n_3}}\!;\;
\Kc_{2:n_1,n_3}\!=\!-\eeee{Z'_{n_3}}{Z_{n_3}}{X'_{n_1}}{X_{n_1}}\!;\;
\Kc_{3:n_1,n_2}\!=\!-\eeee{X'_{n_1}}{X_{n_1}}{Y'_{n_2}}{Y_{n_2}}\!\!,
       \ny \eeq each of the relations in \r{clas77} can be written in the
form of a three-linear Hirota-type equation for the triple of
unknown functions $\;\tau_{\al,\pot}$, $\;\al =1,2,3\,$:
\begin{equation}\label{ttt}
\begin{array}{l}
\ds (X_\alpha-X_\beta)(X'_\beta-X'_\gamma)(X_\gamma-X_\alpha)
\tau_{\alpha,\pot+\mathbf{e}_\beta+\mathbf{e}_\gamma}
\tau_{\beta,\pot} \tau_{\gamma,\pot}\\[2mm]
\ds\hs +\;
(X_\alpha-X'_\beta)(X_\beta-X_\gamma)(X'_\gamma-X_\alpha)
\tau_{\alpha,\pot} \tau_{\beta,\pot+\mathbf{e}_\gamma}
\tau_{\gamma,\pot+\mathbf{e}_\beta} \\[2mm]
\ds \; =\;
(X_\alpha-X_\beta)(X'_\beta-X_\gamma)(X'_\gamma-X_\alpha)
\tau_{\alpha,\pot+\mathbf{e}_\beta}
\tau_{\beta,\pot+\mathbf{e}_\gamma} \tau_{\gamma,\pot} \\[2mm]
\ds\hs +\;
(X_\alpha-X'_\beta)(X_\beta-X'_\gamma)(X_\gamma-X_\alpha)
\tau_{\alpha,\pot+\mathbf{e}_\gamma} \tau_{\beta,\pot}
\tau_{\gamma,\pot+\mathbf{e}_\beta}\;,
\end{array}
\end{equation}
where $\{\al,\beta,\gamma\}$ is any even permutation of the set
$\{1,2,3\}$. The notations in \r{ttt} are related to those in
\r{block1} as follows: \beq
\CX_1=\CX_{n_1},\;\;\CX_2=\CY{\,'}_{\!\!n_2},\;\;\CX_3=\CZ_{n_3}\,,\hs\mbox{where}
\hs\CY{\,'}=(Y,Y')\,.\label{Id1x}\eeq

Now our goal is to describe the general solution to \r{ttt} in the
ring of the "rational $\Theta$-functions" $H$ defined in \r{fay2}.
The main tool to be used will be an identity, which is the
rational limit of the Fay identity ("rational Fay-identity"):

Let $A$, $B$, $C$, $D$ be any pair-wise different complex
parameters. Using the definitions given in
(\ref{fay2}),(\ref{mfsig}), the following identity is valid:
\bea\label{fay3}\fl
\lefteqn{H\lk\lb\fj\rb\rk\;H\lk\lb\fj\,\sj(A,B)\,\sj(C,D)\rb\rk\,=
\eeee{A}{B}{D}{C} H\lk\lb\fj\sj(A,B)\rb\rk
H\lk\lb\fj\sj(C,D)\rb\rk }\ny\\ &&\hs\hs+\; \eeee{A}{D}{B}{C}
H\lk\lb\fj\sj(A,D)\rb\rk\: H\lk\lb\fj\sj(C,B)\rb\rk \eea For a
proof of this identity see the Appendix to the paper
\cite{SP-RTCh}. In order to get the form of \r{ttt}, we combine
two such identities and obtain the more complicated "rational
double-Fay identity" \bea\label{fay33}\fl
\lefteqn{(X-Y)(Y'-Z')(Z-X)\;
H\lk\lb\mf(X)\,\sj(\CY)\,\sj(\CZ)\rb\rk
\;H\lk\lb\mf(Y)\rb\rk\; H\lk\lb\mf(Z)\rb\rk}\ny \\[1mm]
\fl&+&\: (X-Y')(Y-Z)(Z'-X)\; H\lk\lb\mf(X)\rb\rk
\;H\lk\lb\mf(Y)\sj(\CZ)\rb\rk\; H\lk\lb\mf(Z)\sj(\CY)\rb\rk\ny\\[1mm]
\fl&=&(X-Y)(Y'-Z)(Z'-X)\; H\lk\lb\mf(X)\sj(\CY)\rb\rk
\;H\lk\lb\mf(Y)\sj(\CZ)\rb\rk\;H\lk\lb\mf(Z)\rb\rk\ny\\[1mm]
\fl&+&\:(X-Y')(Y-Z')(Z-X) \;H\lk\lb\mf(Z)\sj(\CY)\rb\rk
\;H\lk\lb\mf(Y)\rb\rk \;H\lk\lb\mf(X)\sj(\CZ)\rb\rk\!.\eea\\[-3mm]
This identity involves five complex points $\,X,\,\CY,\,\CZ\,$ and
the sets $\lb f_j\rb$ and $\lb P_j,\:P_j'\rb$. Dividing by
$\:(X-Z')(Y-Z)(X-Y')(Z'-Y)/(Y-Z')\:$ the factors multiplying the
functions $H$ can be written as cross-ratios. The structure of
both \r{fay3} and \r{fay33} is precisely the same as that of the
corresponding identities for $\Theta$-functions, see e.g.
equations (21)--(24) of~\cite{gps2}.

Comparing \r{ttt} and \r{fay33}, we conclude that for arbitrary
$\lb f_j\rb$ the discrete equations \r{ttt} on the 3D cubic
lattice are solved by (we have to use ${\CY\,}'$ instead of $\CY$
in \r{fay33} which explains the inverse in the middle term of
\r{Ikn}):
$$\fl \tau_{1,\pot}\!=H\lk\lb\mf\lk
X_{n_1}\rk \sig_j(\CY_{n_2}) I_{j:\pot}\rb\rk;\hx
\tau_{3,\pot}\!=H\lk\lb\mf\lk Z_{n_3}\rk \sig_j(\CY_{n_2})
I_{j:\pot}\rb\rk, $$ \beq\label{g-s} \tau_{2,\pot}\!=H\lk\lb\mf\lk
Y'_{n_2}\rk \sig_j(\CY_{n_2}) I_{j:\pot}\rb\rk= H\lk\lb\mf\lk
Y_{n_2}\rk I_{j:\pot}\rb\rk\:, \eeq where we introduced
\begin{equation}\label{Ikn}
I_{j:\pot}\:=\: \prod_{m_1=0}^{n_1-1} \sig_j(\CX_{m_1})\;
\prod_{m_2=0}^{n_2-1}\sig_j^{-1}(\CY_{m_2})\;
\prod_{m_3=0}^{n_3-1}\sig_j(\CZ_{m_3})\,;\hq\;\; I_{j:\,{\mathbf
{\vec{0}}}}\;=\;1\,.
\end{equation}
\vspace*{2mm} This is the rational analog to solving trilinear
Hirota equations by use of the double Fay identity for
$\Theta$-functions \cite{S,gps2}.

Inserting \r{g-s} and \r{Ikn} into \r{block1} we get the
parametrization for all variables on the lattice: \bea
\Uc_{1,\pot}&=&-\eps\frac{Y_{n_2}-Z_{n_3}'}{Y_{n_2}-Z_{n_3}}\,
\frac{H\lk\lb\mf\lk Y_{n_2}\rk I_{j:\pot}\rb\rk} {H\lk\lb\mf\lk
Y_{n_2}\rk I_{j:\pot}\sig_j(\CZ_{n_3})\rb\rk}\,; \ny\\[2mm]
\Uc_{2,\pot}&=&-\eps\frac{X_{n_1}-Z_{n_3}'}{X_{n_1}-Z_{n_3}}\,
\frac{H\lk\lb\mf\lk X_{n_1}\rk I_{j:\pot}\sig_j(\CY_{n_2})\rb\rk}
{H\lk\lb\mf\lk
X_{n_1}\rk I_{j:\pot}\sig_j(\CY_{n_2})\sig_j(\CZ_{n_3})\rb\rk}\,; \ny\\[2mm]
\Uc_{3,\pot}&=&-\eps\frac{X_{n_1}-Y_{n_2}'}{X_{n_1}-Y_{n_2}}\,
\frac{H\lk\lb\mf\lk X_{n_1}\rk I_{j:\pot}\rb\rk} {H\lk\lb\mf\lk
X_{n_1}\rk I_{j:\pot}\sig_j(\CY_{n_2})\rb\rk}
 \,.\label{UWH}\eea
The $W_{i:\pot}$ are obtained from the $U_{i:\pot}$ interchanging
the arguments and removing the overall minus signs.

\subsection{The classical BS chain}

\begin{figure}
\setlength{\unitlength}{0.00064in}
{\renewcommand{\dashlinestretch}{30}
\begin{center}
\begin{picture}(8034,3800)(-590,-10)
\put(7040,1735){\makebox(0,0)[lb]{$(\CY_{Q-1};\CZ_0)$}}
\put(7040,3235){\makebox(0,0)[lb]{$(\CY_{Q-1};\CZ_1)$}}
\put(4200,1735){\makebox(0,0)[lb]{$(\CY_{1};\CZ_0)$}}
\put(4200,3235){\makebox(0,0)[lb]{$(\CY_{1};\CZ_1)$}}
\put(1500,1735){\makebox(0,0)[lb]{$(\CY_{0};\CZ_0)$}}
\put(1500,3235){\makebox(0,0)[lb]{$(\CY_{0};\CZ_1)$}}
\put(6250,75){\makebox(0,0)[lb]{$(\CX;\CY_{Q-1})$}}
\put(3550,75){\makebox(0,0)[lb]{$(\CX;\CY_1)$}}
\put(850,75){\makebox(0,0)[lb]{$(\CX;\CY_0)$}}
\put(-750,875){\makebox(0,0)[lb]{$(\CX;\CZ_0)$}}
\put(-750,2375){\makebox(0,0)[lb]{$(\CX;\CZ_1)$}}
\dottedline{45}(4800,1050)(5700,1050)
\dottedline{45}(4800,2550)(5700,2550) \path(6600,450)(6600,3450)
\blacken\path(6630.000,3330.000)(6600.000,3450.000)
(6570.000,3330.000)(6600.000,3294.000)(6630.000,3330.000)
\path(5700,1950)(7500,3150)
\blacken\path(7416.795,3058.474)(7500.000,3150.000)
(7383.513,3108.397)(7370.200,3063.467)(7416.795,3058.474)
\path(5700,450)(7500,1650)
\blacken\path(7416.795,1558.474)(7500.000,1650.000)
(7383.513,1608.397)(7370.200,1563.467)(7416.795,1558.474)
\path(5700,1050)(7500,1050)
\blacken\path(7380.000,1020.000)(7500.000,1050.000)
(7380.000,1080.000)(7344.000,1050.000)(7380.000,1020.000)
\path(5700,2550)(7500,2550)
\blacken\path(7380.000,2520.000)(7500.000,2550.000)
(7380.000,2580.000)(7344.000,2550.000)(7380.000,2520.000)
\path(3900,450)(3900,3450)
\blacken\path(3930.000,3330.000)(3900.000,3450.000)
(3870.000,3330.000)(3900.000,3294.000)(3930.000,3330.000)
\path(3000,450)(4800,1650)
\blacken\path(4716.795,1558.474)(4800.000,1650.000)
(4683.513,1608.397)(4670.200,1563.467)(4716.795,1558.474)
\path(3000,1950)(4800,3150)
\blacken\path(4716.795,3058.474)(4800.000,3150.000)
(4683.513,3108.397)(4670.200,3063.467)(4716.795,3058.474)
\path(300,450)(2100,1650)
\blacken\path(2016.795,1558.474)(2100.000,1650.000)
(1983.513,1608.397)(1970.200,1563.467)(2016.795,1558.474)
\path(300,1050)(4800,1050) \path(300,1950)(2100,3150)
\blacken\path(2016.795,3058.474)(2100.000,3150.000)
(1983.513,3108.397)(1970.200,3063.467)(2016.795,3058.474)
\path(300,2550)(4800,2550) \path(1200,450)(1200,3450)
\blacken\path(1230.000,3330.000)(1200.000,3450.000)
(1170.000,3330.000)(1200.000,3294.000)(1230.000,3330.000)
\put(-480,-200){\makebox(0,0)[lb]{${\large\two}$}}
\put(-680,300){\makebox(0,0)[lb]{${\large\one}$}}
\put(-1380,700){\makebox(0,0)[lb]{${\large\thr}$}}
\put(-1380,-300){\vector(1,0){1000}}
\put(-1380,-300){\vector(0,1){1000}}
\put(-1380,-300){\vector(3,2){800}} {\large
\put(-50,200){\makebox(0,0)[lb]{${U_{1,\zer}}$}}
\put(2750,200){\makebox(0,0)[lb]{${U_{1,\two}}$}}
\put(250,1100){\makebox(0,0)[lb]{${U_{2,\zer}}$}}
\put(2200,650){\makebox(0,0)[lb]{${U_{2,\two}}$}}
\put(2050,1300){\makebox(0,0)[lb]{${U_{1,\one}}$}}
\put(2200,2630){\makebox(0,0)[lb]{${U_{2,{\two+\thr}}}$}}
\put(250,2630){\makebox(0,0)[lb]{${U_{2,\thr}}$}} }
\put(6450,-300){\makebox(0,0)[lb]{$\CL_{Q-1}$}}
\put(3850,-300){\makebox(0,0)[lb]{$\CL_1$}}
\put(1150,-300){\makebox(0,0)[lb]{$\CL_0$}}
\end{picture}
\end{center}
} \caption{\footnotesize Slice of the 3-dimensional lattice
associated with the BS chain. The orientation of
$\large{\one,\;\two,\;\thr}$ is shown in the left lower corner.
The lattice is taken periodic after two steps in the
$\thr$-direction. The classical variables are assigned to the
links as indicated in Fig.~\ref{fig4}. In the Figure we show a few
of these variables where we just indicate the $U_{i,\pot}$
variables, the $W_{i,\pot}$ at the same links are not shown.
Observe that $\tilde{U}_{i,\pot}\;\equiv\;U_{i,\pot+\thr},\;\;
\tilde{W}_{i,\pot}\;\equiv\;W_{i,\pot+\thr}$.} \label{fig3}
\end{figure}
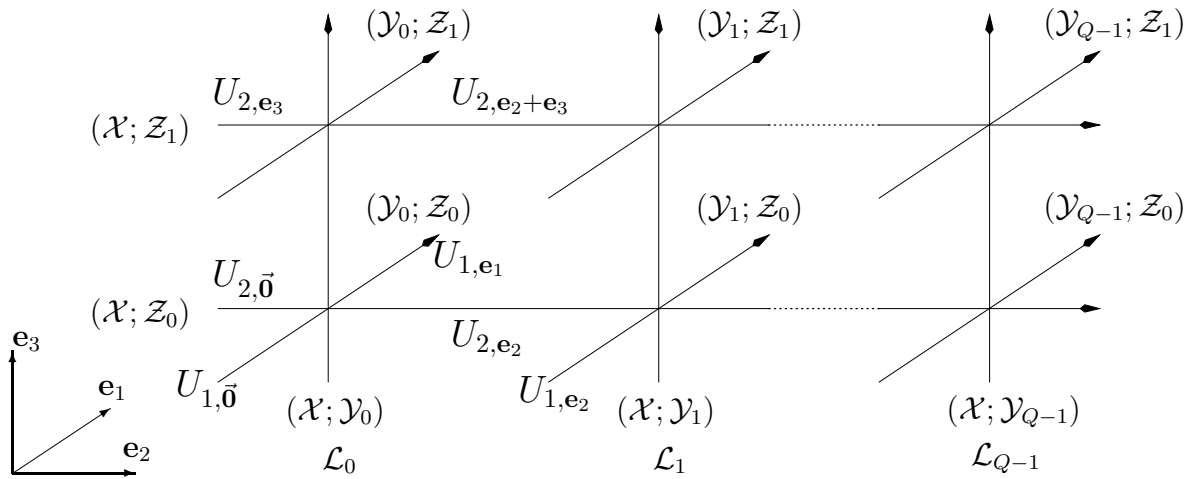

In the last subsection we considered the whole 3D lattice. In the
following, we shall be interested in applying this 3D formalism to
the BS chain defined by the product of operators $\CL_n$. The
chain will be taken in the $\two$-direction, see Fig.~\ref{fig3}.
As in Fig.~\ref{LL}, for each $\CL_n$ we have to consider a pair
of vertices, here on top of each other in the $\thr$-direction,
with periodic boundary conditions after two steps. Just one layer
with open b.c. is needed in the $\one$-direction.

To each line of the 3D lattice we associate two pairs of points as
shown in Fig.~\ref{fig3}: a line in direction $\one$ is labeled by
the two pairs $\CY_i,\:\CZ_j$. Classical variables are associated
to the links as in Fig.~\ref{fig4}. In the notation of \r{clas3}
we have $\tilde{U}_{i,\pot}\;\equiv\;U_{i,\pot+\thr},\;\;
\tilde{W}_{i,\pot}\;\equiv\;W_{i,\pot+\thr}$. The indices take the
values $\,n_1=0$, $\:n_2=0,1,\ldots,Q-1$, $\:n_3=0,1\,$ starting
from the left bottom.

Looking to the allocation of the variables to the links of the
lattice, the Lax operators associated to the lines $\CY_{n_2}$
will be taken to be of the form \r{clas3} and depending on the
index "1"-variables
$\:\Uc_{1,n_2\two},\:\Wc_{1,n_2\two},\:K_{1,n_2}\;$ and their
tilde counterparts. Only $n_2$ will occur. We drop the index $2$
and write $\:\CL_{1,n}\,.\:$ Explicitly:
\beq\label{classq} \fl \CL_{1,n}(\Lam)\:=\: \lk\begin{array}{cc}
\!\!\Wtc_{1,n\two} +\! \Lam
\Uc_{1,n\two}\Utc_{1,n\two}\Kc_{1:n}\Wc_{1,n\two}
& \Uc_{1,n\two}\lk \Wtc_{1,n\two}+\Kc_{1:n} \Wc_{1,n\two}\rk\\[4mm]
\!\!\!\!\Lam \Utc_{1,n\two} \lk\Wc_{1,n\two}+ \Ktc_{1:n}
\Wtc_{1,n\two} \rk &\!\!\Wc_{1,n\two}\! +\Lam
\Ktc_{1:n}\Uc_{1,n\two}\Utc_{1,n\two}\Wtc_{1,n\two}\!\!\!
\end{array}\rk\!\!.\eeq
Since these classical variables are solutions of \r{ttt},
according to \r{UWH} we can write them in terms of the rational
$\Theta$-functions $H$ defined in
\r{fay2}:  \\[-2mm]
\bea\label{cent1} \fl \lefteqn{
U_{1,n\two}\;=\,-\eps\frac{Y_n-Z_0'}{Y_n-Z_0}\,
\frac{H(\{\mf(Y_n)\: I_j(n)\})}
{H(\{\mf(Y_n)\:I_j(n)\sig_j(\CZ_0)\})}\,;\hq
\tilde{U}_{1,n\two}\:=\:\frac{(Y_n-Z_0')(Y_n-Z'_1)}{(Y_n-Z_0)(Y_n-Z_1)}
                 \;\frac{1}{U_{1,n\two}}\,;}\ny \\[2mm] \fl
 \frac{W_{1,n\two}}{\tilde{W}_{1,n\two}}&=&\eeee{Y_n'}{Y_n}{Z_0}{Z_1}
\frac{H(\{\mf(Z_0)I_j(n+1)\})}{H(\{\mf(Z_1)I_j(n)\})}
\frac{H(\{\mf(Z_0)I_j(n)\})}{H(\{\mf(Z_1)I_j(n+1)\})}\,;\ny\\[2mm]
\fl \hs\!\!\Kc_{1:n}&=&-\eeee{Y'_n}{Y_n}{Z'_0}{Z_0}\,;\hs
\tilde\Kc_{1:n}=-\eeee{Y'_n}{Y_n}{Z'_1}{Z_1}.   \eea Here $\:\lb
f_j\rb\:$ is an arbitrary set of $Q$ complex parameters. The
$\:I_j(n)\:$ are given in \r{I-q} and are related to the
$\I_{j:\pot}\;$ of \r{Ikn} by $\;I_j(n)\:=\:\I_{j:n\two}\:$.

We define the classical monodromy matrix, the classical
counterpart of \r{transu}, as: \beq\label{clas-mon}
  \CM_1(\Lam)\,=\,\lk\,\prod_{n'=0}^{Q-1}\tilde{W}_{1,n'\two}^{-1}\!\!\rk
  \CL_{1,0}(\Lam)\;\CL_{1,1}(\Lam)\:\cdots\:\CL_{1,n}(\Lam)\:\cdots\:
  \CL_{1,Q-1}(\Lam). \eeq
%
%
The simplest choice, still compatible with the functional mapping
relating neighboring variables, is to take all $\lb f_j\rb=\lb
0\rb$, resulting in all $H$ being unity. Then $\CM_1\,$ still
depends on the $Q$ pairs $\CY_n$, the two pairs $\CZ_0,\;\CZ_1$
and the spectral parameter $\Lambda$. Since all $\CY_n$ may be
chosen differently, in general also for $\lb f_j\rb=\lb 0\rb$ the
chain will be inhomogenous.

\subsection{Uniformization of the classical BS intertwining mapping.}

We like to establish an isospectrality transformation by commuting
an auxiliary Lax operator through the monodromy $\:\CM_1\:$
\r{clas-mon}. For this we shall first parameterize the
intertwining of {\it two} Lax operators considered in
Sec.~\ref{intertwining}, maps \r{clas6} and \r{clas11}, in terms
of cross ratios and $H$-functions. We use the results \r{UWH}
specialized to a single site, e.g. $\pot\:=\:\vec{0}\,.\:$ In this
subsection this index will be suppressed.

For a compact notation, we introduce the following functions
\bea\label{lin40}
\Uc\lk\lb\fj\rb,\,\CA,\,\CB\rk&=&-\eps\frac{A-B'}{A-B}\;\;
\frac{H(\{\mf(A)\})}{H(\{\mf(A)\,\sj(\CB)\})}\;;\ny\\[2mm]
\Wc(\{\fj\},\,\CA,\,\CB)&=&-\,\Uc\lk\lb\fj\rb,\,\CB,\,\CA\rk\,.\eea
For convenience we write $\CA$ as the argument of
$\:\Uc\:$ although it does not depend on $A'$.\\[-4mm]

Using \r{lin40}, equations \r{UWH} uniformize the mapping
\r{clas6} as follows:
 \bea\label{fun-map0}
 \Uc_1,\Wc_1&=&\Uc,\Wc(\{\fj\},\,\CY,\,\CZ)\;;\hs
 \Uc_2,\Wc_2\;=\;\Uc,\Wc(\{\fj\,\sj(\CY)\},\,\CX,\,\CZ)\;;\ny\\[3mm]
 \Uc_3,\Wc_3&=&\Uc,\Wc(\{\fj\},\,\CX,\,\CY)\;;\ny\\[3mm]
 \Uc'_1,\Wc'_1&=&\Uc,\Wc(\{\fj\,\sj(\CX)\},\,\CY,\,\CZ)\;;\hs\!\!\!
 \Uc'_2,\Wc'_2\;=\;\Uc,\Wc(\{\fj\},\,\CX,\,\CZ)\;;\ny\\[3mm]
 \Uc'_3,\Wc'_3&=&\Uc,\Wc(\{\fj\,\sj(\CZ)\},\,\CX,\CY)\;,
\eea with \beq \Kc_1=-\eeee{Y'}{Y}{Z'}{Z};\hq
\Kc_2=-\eeee{X'}{X}{Z'}{Z};\hq
\Kc_3=-\eeee{X'}{X}{Y'}{Y}.\label{Kdrei} \eeq

To do the same for the map \r{clas11} which describes the
intertwining of the $\CL$, we go back to
\r{clas8},\r{clas9},\r{clas10} and express these in terms of
\r{fun-map0}. We shall write only the equations for the
$\,U$-variables, the $\,W\,$ are
analogous, see \r{lin40}.\\[-4mm]

We describe the mapping $\Rop_{123}^{(f)}$ \r{clas8} by
\r{fun-map0},\,\r{Kdrei} with the parameters $\lb
f_j\rb,\,\CX,\,\CY,\,\CZ_0$: \bea\label{gun0}
 \fl&&\Uc_1\;=\;\Uc(\{\fj\},\,\CY,\,\CZ_0)\,;\hx\;\;
 \Uc_2\;=\;\Uc(\{\fj\,\sj(\CY)\},\,\CX,\,\CZ_0)\,;\hx
 \Uc_3\;=\;\Uc(\{\fj\},\,\CX,\,\CY)\,;\ny\\[3mm]
 \fl&&\Uc^\star_1\;=\;\Uc(\{\fj\,\sj(\CX)\},\,\CY,\,\CZ_0)\,;\hx
 \!\Uc^\star_2\;=\;\Uc(\{\fj\},\,\CX,\,\CZ_0)\,;\hx
 \!\Uc'_3\;=\;\Uc(\{\fj\,\sj(\CZ_0)\},\,\CX,\CY);\ny\\[2mm]
\fl &&\Kc_1=-\eeee{Y'}{Y}{Z_0'}{Z_0};\hq
\Kc_2=-\eeee{X'}{X}{Z_0'}{Z_0};\hq \Kc_3=-\eeee{X'}{X}{Y'}{Y}.
\eea The corresponding parameters for
$\Rop_{\widetilde{1}\,\widetilde{2}\,3}^{(f)}$ will be called $\lb
g_j\rb,\,\CX,\,\CY,$ and $\CZ_1$: \bea\label{gun1}
 \fl&&\Utc_1=\Uc\lk\lb g_j\rb,\CY,\CZ_1\rk\,;\hx
 \Utc_2=\Uc\lk\lb g_j\,\sj(\CY)\rb,\CX,\CZ_1\rk\,;\hx
 \Uc'_3=\Uc\lk\lb g_j\rb,\CX,\CY\rk\,;\ny\\[4mm]
 \fl&&\Utc^\star_1=\Uc\lk\lb g_j\,\sj(\CX)\rb\!,\CY,\CZ_1\rk;\hx
 \!\Utc^\star_2=\Uc\lk\lb g_j\rb,\CX,\CZ_1\rk\!;\!\hx
 \!\Uc^\star_3=\Uc\lk\lb
 g_j\,\sj(\CZ_1)\rb\!,\CX,\CY\rk;\ny\\[2mm]
\fl && \Ktc_1=-\eeee{Y'}{Y}{Z_1'}{Z_1};\hq
\Ktc_2=-\eeee{X'}{X}{Z_1'}{Z_1};\hq \Ktc_3=\Kc_3. \eea

Since $\:\Uc'_3\:$ of the first map \r{gun0} is the initial
variable in the second map \r{gun1} we must take \beq
g_j\;=\;f_j\:\sj(\CZ_0)\,.\label{gk}\eeq  Now the periodic
condition \r{clas10} $\;\Uc_3^\star\:=\:\Uc_3$ requires that
 \beq
g_j\:\sj(\CZ_1)\:=\:f_j\hs\mbox{or}\hs
\sj(\CZ_0)\,\sj(\CZ_1)\:=\:1\,,\label{szsz} \eeq which is equation
\r{psys2}. Inserting \r{gk} into
 \r{gun1} we get the uniformization of the composite map
$\;S_{12}^{(f)}\;$ \r{clas11} which solves the classical
intertwining relations \r{clas5}\bdm
\CL_2(\Lam)\:\CL_1(\Lam)\;=\;\CL^\star_1(\Lam)\:\CL^\star_2(\Lam)\,.\edm

\subsection{Auxiliary classical $L$-operator,
Isospectrality}\label{Iso-class}

We now introduce an auxiliary classical operator
$\,\CL_0^{aux}(\Lam)\,$ which by successive intertwining through
the monodromy $\,\CM_1\,$ \r{clas-mon} and imposing a periodic
condition will lead to an isospectral transformation of the
classical transfer matrix ${\rm tr}\: \CM_1(\Lam)$. Using the
notation \r{lin40} the arguments of its operators $\,\CL_{1,n}\,$
are
\beq  U_{1,n}\;=\;U\lk\,\lb\,
f_j\;\textstyle{\prod_{i=1}^n}\:\sig_j^{-1}(\CY_i)\,\rb,
\:\CY_n,\:\CZ_0\,\rk,\label{m1}\eeq
analogously $\;W_{1,n}\,,$ see \r{lin40}.
$\,\tilde{U}_{1,n},\,\tilde{W}_{1,n}\,$ are obtained replacing
$\,\CZ_0\rightarrow \CZ_1\,$ and $\,f_j\,\rightarrow\,
f_j^\star\:=\:f_j\sig_j(\CZ_0)\,.$
\\[2mm]
We start writing (omitting the argument $\Lam$ which will always
remain the same) \beq
\CL_0^{aux}\;\:\CL_{1,0}\;\CL_{1,1}\,\ldots\:\CL_{1,Q-1}
\;=\;\CL_{1,0}^{\star}\;\CL_1^{aux}\:\CL_{1,1}\ldots\:\CL_{1,Q-1}\,.
              \label{LLLL} \eeq
We shall achieve our goal using as auxiliary operator the operator
$\CL_{2,n}$, which is $\CL_{1,n}$ of \r{classq} with just the
first indices "1" replaced by "2" (the second index $n\two$ is not
modified). Then the intertwining in \r{LLLL} is the same as that
considered in the previous subsection and the arguments of the
various Lax operators are as follows (again, as in \r{m1}, we
write here only the $\,U_{i,j}\,$):
\bea\label{clas-para} \fl\CL_{1,0}:&\hq U_{1,0} \:=\:U\lk\,\lb
f_j\rb,\CY_0,\CZ_0\,\rk;\hq&\CL_{1,0}^\star:\hq\hx\,
\Uc_{1,0}^{\star}\:=\:U\lk\,\lb f_j\,\sig_j(\CX)\rb,\CY_0,\CZ_0\rk;\ny\\
\fl\CL_0^{aux}:&\hq \Uc_{2,0}        \:=\:U\lk\,\lb
f_j\sig_j(\CY_0)\rb,\CX,\CZ_0\,\rk;\hq
   &\CL_1^{aux}:\hq\;\; \Uc_{2,0}^\star  \:=\:\;U\lk\,\lb
   f_j\rb,\CX,\CZ_0\,\rk.\eea
Moving $\:\CL_1^{aux}(\Lam)\:$ one step further through the
monodromy matrix \r{clas-mon} we write
\bdm
\CL_{1,0}^\star\;\:\CL_1^{aux}\;\:\CL_{1,1}\;\CL_{1,2}\;\ldots\;\CL_{1,Q-1}
\;=\;\CL_{1,0}^{\star}\;\:\CL_{1,1}^\star\;\:\CL_2^{aux}\;\:\CL_{1,2}
\ldots\:\CL_{1,Q-1}\,.     \edm
Again using ~\r{gun1} and \r{gk}, the variables of the new
operators are: \bdm \fl \CL_{1,1}^\star:\:U_{1,\two}^{\star}=
U\lk\lb f_j\sig_j(\CX)/\sig_j(\CY_1)\rb,\CY_1,\CZ_0\rk\!;\;\;\;\;
\CL_2^{aux}:\Uc_{2,\two}^\star=U\lk\lb
f_j/\sig_j(\CY_1)\rb,\CX,\CZ_0\rk. \edm Continuing this way, we
finally arrive at \beq\label{Bac}
\CL_0^{aux}(\Lam)\:\CM_1(\Lam)\;=\;\CM_1^\star(\Lam)\:\CL_Q^{aux}(\Lam)\,,\eeq
where \\[-9mm]
\beq
\CM_1^\star\;=\;\lk\:\prod_{n'=0}^{Q-1}\;\tilde{W}_{1,n'\two}^{-1}\!\rk\,
\CL_{1,0}^\star\;\:\CL_{1,2}^\star\:\ldots\:
\CL_{1,n}^\star\:\ldots\:\CL_{1,Q-1}^\star\,\label{M1star}\eeq
and the operators in \r{Bac}, \r{M1star}~ have the following
arguments: \bea \CL_{1,n}^\star:\hs U_{1,n\two}^\star&=&U\lk\lb
f_j\:\sig_j(\CX)\:\textstyle{\prod_{i=1}^n}\:\sig_j^{-1}(\CY_i)\rb\!,
  \:\CY_n,\:\CZ_0\rk;\label{isos}\\[2mm]
\CL_{n}^{aux}:\hs U_{2,n\two}^\star&=&U\lk\lb
f_j\:\textstyle{\prod_{i=0}^{n-1}}\:\sig_j^{-1}(\CY_i)\rb\!,
  \:\CX,\:\CZ_0\rk.\label{isosp}\eea
We demand a periodic boundary condition for the auxiliary operator
$\CL_n^{aux}$: \beq\label{per-au}
\CL_Q^{aux}(\Lam)\;=\;\CL_0^{aux}(\Lam). \eeq This is fulfilled if
the constraint \r{psys1} is imposed, i.e. if \bdm
I_j(Q-1)\:\equiv\:\prod_{i=0}^{Q-1}\sig_j^{-1}(\CY_i)\;=\;1\,.\edm
Since the $\:\CL_n^{aux}\;$ depend on the points $\:\CY_n\:$ only
via $\:\lb f_j/\sig_j(\CY_{n-1})\rb\:$ and have the two other
arguments fixed to be $\CX,\;\CZ_0$ resp. $\CX,\;\CZ_1$, the
parameters $\Kc_2,\:\Ktc_2$ in the $\:\CL_n^{aux}$ are constant
along the chain and are the same as in \r{gun0} and \r{gun1}.

Taking the trace of \r{Bac} and using \r{per-au}, we see that the
transfer matrices
\beq {\rm Tr}\;\CM_1(\Lam)\:\hs \mbox{and}\hs {\rm
Tr}\;\CM_1^\star(\Lam)\:\hs \mbox{are~ isospectral}.
\label{isospec} \eeq
$\CM_1\:$ is composed of the Lax operators with the arguments
\r{m1}, the arguments of $\,\CM_1^\star$ are given in \r{isos}.~
The classical integrals of motion $\:\mathsf{T}_j\;$ given by the
generating series \beq \sum_{j=0}^Q\:\mathsf{T}_j\:\Lam^j={\rm
Tr}\;\: \CM_1(\Lam)\;=\; {\rm Tr}\;\,\CM_1^\star(\Lam)\eeq are
invariants of the isospectral transformation.

Summarizing: equation \r{psys2} results from the periodicity in
the $\thr$-direction, while \r{psys1} guarantees the periodicity
of the auxiliary operator in the $\two$-direction. Both together
determine the parameters $\:P_j\,$ and $\:P_j'\,.$
Due to the substitution $\:f_j\:\rightarrow\;f_j^\star\:=\:
f_j\sig_j(\CX)\:$ in the variables, compare \r{m1} with \r{isos},
there is a non-trivial isospectrality if some $\:f_j\,\neq 0\,.$

\subsection{Isospectral transformation of the BS quantum chain}

Finally, we generalize the isospectrality \r{isospec} of the
classical BS-model to the quantum BS-model defined by the
monodromy $\;\mathbf{M}(\xi)\:$ and transfer matrix
$\:\mathbf{T}(\xi)\;$ of \r{transu}. For this we have to find
explicit expressions for the operators $\;\mathbf{Q}\,,\:\;L_{\rm
aux}\:$ and $\:\mathbf{M}^\star\:$ in
\beq\label{qBac} \mathbf{Q}\;L_{\rm aux}(\xi)\cdot
\mathbf{M}(\xi)=\mathbf{M}^\star(\xi)\cdot L^\star_{\rm
aux}(\xi)\: \mathbf{Q}\,. \label{QLM}\eeq
Once this has been found, imposing periodicity $\;L_{\rm
aux}(\xi)\:=\:L^\star_{\rm aux}(\xi)\:$ and taking the trace over
both spaces $\CC^2$ and $\CC^N$, we will have the intertwining
relation
\beq\label{qBac1} \mathbf{K}\;
\mathbf{T}(\xi)\;=\;\mathbf{T}^\star(\xi)\: \mathbf{K}\,;\hs
\mathbf{T}\:=\:\mbox{Tr}_{\mathbb{C}^2}\;\mathbf{M}\,;\hs
\mathbf{K}\:=\:\Tr_{\CC^N}\, \mathbf{Q}\,.\eeq

We start considering the intertwining of {\it two} quantum Lax
operators \r{LUW}: \bea \label{QSLL} \lefteqn{
\mathbf{S}_{m\,n}\;L_m(\xi;\,u_m,\,\tilde{u}_m,\ldots,\,\tilde{\kappa}_m)\cdot
L_n(\xi;\,u_n,\,\tilde{u}_n,\ldots,\,\tilde{\kappa}_n)}\ny\\[1mm]
&&\hspace*{23mm}=\;
  L_n(\xi;\,u_n^\star,\,\tilde{u}_n^\star,\ldots,\,\tilde{\kappa}_n)\cdot
  L_m(\xi;\,u_m^\star,\,\tilde{u}_m^\star,\ldots,\,\tilde{\kappa}_m)\;\mathbf{S}_{m\,n}. \eea
The Lax operators are matrices both in $\:\CC^2\:$ and $\:\CC^N$.
Written in components, ~\r{QSLL} takes the same form as ~\r{SLL}.
We also remark that in ~\r{QSLL}, instead of $\;L_n\:$ of ~\r{LUW}
we could also use $\:L(\la;\,q_n,q_n')\:$ of ~\r{BS-L-oper} since
according to ~\r{rescinv} the gauge transformations of ~\r{Lxila}
cancel: $\;P_n^{-1}P_m\;=\;1\,.$

We want to find an explicit expression for $\;\mathbf{S}_{m\,n}\;$
for the case that the variables $\:u_m,\:u_n,\,etc.\;$ on the left
of \r{QSLL} are related to those on the right,
$\:u_m^\star,\:u_n^\star,\:etc.\,,\:$ by the functional mapping
\r{clas11}, i.e. by \r{gun0}--\r{szsz}, where we take
$\;m\leftrightarrow\, 2,\;n\leftrightarrow\, 1.$

If the functional mapping is chosen to be trivial, \r{QSLL}
reduces to the Bazhanov-Stroganov intertwining relation \r{SLL}
with $\:\mathbf{S}_{m\,n}\;$  equal to $\:\mathsf{S}\:$ as given
in \r{BSS},~\r{S-mat},~\r{reduc}. Then in the cross-ratio
parametrization the relation depends on the four pairs
$\;\CX,\:\CY,\:\CZ_0,\:\CZ_1\,.$ In the CP-parametrization one has
to use \r{potts-divisors}.
It appears that the parametrization of the intertwining operator
$\mathbf{S}$ in terms of the CP-functions $x_p$, $x_q$, $x_{p'}$
and $x_{q'}$ becomes inconvenient in the dynamical case since
\r{potts-divisors} and \r{modules} seem to have no simple
generalization to the case $\lb f_j\rb\neq \lb 0\rb$.

However, formulas \r{Fermat-div} for the Fermat points have a nice
dynamical extension. This can be used if we construct the quantum
intertwiner $\,\mathbf{S}\,$ from the 3D-operator
$\:\mathcal{R}_{123}\:$ imposing periodical boundary conditions.

Now we can generalize the derivation of section~\ref{CLIT} to the
quantum case. In \r{frfr} we have seen that since $\,\om^N=1\,,$
the map $\mathcal{R}_{123}$ acting in the space of rational
functions of Weyl operators $\,\Psi\,$ decomposes into the
functional mapping $\;\Rop_{123}^{(f)}\;$ and a similarity
transformation by the matrix $\mathbf{R}_{123}\,.$ The composition
of two such maps is
\begin{equation}\label{comp1}
\begin{array}{l}
\mathcal{R}_{\tilde{1}\tilde{2}3}\:\mathcal{R}_{123}\circ \Psi
\;=\;\mathcal{R}^{(f)}_{\tilde{1}\tilde{2}3}\circ\left(
\mathbf{R}_{\tilde{1}\tilde{2}3}\cdot
\mathcal{R}^{(f)}_{123}\circ\left( \mathbf{R}_{123} \cdot
\Psi\cdot \mathbf{R}_{123}^{-1} \right)\cdot
\mathbf{R}_{\tilde{1}\tilde{2}3}^{-1}
\right)\\
\quad=
\mathcal{R}^{(f)}_{\tilde{1}\tilde{2}3}\:\mathcal{R}^{(f)}_{123}\circ\left(
\left(\mathcal{R}^{(f)}_{123}\circ
\mathbf{R}_{\tilde{1}\tilde{2}3}\right)\cdot \mathbf{R}_{123}
\cdot \Psi \cdot \mathbf{R}_{123}^{-1}\cdot
\left(\mathcal{R}^{(f)}_{123}\circ
\mathbf{R}_{\tilde{1}\tilde{2}3}^{-1}\right) \right)
\end{array}
\end{equation}
where the conjugation matrices map
\begin{equation}\label{map8}
\R_{123}\;:\;\uop_1,\wop_1^{},\uop_2^{},\wop_2^{},\uop_3^{},\wop_3^{}
\;\;\mapsto\;\;
\uop_1^\star,\wop_1^\star,\uop_2^\star,\wop_2^\star,\uop'_3,\wop'_3\;
\end{equation}
and
\begin{equation}\label{map9}
\Rop_{123}^{(f)}\circ \R_{\tilde{1}\tilde{2}3}
\;:\;\uopt_1,\wopt_1^{},\uopt_2^{},\wopt_2^{},\uop'_3,\wop'_3
\;\;\mapsto\;\;
\uopt_1^\star,\wopt_1^\star,\uopt_2^\star,\wopt_2^\star,
\uop_3^\star,\wop_3^\star\;.
\end{equation}
Imposing the periodic boundary conditions
\begin{equation}\label{qpbc}
\uop_3^\star=\uop_3\,;\hs \wop_3^\star=\wop_3\,,
\end{equation}
which imply the classical ones \r{clas10}, we define the quantum
analog of the classical map \r{clas11} by
\begin{equation}\label{S-dyn}
\mathbf{S}_{12}={\rm tr}_3\ \left(\mathcal{R}^{(f)}_{123}\circ
\mathbf{R}_{\tilde{1}\tilde{2}3}\right)\cdot \mathbf{R}_{123}\,.
\end{equation}
An expression $\left(\mathcal{R}^{(f)}_{123}\circ
\mathbf{R}_{\tilde{1}\tilde{2}3}\right)$ means that the Fermat
point parameters which enter the matrix elements of the matrix
$\mathbf{R}_{\tilde{1}\tilde{2}3}$ in the third quantum space
should be taken after the functional mapping
$\;\mathcal{R}^{(f)}_{123}\;$ of \r{clas7} has been applied.

Formula \r{S-dyn} provides a generalization of the
Bazhanov-Stroganov intertwining operator \r{S-mat}. It performs a
canonical map of normalized Weyl operators
\beq\label{dyn-tran}
\frac{\uop_1^\star}{\uc_1^\star}\,=\,\mathbf{S}_{12}\cdot\frac{\uop_1}{\uc_1}
\cdot\mathbf{S}_{12}^{-1}\,;\hs
\frac{\uopt_1^\star}{\utc_1^\star}\,=\,\mathbf{S}_{12}\cdot\frac{\uopt_1}{\utc_1}
\cdot\mathbf{S}_{12}^{-1}\,;\quad\mbox{etc.} \eeq
The Fermat points $\:x_1,\,x_2,\,x_3$ determining the matrix
$\mathbf{R}_{123}$ in \r{S-dyn} are calculated from \r{fer-utt}
raised to the $N$-th power and then using \r{gun0}. Analogously
for $\:\mathbf{R}_{\tilde{1}\tilde{2}3}\:$ take the $N$-th power
of \r{tfer-u} and use \r{gun1},~\r{gk}.
The result is \bea\label{fer-soliton}
x_1^N&=&\eeee{X}{Y'}{Z_0'}{Z_0} \frac
{H\lk\lb\mf(X)\sig_j(\CY)\rb\rk\;H\lk\lb\mf(Y)\sig_j(\CZ_0)\rb\rk}
{H\lk\lb\mf(Y)\rb\rb\;H\lk\lb\mf(X)\sig_j(\CY)\sig_j(\CZ_0)\rb\rk}\,;
\ny\\ \tilde{x}_1^N&=&\ds
\eeee{Y'}{X}{Z_0}{Z_0'}\,\eeee{Y'}{X}{Z_1}{{Z_1}'} \;x_1^{-N},\hs
etc.\eea The $\,\mathbf{R}\,$ are determined by the $\,x_j\,$
rather than by the $x_j^N.\,$ So we must take $N$-th roots. The
possible choices of phases in this step have been discussed in
\cite{gps}.

The cancellation of the $\,H$-functions in the product
$x_1^N\tilde{x}_1^N$ arises as follows: The matrix
$\;\mathcal{R}^{(f)}_{123}\circ
\mathbf{R}_{\tilde{1}\tilde{2}3}\;$ has its Fermat parameters
given by the same formulas as $\mathbf{R}_{123}$, just with
$\:\CZ_0\:$ replaced by $\:\CZ_1\:$, and the vector $\:\fj\:$
replaced by $\;\tilde{\fj}=\fj\,\sj(\CZ_0).\;$ Then take into
account the periodicity
$\;\sig_j(\CZ_0)\,\sig_j(\CZ_1)\,=\,1\,,\;$ equation \r{psys2}.

Given $\:g,\;\lb f_j\rb$ and the four pairs
$\:\CX,\:\CY,\:\CZ_0,\:\CZ_1\,,$ from \r{fer-soliton} we obtain
the Fermat points which are then used to calculate the operator
$\mathbf{S}_{12}$ from \r{S-mat2} with \r{idi}. Suppressing to
indicate $\:g,\:\CZ_0,\:\CZ_1\:$ explicity, we shall write
\beq \mathbf{S}_{12}\;\equiv\;\mathbf{S}(\,\lb f_j\rb,\CX,\CY\,).
\label{SXY} \eeq

Finally, we consider the BS-quantum chain of length $Q$. The Lax
operators forming the monodromy are taken to be of the same form
as the operator $\,L_n\:$ in \r{QSLL}, with the scalar variables
as in \r{m1}. So, indicating the parameters, we write, recalling
\r{I-q}: \beq
L_n(\xi;\,u_n,\,\tilde{u}_n,\frac{w_n}{\tilde{w}_n},\,\kappa_n,\,
\tilde{\kappa}_n)\;\equiv\;L(\xi;\,\lb
 f_j\:I_j(n-1)\rb,\CY_n,\CZ_0,\CZ_1).\eeq
so that the monodromy is
 \bea \fl \mathbf{M}(\xi)\;&=&\;
L(\xi;\lb
 f_j\rb,\CY_0,\CZ_0,\CZ_1)\;\cdot\;L(\xi;\lb
f_j/\sig_j(\CY_1)\rb,\CY_1,\CZ_0,\CZ_1)\;\cdot\;\ldots\;\ny\\ \fl
&&\hspace*{50mm} \ldots\;\cdot\; L(\xi;\lb
f_j\,I_j(Q-1)\rb,\CY_{Q-1},\CZ_0,\CZ_1)\,.\label{Mbf} \eea
The auxiliary operator $\:L_{aux}\:$ is chosen to be of the same
form, like $\,L_m\,$ in \r{QSLL}, with the scalar variables taken
as for $\:\CL_0^{aux}\:$ in \r{clas-para}:
$$\;L_{aux}(\xi)\;\equiv\;L(\xi;\lb f_j\sig_j(\CY_0)\rb,\CX,\CZ_0,\CZ_1)\,.\:$$
Commuting the auxiliary operator through the chain proceeds in
analogy to the classical case of section \ref{Iso-class}, only in
each step there is also the conjugation by a matrix
$\:\mathbf{S}(\,\lb f_j I(n-1)\rb,\CX,\CY_n)\,.$ So
$\,\mathbf{M}^\star(\xi)\:$ is given substituting on the right
hand side of \r{Mbf} $\;f_j\rightarrow f_j\,\sig_j(\CX)\,,$ and we
have
\beq \fl \mathbf{Q}\:=\:\mathbf{S}(\lb
\,f_j\rb,\CX,\CY_0)\,\cdot\,\mathbf{S}(\lb
\,f_j/\sig_j(\CY_1)\rb,\CX,\CY_1)\,\cdot\ldots\,\cdot\,
\mathbf{S}(\,\lb f_j I(Q-1)\rb,\CX,\CY_{Q-1})\,.\label{QSQ}\eeq

Now all information necessary for the proof of Proposition
\ref{P2} has been collected: The matrix conjugation by
$\mathbf{K}$ does not change the spectrum so that we have to
consider just the functional transform. Indeed, as seen in
\r{cent1} $u_1^N\tilde{u}_1^N$ is independent of the $\lb f_j\rb$
and $u_1\Xop$ has the same $\lb f_j\rb$-dependence as
$\tilde{u}_1\Xop^{-1}$ and this is written in the first equation
of \r{inhom}. Also, since $\Zop_n$ appears only in the combination
$(w_n/\tilde{w}_n)\,\Zop_n\,,$ from \r{cent1} we confirm the
second line of \r{inhom}. The two conditions \r{psys1} and
\r{psys2} fixing the $P_j,\;P_j'$ have already been encountered in
\r{per-au} and \r{szsz}.

The operator $\mathbf{K}$ explicitly given by
\r{qBac1},\r{QSQ},\r{S-dyn} performs the isospectral
transformation of the BS quantum transfer matrix since according
to \r{qBac1} the spectrum of the transfer matrices
$\;\mathbf{T}(\xi;\{\,f_j\})\;$ and
$\;\mathbf{T}(\xi;\{\,f_j\,\sig_j(\CX)\,\})\;$ is the same.

\section{Conclusion}\label{discus}

This paper has been devoted to describe the Bazhanov-Stroganov
quantum chain using the tools of the 3D integrable generalized
Zamolodchikov-Baxter-Bazhanov model in the vertex formulation of
\cite{S}. The BS-Lax operator is constructed from the Linear
Problem \r{lipro} imposing periodicity after two layers. The BS
quantum intertwiner $\,\mathsf{S}\,,$ which is a product of four
chiral Potts Boltzmann weights, is obtained applying twice the
matrix conjugation part $\,\R_{123}\,$ of the 3D mapping operator
$\,\Rop_{123}$. The corresponding functional operator
$\,\Rop_{123}^{(f)}\,$ is used to solve the intertwining of two
classical BS Lax-operators, a relation which would be difficult to
obtain without the insight from 3D.

There are many possible parametrization of the 3D mapping operator
$\,\Rop_{123}$. These give rise to different more and less
convenient parameterizations of the BS-transfer matrix. The
parametrization in terms of cross-ratios and rational
$\Theta$-functions $H$ turns out to be the most useful and is
adopted for the explicit description of the isospectral
transformations of the classical and quantum BS-transfer matrices.
Whether our results form a sufficient preparation for solving the
problem of separation of variables for the BS-model, similarly to
what has been tried for other models in the papers
\cite{SP-RTCh,S-ZBB}, is subject of further work in progress.

\section*{Acknowledgments}
This work was partially supported by the grant
INTAS-OPEN-03-51-3350, CRDF grant RM1-2334-MO-02 and the
Heisenberg-Landau program. S.P. acknowledges the support from the
Japan Society for the Promotion of Sciences and the hospitality of
the Mathematical Department of Kyushu University and the
Max-Planck Institut f\"ur Mathematik (Bonn) where this work was
partially done. G.v.G. thanks the Department of Theoretical
Physics of the ANU Canberra for kind hospitality. The research of
S.S. was supported by the Australian Research Council.

\appendix
\setcounter{section}{1}
\section*{Appendix: Alternative definitions of the
Bazhanov-Stroganov $\;L\:$-operator.}

In this appendix we give the relation of two alternative
formulations of the BS model to our definitions \r{BS-L-oper}.

\subsection{Bazhanov-Stroganov model}
In the Bazhanov-Stroganov original paper \cite{BS} the
$\,L$-operator has been defined by $$  L^{BS}(p;q,q')=\rho_1 \lk
\BAR{cc}\!\!\!
  -c_pd_pb_qb_{q'}\Zop^{-\rho}+\om a_pb_pd_qd_{q'}\Zop^{\,\rho} \;&
  b_pd_p\lk-\om\,a_qd_{q'}\Zop^{\,\rho}+
  c_qb_{q'}\Zop^{-\rho}\rk \Xop^{-1}\!\!\!\!\!\!\\
  \om\,a_pc_p\lk d_qa_{q'}\Zop^{\,\rho}- b_qc_{q'}\Zop^{-\rho}\rk \Xop &
  \;\:\om \lk -c_pd_pa_qa_{q'}\Zop^{\,\rho}+ a_pb_pc_qc_{q'}\Zop^{-\rho}\rk
  \EAR\rk $$
where the Chiral Potts variables
$\;a_p,\,b_p,\,c_p,\,d_p,\;etc.\:$ are used, from which we obtain
the variables of \r{Wpq},~\r{Bxteq} by  $$ x_p=a_p/d_p;\hq
y_p=b_p/c_p;
 \hq \mu_p=d_p/c_p\,,\hq
\mbox{analogously with indices}\;\;\; p',\;q,\;q'\:.$$  For the
operators we use our notation $\:\Xop,\,\Zop\:$ of \r{Nrep}, so
that $\:\Xop\,\Zop\:=\,\om\,\Zop\,\Xop\,.$ In \cite{BS} the
$\ZZ_N$ operators are written $X$ and $Z$, which correspond to our
$\,\Zop^{-1}$ and $\Xop^{-1},$ respectively. Furthermore,
$\rho=(N-1)/2,$ and $\,\rho_1$ is a constant. Extracting some
factors, we get \beq \fl L^{BS}(p;\,q,q')=\:\rho_1\om
a_pb_pd_qd_{q'}\: \Zop^{\,\rho}\!\lk
 \BAR{cc} 1\;+\;\la_1\la_2\frac{y_qy_{q'}}{\mu_q\mu_{q'}}\Zop \, &
   \: \la_1 \Xop^{-1}\lk x_q -\frac{y_{q'}}{\mu_q\mu_{q'}}\:\Zop\rk \\
 \!\!\!\la_2 \Xop\lk \om x_{q'} -\frac{y_q}{\mu_q\mu_{q'}}\:\Zop\rk  &
 \,\la_1\la_2\,\om\,x_q x_{q'}\:
 +\frac{1}{\mu_q\mu_{q'}}\,\Zop\!\!
   \EAR\rk\eeq
where \\[-9mm] \beq \la_1\;=\;-\frac{1}{x_p};\hs
\la_2\;=\;\frac{1}{\om\,y_p}\,.\label{la12}\eeq
A simple gauge transformation by
$\;P=diag(\la_2^{1/2},\la_2^{-1/2})\;$ gives us the relation \beq
L^{BS}(p;\,q,q')\;=\:\rho_1\,\om a_pb_pd_qd_{q'}\;
\Zop^{\,\rho}\:P\;L(\la_1\la_2;\,q,q')\;P^{-1} \eeq with our
$\;L(\la;\,q,q')\;$ of \r{BS-L-oper}.

\subsection{$\tau^{(2)}(t_q)$-model}
In equation (5.33) of \cite{BS} a transfer matrix $\tau^{(2)}_k$
is introduced as the starting object for a fusion procedure.
$\tau^{(2)}_k$, which essentially is the transfer matrix of the BS
model \r{transfer}, has played a major role in solving the Chiral
Potts model \cite{BBP,CPM-sol,CPM-sol1,Baxt-sol}. Since in many
papers the definitions of \cite{BBP}
have been used, we give the relation to our \r{transfer}. \\
Equation (3.44)~ of \cite{BBP} is (as in \r{transfer}, we denote
the length of the system by $\,Q\,$): \bea
\Lb\tau_{k,q}^{(j)}\Rb_{\sig,\sig'}&=&\sum_{m_0,\ldots,m_{Q-1}=0}^{j-1}\;\:
  \prod_{J=1}^L\lb\om^{-m_J\lk\sig_{J+1}-\sig_J'+k\rk}\;
  \frac{\eta_{q,j,\sig_J-\sig_J'+k}}{\eta_{q,j,m_J}}\;\times\right.\ny\\[2mm]
 &\times &\left.
 F_{pq}(j,\sig_J-\sig_J'+k,m_J)\;F_{p'q}(j,\sig_{J+1}-\sig_{J+1}'+k,m_J)\rb\,;\ny\\[2mm]
\Lb\tau_{k,q}^{(j)}\Rb_{\sig,\sig'}&=&0\hs \mbox{if}\hq
j-k\le\sig_J-\sig_J'<N-k\hs\mbox{for any}\;\;J.
 \label{tau-BBP}\eea
The $\tau_{k,q}^{(j)}$ are transfer matrices leading from the
initial $Z_N$ spins $\;\sig=\lb\sig_0,\ldots,\sig_{Q-1}\rb\;$ to
the final spins $\;\sig'=\lb\sig_0',\ldots,\sig_{Q-1}'\rb\;$.
There is a fusion hierarchy $\;j=0,\ldots,N\;$ of the
$\;\tau_{k,q}^{(j)}\,$ which is exploited in the applications to
the CP. Of direct interest here is the case $\:j=2\,,\;N\ge 2\,$.
The index $k$ labels a redundancy, see (3.51) of \cite{BBP}, we
shall take $k=0$. Then the second part of \r{tau-BBP} tells us
that the $\Lb\tau_{0,q}^{(2)}\Rb_{\sig,\sig'}$ are non-vanishing
only if for all $J$ we have $\sig_J-\sig_J'=0\;\;\mbox{or}\;\:1.$
So at $J$ we can express the spin dependence in terms of the unit
matrix and the two standard matrices $\Xop_J$ and $\Zop_J$.
\\[1mm]\indent
Equation (3.37) of \cite{BBP} defines \\[-6mm]  \bdm \eta_{q,j,\al}\;=\;
\om^{j\,\al}\:t_q^{\,\al}\;\prod_{\ell=\al+1}^{\al+N-j}(1-\om^\ell)\hs\mbox{with}\hs
t_q=x_q\,y_q\, \edm from which here we need only
$\hq\eta_{q,2,1}/\eta_{q,2,0}\:=\:-\om\,t_q\,.\;$ Equation (3.38)
of \cite{BBP} gives   \bdm
F_{pq}(j,\al,m)\,=\,\mu_p^\al\,y_p^{-\al-m}\:\Phi(t_p,\om^\al
t_q)_m^{\al,\,j-\al-1} \hq\mbox{for}\hq 0\le\al,\:m<j\le\,N,\edm
where $\;\Phi(x,y)_i^{m,n}\:$ is a polynomial in $x,\:y$ which is
expressible in terms of Gauss polynomials. We need only (equations
(3.48) of \cite{BBP}): \bdm \fl F_{pq}(2,0,0)=1;\;\;
F_{pq}(2,0,1)=-\om\frac{t_q}{y_p}; \;\;
F_{pq}(2,1,0)=\frac{\mu_p}{y_p};\;\;
F_{pq}(2,1,1)=-\om\frac{x_p\mu_p}{y_p}\,.\edm Let us write \bdm
s_J=\sig_J-\sig_J';\hs \eta_i\:\equiv\:\eta_{q,2,i}\:=\:\lb
{1\atop-\om\,t_q}\right.\hq {i=0\atop \,i=1.} \edm Then, omitting
the first argument $\:j=2\:$ of $\,F(j,\ldots,\ldots)\,$ and the
index $\:k=0\:$ of $\;\:\tau_{k,q}^{(2)}$: \bea
\fl\Lb\tau^{(2)}_q\Rb_{\sig,{\sig}'}&=&\!\!\!\!\!
   \sum_{m_0,m_1,m_2\ldots=0,1}\!\!\!\!\!\!
\om^{m_0(\sig_0'-\sig_1)}
\frac{\eta_{s_0}}{\eta_{m_0}}\,F_{pq}(s_0,m_0)\:F_{p'q}(s_1,m_0)\,
\om^{m_1(\sig_1'-\sig_2)}
\frac{\eta_{s_1}}{\eta_{m_1}}\,F_{pq}(s_1,m_1)\times \ny\\ \fl
&&\hs\hs\times\:F_{p'q}(s_2,m_1)\:
 \om^{m_2(\sig_2'-\sig_3)}\;\frac{\eta_{s_2}}{\eta_{m_2}}
\:F_{pq}(s_2,m_2)\:F_{p'q}(s_3,m_2)\:\times\ldots\ny\eea We
collect all terms containing the spins $\:\sig_J,\;\sig_J'\;$ into
a $\:2\times 2$-matrix $\;\tau_J\,$:
\beq(\tau_J)_{m_{J-1},m_J}=\om^{m_J\sig_J'-m_{J-1}\sig_J}
      \frac{\eta_{s_J}}{\eta_{m_J}}F_{p'q}(s_J,m_{J-1})F_{pq}(s_J,m_J)\,,\label{tau-L}\eeq
so that for periodic boundary conditions the transfer matrix is
\beq \Lb\tau^{(2)}_q\Rb_{\sig,{\sig}'}= {\rm
Tr}\;\:\tau_0\:\tau_1\:\tau_2\,\ldots\,\tau_{Q-1}.\eeq
We now write \r{tau-L} in matrix form: \bea
\lk\tau_J\rk_{m_{J-1},m_J}
&=&\lk\BAR{cc} 1 & \om^{\sig_J'}/y_p\\ -\om^{-\sig_J}\om
t_q/y_{p'}&\;\;\; -\om t_q/(y_p
y_{p'})\EAR\rk_{m_{J-1},m_J}\!\!\!\delta_{\sig_J,\sig_J'} \ny\\[2mm]
&+&\om\,\frac{\mu_p\mu_{p'}}{y_p y_{p'}} \lk\BAR{cc}-t_q&-\om^{\sig_J'}x_p\\
\om^{-\sig_J}\om x_{p'}t_q&\;\;\;\om^{-s_J}\om x_p
x_{p'}\EAR\rk_{m_{J-1},m_J}\!\!\!\!\!\delta_{\sig_J',\sig_J-1}
\ny\\[3mm]
\fl&&\hspace*{-26mm}=\;\frac{1}{\la}\;\frac{\mu_p\mu_{p'}}{y_py_{p'}}
\lk\BAR{cc}1+\la\frac{y_py_{p'}}{\mu_p\mu_{p'}}\Zop_J;&\!\!\!
\!\!\!-\la\,\Xop_J\lk\om x_p-\frac{y_{p'}}{\mu_p\mu_{p'}}\Zop_J\rk\\
\!\!-\Xop_J^{-1}\!\!\lk
x_{p'}-\frac{y_{p}}{\mu_p\mu_{p'}}\Zop_J\rk;&\!\!\!\!\! \la\,\om\,
x_px_{p'}+\frac{1}{\mu_p\mu_{p'}}\Zop_J\EAR\!\!
\rk_{\!\!\!m_{J-1},m_J}\hspace*{-3mm}\Zop_J^{-1} \!,  \ny \eea
where we put $\;\la=-1/(\om t_q)\,,\;$ compare this with
$\:\la_1\la_2\:$ in \r{la12}. So (\,$L^T\:$ denotes the $\:2\times
2\:$ matrix transpose of $\;L\,$) \beq\fl
\lk\tau_J\rk_{m_{J-1},m_J}\,=\;\frac{1}{\la}\;\frac{\mu_p\mu_{p'}}{y_py_{p'}}
\;P\;L_J^T\lk\frac{-1}{\om t_q};\,p',\,p\,\rk
\,P^{-1}\Zop_J^{-1};\hs P\!=\!\lk\BAR{cc} \!\!i\la^{-1/2}&0\\
0&-i\la^{1/2}\!\!\EAR\rk\!\!.\label{BS-tau}\eeq Observe that
Baxter's variable $t_q$ corresponds to our spectral parameter,
hence the polynomial dependence of the transfer matrix on $t_q$.

\section*{References}
\bibliographystyle{amsplain}

\end{document}